\documentclass{article}
\usepackage{epsfig}
\usepackage[dvips]{color}
\usepackage{geometry}
\begin{document}

\newcommand{\ncd}{\newcommand}
\ncd{\x}{$\bullet\,\,\,\,$}
\ncd{\oo}{$\mbox{}\,\,\,\,\,\,\,$}
\ncd{\nil}{$\bigcirc$}
\ncd{\pu}{$\bullet$}
\ncd{\ua}{$\uparrow$}
\ncd{\ra}{$\rightarrow$}
\ncd{\ds}{\displaystyle}
\ncd{\dummy}{\mbox{\tiny{\textcolor[cmyk]{0.5,0,0,0}{+}}}}
\ncd{\CNOT}{\mbox{CNOT}}
\ncd{\QC}{$\mbox{QC}_{\cal{C}}\,\,$}
\ncd{\QCns}{$\mbox{QC}_{\cal{C}}$}
\ncd{\fc}{\mbox{fc}}
\ncd{\bc}{\mbox{bc}}

\newtheorem{definition}{Definition}

\title{Computational model underlying the one-way quantum computer}

\author{Robert Raussendorf \thanks{email:
    raussen@theorie.physik.uni-muenchen.de} \mbox{ }and 
    Hans J. Briegel \thanks{email:
    briegel@theorie.physik.uni-muenchen.de} \\
    Ludwig-Maximilians-Universit{\"a}t M{\"u}nchen, Germany}
  
\maketitle

\begin{abstract}
   In this paper we present the computational model underlying the
   one-way quantum computer which we introduced
   recently [Phys. Rev. Lett. {\bf{86}}, 5188 (2001)]. The one-way
   quantum computer has the property that any quantum logic network
   can be simulated on it. Conversely, not all ways of quantum
   information processing that are possible with the one-way quantum
   computer can be understood properly in  network model terms. 
   We show that the logical depth is, for certain algorithms,
   lower than has so far been known for networks. For example, every
   quantum circuit in the Clifford group can be performed on the one-way
   quantum computer in a  single step.  
\end{abstract}

\section{Introduction}
\label{Introduction}

Quantum computation has been formulated within different frameworks
such as the quantum Turing machine \cite{QTM} or the quantum logic
network model \cite{QLNW}. In the latter it is particularly easy to
establish a connection between physics and the processing of quantum
information, since the building blocks of the quantum logic network
--the quantum gates-- are unitary transformations
generated by suitably tailored Hamiltonians.
More recently, the capability of projective (von Neumann) measurements
to drive a quantum computation has been investigated
\cite{QFT}--\cite{Leu}. 

In \cite{QCmeas} we have shown that universal 
quantum computation can be entirely built on one-qubit measurements on
a certain class of highly entangled multi-qubit states, the cluster
states \cite{BR}. In this scheme, a cluster state forms a resource
for quantum computation and the set of 
measurements form the program. This scheme we called the ``one-way
quantum computer'' since the entanglement in a cluster state is
destroyed by the one-qubit measurements and therefore the cluster state
can be used only once. To stress the importance of the cluster state
for the scheme,  we will use in the following the abbreviation \QC for
``one-way quantum computer''. 

If a quantum logic network is simulated 
on a \QCns, for many algorithms the number of computational steps
scales  more favorably with the input size than the number 
of steps in the original network does. To be 
specific, circuits which realize transformations in the Clifford group
--which is generated by all the  CNOT-gates,
Hadamard-gates and $\pi/2$-phase shifts-- can be performed
by a \QC in a single time step, i.e. all the measurements to implement
such a circuit can be carried out at the same time. Generally, in a simulation 
of a quantum logic network by a one-way quantum computer, the temporal
ordering of the gates of the network is transformed into a spatial
pattern of measurement bases for the individual qubits on the resource cluster
state. For the temporal ordering of the measurements there is no
counterpart in the network model. Therefore, the question of
complexity of a quantum computation must be possibly revisited.
 
It should be understood that both the quantum logic
network computer and the \QC can simulate each other efficiently. The
fact that
each quantum logic network can be simulated on the \QC has been shown
in \cite{QCmeas}. The converse is also true because a resource cluster
state of arbitrary size can be created by a quantum logic network of
constant logical depth. Furthermore, the subsequent one-qubit measurements
are within the set of standard tools employed in the network
scheme of computation. In this sense, the \QC does not add physical means to
quantum computation. However, while the network model can describe the
means that are used in a 
computation on a \QCns, it cannot describe {\em{how}} they
have to be used. First, the
network for the creation of the cluster state does not tell
anything about the computational process 
itself, since the cluster state is a universal resource. 
Second, to the description of the 
computational process there belongs the temporal order in which the
measurements are performed. But the temporal ordering of measurements
to simulate quantum gates on the \QC is not pre-imposed by the
temporal ordering of these gates in the corresponding quantum logic
network. For example, in the network 
model two gates cannot be performed in parallel if
they do not commute. In the \QCns-realization they
still can if they both belong to the Clifford group. 
As a consequence, all circuits in the Clifford group can be parallelized
to logical depth $D=1$ on the \QCns, as mentioned before. Similarly,
all circuits which  
consist of CNOT-gates and either $x$- or $z$-rotations with variable
angle, can be parallelized to logical depth $D=2$, independent of the number
of logical qubits or gates. The best networks that have
been found  for the described circuits or at least for special cases
thereof have logarithmic depth in the number of qubits or gates, respectively
\cite{M&N}. 

These observations give rise to two questions: I) ``What is the adequate
computational model for the one-way quantum computer ?'', II) ``How
can the complexity of computations with the \QC be calculated?''. The
present  paper deals with these two questions. It 
describes the computational model underlying the \QC and provides the tools
by which the logical depth of algorithms on the \QC can be discussed
quantitatively. 

The paper is organized as follows.  Section \ref{summary}
contains a summary of the \QCns, as described in
\cite{QCmeas}. In Section \ref{beyond} the terminology required to
describe the computational 
model for the \QC is developed.  The central part of this paper is
Section \ref{about} where the 
computational model underlying the one-way quantum computer is
presented. In Section
\ref{depth}, the non-network character of the \QC is illustrated at the
hand of the temporal complexity for the above mentioned two special
classes of circuits. Section \ref{discussion} is the discussion of the
results, Section \ref{conclusion} the conclusion.

\section{Network picture of the \QC}
\label{summary}

Before starting the description of the objects relevant for the
computational model we would like to give a short summary of the
one-way quantum computer. In \cite{QCmeas} it was proved that the \QC is
universal by showing that it can simulate any quantum logic network. 

For the one-way quantum computer, the entire resource for the quantum
computation is provided  initially in the form of a specific entangled
state --the cluster state \cite{BR}-- 
of a large number of qubits. Information is then written onto the 
cluster, processed, and read out from the cluster by one-particle 
measurements only. The entangled state of the cluster thereby serves as a 
universal ``substrate'' for any quantum computation. Cluster states can be created 
efficiently in any system with a quantum Ising-type interaction (at very low 
temperatures) between two-state particles in a lattice
configuration. More specifically, to create a cluster state
$|\Phi\rangle_{\cal{C}}$ the qubits on a cluster ${\cal{C}}$ 
are at first all prepared in an individual state $|+\rangle = 1/\sqrt{2}
(|0\rangle + |1\rangle)$ and then brought into a cluster state by
switching on the Ising-type interaction $H_{\mbox{\footnotesize{int}}}$ for an
appropriately chosen finite time span $T$. The time evolution operator
generated by the Ising-type Hamiltonian which takes the initial
product state to the cluster state is denoted by $S$. The unitary
transformation $S$ on a two-dimensional array of qubits, as used in
this paper, has the form 
\begin{equation}
    \label{Sform}
    \begin{array}{rcl}
        S &=& e^{-i\pi \sum_{k\in {\cal{C}}} |0\rangle_k\langle0|
        \otimes |1\rangle_{r(k)}\langle1| + |0\rangle_k\langle0|
        \otimes |1\rangle_{u(k)}\langle1| }\\
          &=& \left( \prod \limits_{k \in {\cal{C}}}
        e^{-i\pi/2\,\sigma_z^{(k)}} 
        e^{i\pi/4\,\sigma_z^{(r(k))}} e^{i\pi/4\,\sigma_z^{(u(k))}}
        \right)  e^{-i\pi/4 \sum \limits_{k \in {\cal{C}}}
        \sigma_z^{(k)} \sigma_z^{(r(k))} +  \sigma_z^{(k)}
        \sigma_z^{(u(k))}}, 
    \end{array}
\end{equation}
where $r(k)$ denotes the site of the right neighbor of qubit $k$,
i.e. the site next following $k$ in the $x$-direction and $u(k)$ the
site of the upper
neighbor of $k$, i.e. the site next following $k$ in the
$y$-direction. The interaction part of $S$ is generated by an Ising
Hamiltonian. In (\ref{Sform}) there appear additional posterior local
unitary transformations 
which have no influence on the entanglement properties of the
states generated under $S$.
 
The quantum state $|\Phi\rangle_{\cal{C}}$, the cluster state of a 
cluster $\cal C$ of neighbouring qubits provides in advance all
entanglement that is involved in the subsequent quantum  
computation. It has been shown \cite{BR} 
that the cluster state $|\Phi\rangle_{\cal C}$ is characterized by a
set of eigenvalue  equations
\begin{equation}
    \sigma_x^{(a)} \bigotimes_{a'\in ngbh(a)}\sigma_z^{(a')} 
    |\Phi\rangle_{\cal C} = {(-1)}^{\kappa_a} |\Phi\rangle_{\cal C}, 
\label{EVeqn}
\end{equation} 
where $ngbh(a)$ specifies the sites of all qubits 
that interact with the qubit at site $a\in {\cal C}$. The eigenvalues
are specified by 
the distribution of the qubits on the lattice and encoded in
$\big\{\kappa_a \in \{0,1\}, a=1,..,\|{\cal{C}}\| \big\}$. They
can be altered e.g. by applying phase-flips $\sigma_z$ before or after
the Ising interaction. For the special case of ${\cal{C}}=
{\bf{Z}}^2$, that is the case of an infinitely extended cluster,
$\kappa_k=0 \; \forall k \in {\cal{C}}$. The equations
(\ref{EVeqn}) are 
central for the described computation scheme.  It is important to realize
that information processing is possible even though the result of
every individual measurement in any direction of the Bloch sphere is completely
random. The reason for the randomness of the measurement results is
that the reduced density operator for 
each qubit in the cluster state is $\frac{1}{2}{\bf{1}}$. While the
individual measurement results are irrelevant for the computation, the
strict correlations between measurement results inferred from 
(\ref{EVeqn}) are what makes the processing of quantum information
 on the \QC possible.

Let us for clarity emphasize that in the scheme of the \QC  we
distinguish between physical cluster qubits in 
${\cal{C}}$  which are measured in the process of computation, and the
logical qubits. The  
logical qubits constitute the quantum information being processed while
the cluster qubits in the initial cluster state form an entanglement resource.
Measurements of their individual one-qubit state drive the
computation.

To process quantum information with this cluster, it suffices to
measure its particles in a certain order and in a certain basis. Quantum 
information is thereby propagated through the cluster and
processed. Measurements of $\sigma_z$-observables effectively remove the
respective lattice qubits from the cluster. Measurements of $\sigma_x$
are used for ``wires'' i.e. to propagate logical quantum bits through the
cluster, and for the CNOT-gate between two logical qubits. Observables
of the form $\cos(\varphi)\,\sigma_x + \sin(\varphi)\, \sigma_y$ are
measured to realize arbitrary rotations of logical qubits. Here, the
angle $\varphi$ specifies the measurement direction.
For the one-qubit rotations, the basis in  which a
certain qubit is  measured depends on the results of preceding
measurements. This introduces a temporal ordering in which the
measurements have to be performed. The processing is finished once all qubits 
except a last one on each wire have been measured. The remaining
unmeasured qubits form the quantum register which is now ready for the
readout. At this point, the
results of previous measurements  
determine in which basis these ``output'' qubits need to be measured for the 
final readout, or if the readout measurements are in the $\sigma_x$-,
$\sigma_y$- or $\sigma_z$-eigenbasis, how the readout measurements have to be
interpreted. Without loss of generality, we assume in this paper that
the readout measurements are performed in the $\sigma_z$-eigenbasis. 

For illustration and later reference we review two points of the
universality proof for the \QCns. First, 
the realization of the arbitrary one-qubit rotation and of the
CNOT-gate as the elements of the
universal set of gates.  And second, the effect of the randomness of
the individual measurement results and how to account for them. 
  
\begin{figure}
    \begin{center}
        \begin{tabular}{cc}
            \multicolumn{2}{c}{{\large{(a)}} \hspace{0.3cm} \parbox{5.7cm}{
            \epsfig{file=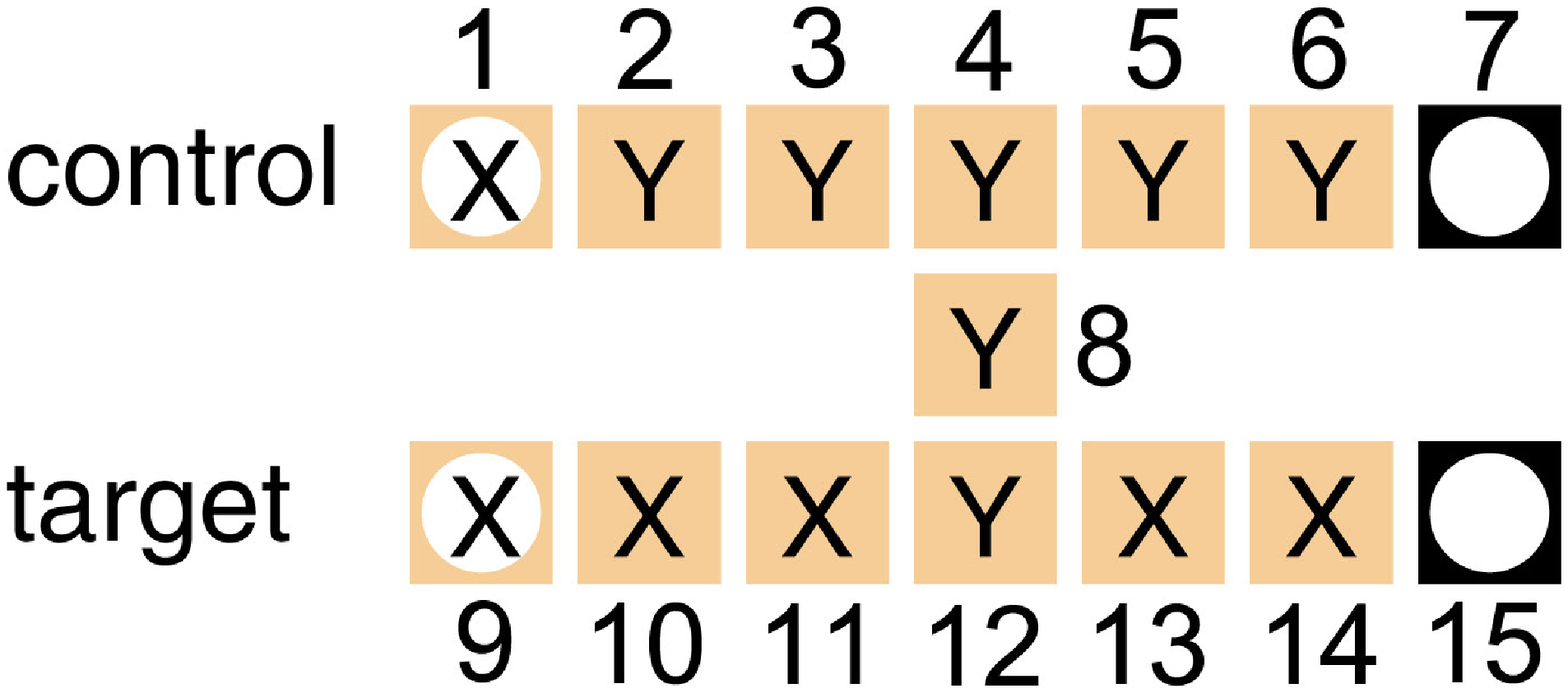, 
            width=5.7cm}}} \\ \multicolumn{2}{c}{CNOT-gate} \\ & \\
            \parbox{3.2cm}{{\large{(b)}} \hspace{0.3cm}
            \epsfig{file=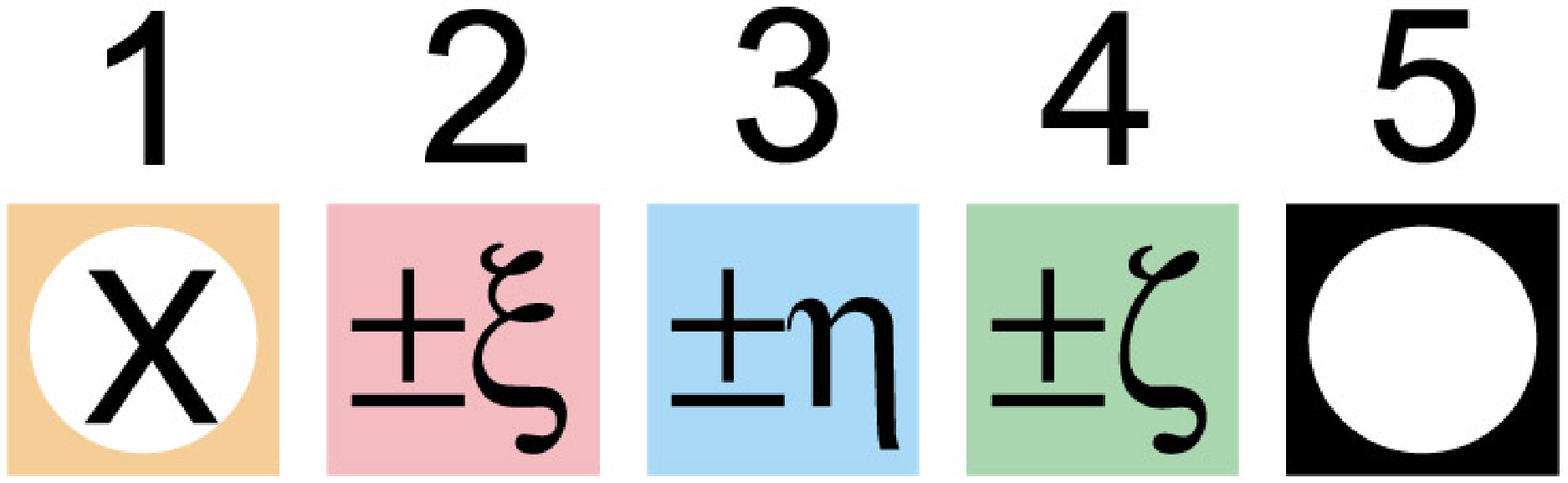,width=3.2cm}} & 
            \parbox{3.2cm}{{\large{(c)}} \hspace{0.3cm}
            \epsfig{file=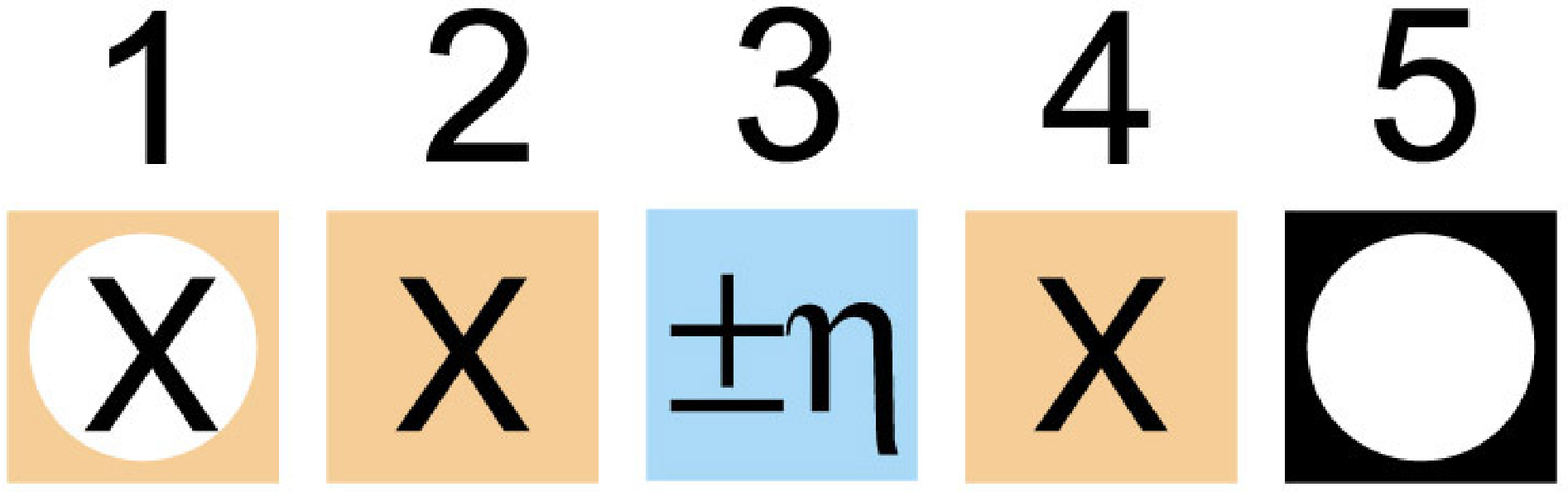,width=3.2cm}} \\ 
            general rotation & $z$-rotation \\ & \\
            \parbox{3.2cm}{{\large{(d)}} \hspace{0.3cm}
            \epsfig{file=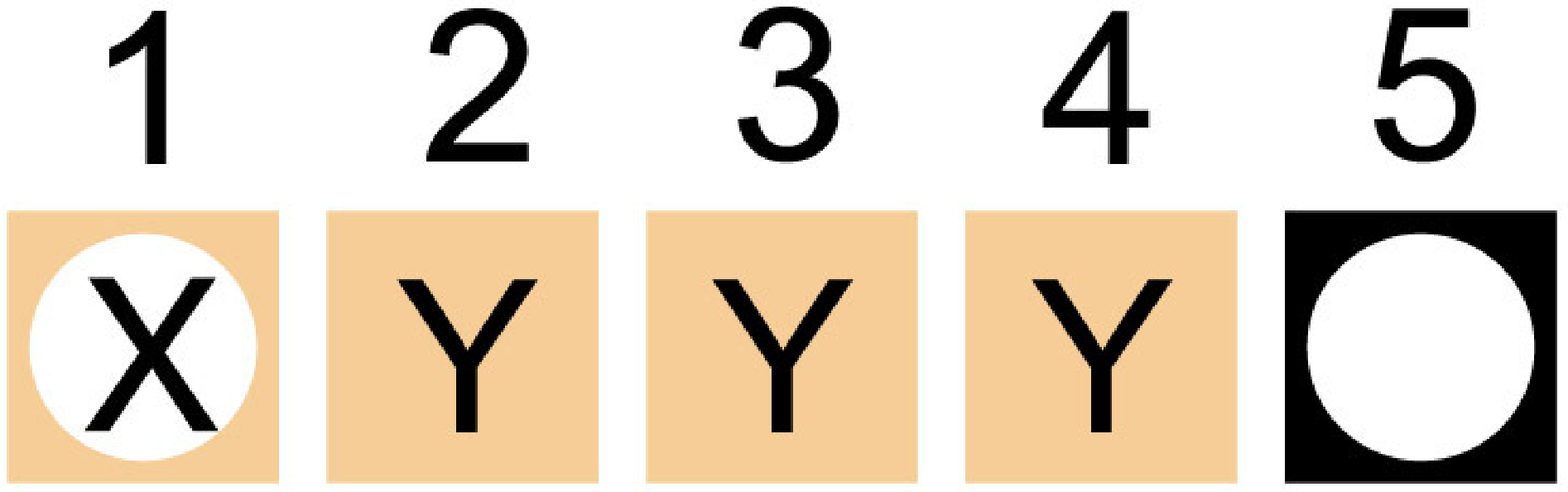,width=3.2cm}} & 
            \parbox{3.2cm}{{\large{(e)}} \hspace{0.3cm}
            \epsfig{file=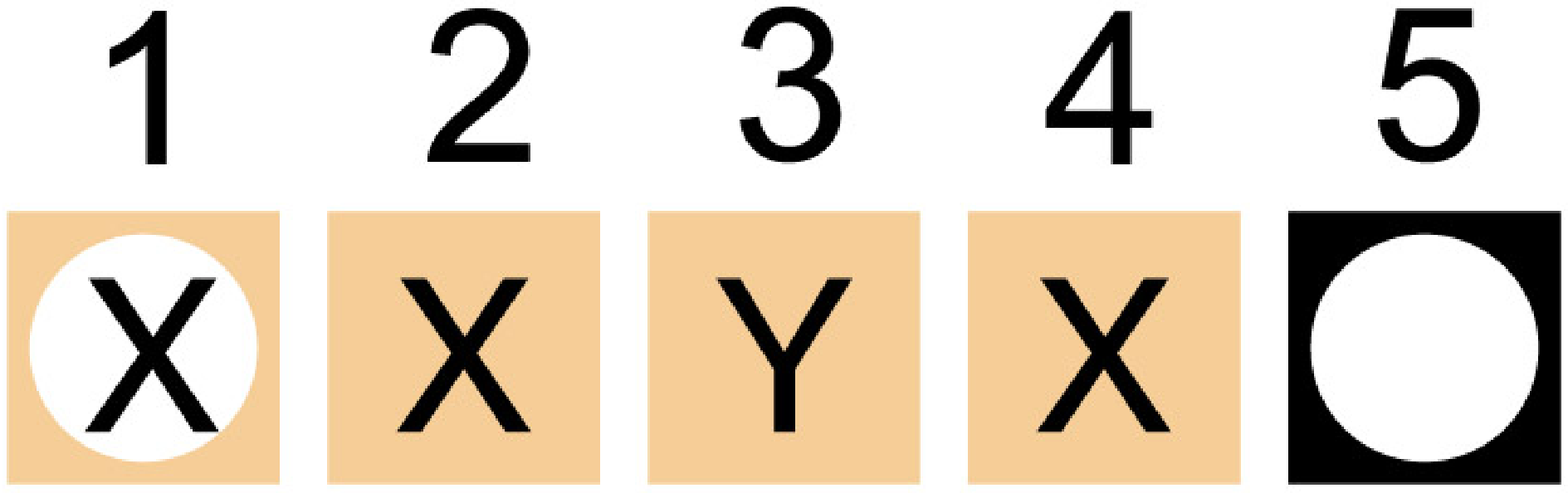,width=3.2cm}} \\ 
            Hadamard-gate & $\pi/2$-phase gate \
        \end{tabular}
        \parbox{0.8\textwidth}{\caption{\label{Gates}\small{Realization of 
              elementary quantum gates on the \QCns. Each square
              represents a lattice qubit. The 
    squares in the extreme left column marked with white circles
    denote the input qubits, they on the right-most column denote the
    output qubits. Note that blank squares can represent either
    $\sigma_z$ measurements or  empty lattice sites.}}}
    \end{center}
\end{figure}

An arbitrary rotation $U_{Rot} \in SU(2)$ can be achieved in a
chain of 5 qubits. Consider a
rotation in its Euler representation 
\begin{equation}
    \label{Euler}
    U_{Rot}(\xi,\eta,\zeta) = U_x(\zeta)U_z(\eta) U_x(\xi),
\end{equation}
where the rotations about the $x$- and $z$-axis are 
\begin{equation}
\label{XZrots}
\begin{array}{rcl}
U_x(\alpha) &=&
\displaystyle{\mbox{exp}\left(-i\alpha\frac{\sigma_x}{2}\right)}\\
U_z(\alpha) &=& \displaystyle{\mbox{exp}\left(-i\alpha 
\frac{\sigma_z}{2}\right)}.
\end{array}
\end{equation}
Initially, the
first qubit is in some state
$|\psi_{\mbox{\footnotesize{in}}}\rangle$, which is to be rotated, and
the other qubits are in 
$|+\rangle$. After the 5 qubits are entangled by the time
evolution operator $S$ generated by the Ising-type Hamiltonian, the
state $|\psi_{\mbox{\footnotesize{in}}}\rangle$ can be rotated by
measuring qubits 1 to 
4. At the same time, the state is also swapped to site 5. The qubits $1
\dots 4$ are measured in appropriately  chosen bases, {\em{viz.}}
\begin{equation}
    \label{Measbas}
    {\cal{B}}_j(\varphi_{j,\mbox{\footnotesize{meas}}}) = \left\{
        \frac{|0\rangle_j+e^{i \varphi_{j,\mbox{\footnotesize{meas}}}}
        |1\rangle_j}{\sqrt{2}} ,\, 
        \frac{|0\rangle_j-e^{i \varphi_{j,\mbox{\footnotesize{meas}}}}
        |1\rangle_j}{\sqrt{2}} 
    \right\}
\end{equation} 
whereby the measurement outcomes  $s_{j} \in \{ 0,1 \}$ for
$j=1\dots 4$ are obtained. Here, $s_{j}=0$ means that qubit $j$ is projected 
into the first state of
${\cal{B}}_j(\varphi_{j,\mbox{\footnotesize{meas}}})$. In
 (\ref{Measbas}) the basis states of all possible measurement bases
lie on the equator of the Bloch sphere, i.e. on the intersection of
the Bloch sphere with the $x$-$y$-plane. Therefore, the measurement
basis for qubit $j$ can be specified by a single parameter, the
measurement angle $\varphi_{j,\mbox{\footnotesize{meas}}}$. The
measurement direction of qubit $j$ is the vector on the Bloch sphere
which corresponds 
to the first state in the measurement basis
${\cal{B}}_j(\varphi_{j,\mbox{\footnotesize{meas}}})$. Thus, the
measurement angle $\varphi_{j,\mbox{\footnotesize{meas}}}$ is equal to the
angle between the measurement direction at qubit $j$ and the positive
$x$-axis. For all of
the so far constructed gates, the cluster qubits are either 
--if they are not required for the realization of the
circuit-- measured in $\sigma_z$, or --if they are required-- measured
in some measurement direction in the $x$-$y$-plane. 
In summary,
the procedure to implement an arbitrary rotation $U_R(\xi,\eta,\zeta)$, 
specified by its Euler angles $\xi,\eta,\zeta$, is this:
\begin{equation}
    \label{Rotproc}
    \begin{array}{rl}
    1. & \mbox{measure qubit 1 in} \; {\cal{B}}_1(0)\\ 
    2. & \mbox{measure qubit 2 in} \;
            {\cal{B}}_2\left( -\xi\,(-1)^{s_1+\kappa_{1,I}^\prime}
            \right)\\
    3. & \mbox{measure qubit 3 in} \; {\cal{B}}_3\left(
        -\eta\,(-1)^{s_2+\kappa_2^\prime}  \right) \\  
    4. & \mbox{measure qubit 4 in} \; {\cal{B}}_4\left(
            -\zeta\,(-1)^{s_1+s_3+\kappa_{1,I}^\prime+\kappa_3^\prime
              } \right)
    \end{array}   
\end{equation}
If the 5-qubit cluster state in
Fig.~\ref{Gates} is
created from a product state of all qubits in $|+\rangle$
via the interaction $S$ of eq. (\ref{Sform}), we have the specific 
values $\kappa_{1,I}^\prime=0, \kappa_2^\prime = \kappa_3^\prime =
\kappa_4^\prime = \kappa_{5,O}^\prime =1$. Please note that in
(\ref{Rotproc})  we used the set 
$\big\{\kappa_{1,I}^\prime, \kappa_2^\prime,  \kappa_3^\prime,
\kappa_4^\prime, \kappa_{5,O}^\prime \big \}$ instead of
$\big\{\kappa_l, l=1,..,5 \big \}$ to specify a cluster state
$|\phi\rangle_5$ on a chain of 5 qubits. We will use primed $\kappa$,
$\{\kappa_a^\prime \}$ whenever we describe a cluster state
$|\phi\rangle_{{\cal{C}}_N}$ on some cluster ${\cal{C}}_N$ which
consists only of those cluster qubits that are necessary for the
implementation of some circuit. Further, the $\kappa^\prime$
associated with the input qubit 1 we denote by $\kappa^\prime_{1,I}$
instead of  $\kappa^\prime_1$, and the $\kappa^\prime$
associated with the output qubit 5 we denote by $\kappa^\prime_{5,O}$
instead of  $\kappa^\prime_5$. The reason for this notation will be
discussed in Section~\ref{Ivonsk}. The unprimed $\kappa$,
$\{\kappa_a\}$, are reserved for the description of the cluster state
$|\phi\rangle_{\cal{C}}$ on the whole cluster ${\cal{C}}$.

Via the procedure (\ref{Rotproc}) the
rotation $U_{Rot}^\prime$ is realized:
\begin{equation}
    \label{Rotprime}
    U_{Rot}^\prime(\xi,\eta,\zeta) =  U_{\Sigma,Rot} \,U_{Rot}(\xi,\eta,\zeta).
\end{equation}
Therein, the random byproduct operator has the form
\begin{equation}
    \label{Byprod1}
    U_{\Sigma,Rot}=\sigma_x^{s_2+s_4+\kappa^\prime_2+\kappa^\prime_4}
    \sigma_z^{s_1+s_3+\kappa^\prime_{1,I}+\kappa^\prime_3+
    \kappa^\prime_{5,O}}.   
\end{equation} 
It can be corrected for at the end of the computation, as explained below.  

A CNOT-gate can be implemented on a cluster state of 15 qubits, as shown
in Fig.~\ref{Gates}. All measurements can be performed
simultaneously. Depending on the measurement results, the following
gate is thereby realized:
\begin{equation}
    \label{CNOT}
    U^\prime_{CNOT} = U_{\Sigma,CNOT}\, CNOT(c,t).
\end{equation} 
Therein the byproduct operator $U_{\Sigma,CNOT}$ has the form
\begin{equation}
    \label{CNOTbyprop}
    \begin{array}{rcl}
    U_{\Sigma,CNOT}&=&{\sigma_x^{(c)}}^{\gamma_x^{(c)}}
       {\sigma_x^{(t)}}^{\gamma_x^{(t)}}
       {\sigma_z^{(c)}}^{\gamma_z^{(c)}}
       {\sigma_z^{(t)}}^{\gamma_z^{(t)}},\;\mbox{with}\\ & \\
       {\gamma_x^{(c)}}&=&s_2+s_3+s_5+s_6+\kappa_2^\prime+\kappa_3^\prime
       +\kappa_5^\prime +\kappa_6^\prime\\
       {\gamma_x^{(t)}}&=&s_2+s_3+s_8+s_{10}+s_{12}+s_{14}+\kappa_2^\prime
       +\kappa_3^\prime+\kappa_8^\prime
       +\kappa_{10}^\prime+\kappa_{12}^\prime +\kappa_{14}^\prime \\
       {\gamma_z^{(c)}}&=&s_1+s_3+s_4+s_5+s_8+s_9+s_{11}+\\
       & &\kappa_{1,I}^\prime
       +\kappa_3^\prime+\kappa_4^\prime
       +\kappa_5^\prime+\kappa_{7,O}^\prime
       +\kappa_8^\prime+\kappa_{9,I}^\prime +\kappa_{11}^\prime+1  \\
       {\gamma_z^{(t)}}&=&s_9+s_{11}+s_{13}+\kappa_{9,I}^\prime
       +\kappa_{11}^\prime+\kappa_{13}^\prime +\kappa_{15,O}^\prime.
    \end{array}
\end{equation}
Please note the constant offset for $\gamma_z^{(c)}$ and further that for both
the general rotation and the CNOT-gate the byproduct operators depend
only on the combinations 
$s_k + \kappa_k^\prime$ if $k$ is not an output qubit of the
gate. The latter is a general feature of the relevant qubits, as will be
shown in Section~\ref{Ivect}.  

The randomness of the measurement results does not
jeopardize the function of the  circuit. Depending on
the measurement results, extra rotations $\sigma_x$ and $\sigma_z$ act on
the output qubit of every implemented gate, as in (\ref{Rotprime}), for
example. By use of the propagation
relation for general one-qubit rotations
\begin{equation}
    \label{Rotprop}
    U_R(\xi,\eta,\zeta) \, \sigma_z^{s} \sigma_x^{s'} =   
    \sigma_z^{s} \sigma_x^{s'} \,
    U_R((-1)^{s}\xi,(-1)^{s'}\eta,(-1)^{s}\zeta),
\end{equation}
and the one for the CNOT-gate, as in \cite{Proprelref},
\begin{equation} 
    \label{CNTprop}
    \mbox{CNOT} (c,t) \, {\sigma_z^{(t)}}^{s_t} {\sigma_z^{(c)}}^{s_c}
    {\sigma_x^{(t)}}^{s_t'} {\sigma_x^{(c)}}^{s_c'} 
    =  {\sigma_z^{(t)}}^{s_t} {\sigma_z^{(c)}}^{s_c+s_t}
    {\sigma_x^{(t)}}^{s_c'+s_t'} {\sigma_x^{(c)}}^{s_c'} \, 
    \mbox{CNOT}(c,t) ,
\end{equation}  
these extra rotations can be propagated
through the network to act upon the output state. By this we mean that
the extra random rotations need not be corrected for after a
gate. Instead, one just needs to keep track of them and delay
correction until the end of the computation. Then, the extra rotations can be
accounted for by properly interpreting the $\sigma_z$-readout
measurement results.

The propagation relations  (\ref{Rotprop}) for the arbitrary rotation
and (\ref{CNTprop}) for the CNOT-gate differ with respect to which
of the two unitary transformations  --the gate or the byproduct operator
$U_\Sigma$- is modified on the right hand side of  (\ref{Rotprop})
and (\ref{CNTprop}). In the case of the
propagation of a byproduct operator through a rotation
(\ref{Rotprop}), the gate is 
changed and the byproduct operator remains unchanged. It passes just through.
Conversely, in the case of the propagation of a
byproduct operator $U_\Sigma$ through a CNOT-gate (\ref{CNTprop}), it
is the byproduct operator which is modified and the gate remains unchanged.  

The measurement bases ${\cal{B}}(\varphi)$ and ${\cal{B}}(-\varphi)$
in (\ref{Measbas})
coincide for angles $\varphi=0$ and for $\varphi=\pm\pi/2$. For
$\varphi=0$ the measurement basis ${\cal{B}}(\varphi)$ is the eigenbasis of $\sigma_x$, and
for $\varphi=\pm\pi/2$ the measurement basis ${\cal{B}}(\varphi)$ is
the eigenbasis of 
$\sigma_y$. In these cases, the choice of the measurement basis is not
influenced by the results of measurements at other qubits. Therefore,
rotations whose Euler angles $\xi,\eta,\zeta$ are in the set $\{0,\pm \pi/2\}$
can be realized simultaneously in the first round of measurements, that is
no other cluster qubits need 
to be measured before. Among these rotations are the Hadamard gate and
the $\pi/2$ phase shift. As displayed in Fig.~\ref{Gates}, the Hadamard 
gate and the $\pi/2$-phase shift are both realized by performing a
pattern of $\sigma_x$- and  $\sigma_y$-measurements on the cluster
${\cal{C}}$. The byproduct operators which are thereby created are
\begin{equation}
    \label{Hadabyprop}
    \begin{array}{rcl}
          U_{\Sigma,H} &=&
          \sigma_x^{s_1+s_3+s_4+\kappa_{1,I}^\prime+\kappa_3^\prime 
          +\kappa_4^\prime}\, \sigma_z^{s_2+s_3+\kappa_2^\prime+\kappa_3^\prime
          +\kappa_{5,O}^\prime}\\
          U_{\Sigma,U_z(\pi/2)} &=&
          \sigma_x^{s_2+s_4+\kappa_2^\prime+\kappa_4^\prime}\,
          \sigma_z^{s_1+s_2+s_3+\kappa_{1,I}^\prime+\kappa_2^\prime 
          +\kappa_3^\prime+\kappa_{5,O}^\prime}.
    \end{array}      
\end{equation}

Owing to the special Euler  angles for the Hadamard- and
the $\pi/2$-phase gate,  the propagation relations for these rotations
can also be written in a form resembling the
propagation relation (\ref{CNTprop}) for the CNOT-gate 
\begin{equation}
    \label{Hadaprop}
    \begin{array}{rcl}
        H\,{\sigma_x}^{s_x}{\sigma_z}^{s_z} &=&
        {\sigma_x}^{s_z}{\sigma_z}^{s_x}\, H, \\
        U_z(\pi/2)\,{\sigma_x}^{s_x}{\sigma_z}^{s_z} &=&
        {\sigma_x}^{s_x}{\sigma_z}^{s_x+s_z}\, U_z(\pi/2).
    \end{array}
\end{equation}
Under propagation --via the propagation  relations (\ref{Rotprop}),
(\ref{CNTprop}) and (\ref{Hadaprop})-- the byproduct operators
resulting from the implementation of the universal gates generate a
subgroup of the group 
\begin{equation}
    \label{byprodgroup}
    {\cal{U}}^{\mbox{\tiny{local}}} = \left\{1, \sigma_x^{(i)}, \sigma_z^{(i)}, i=1..n,
    \;\,\mbox{and products thereof}  \right\}
\end{equation}
of all possible byproduct operators. ${\cal{U}}^{\mbox{\tiny{local}}}
\subset SU(2)^{\otimes n}$ is a subset of the set of all multi-local unitary
operations. Hence, it can be compensated for at the end of the
computation by a local change of the measurement bases.

To summarize, any quantum logic network can be
simulated on a one-way quantum computer. A set of universal gates can
be realized by one-qubit measurements and the gates can be combined to
circuits. Due to the randomness of the results of the individual
measurements, unwanted byproduct operators are introduced. These
byproduct operators can be accounted for by 
adapting measurement directions throughout the process of
computation. In this way, a 
subset of qubits on the cluster ${\cal{C}}$ is prepared as the output
register. The quantum state on this subset of qubits equals
that of the quantum register of the simulated network up to the action
of an 
accumulated byproduct operator. The byproduct operator determines how the
measurements on the output register are to be interpreted. 

\section{Beyond the network picture}
\label{beyond}

In the previous section we have described the \QC in a network
terminology, which has been useful to prove the universality of the
scheme. On the other hand, the cluster qubits do not have to be
measured in the order prescribed by the order of the gates in the
corresponding network. This observation indicates that the network
picture does not describe the \QC in every respect.

\subsection{The sets $Q_t$ of simultaneously measurable qubits}
\label{sets}

The cluster qubits which we have chosen to take the role of the 
readout register, for example,  are just qubits  like any other
cluster qubits. It turns out that, in a more efficient way of running
the \QCns, the ``readout'' qubits are not the last ones to be measured
but among the first.
It is advantageous to
forget about the network altogether and to view
the \QC as a set of one-qubit measurements on a resource quantum
state, the cluster state. These
measurements have to be performed in a certain order and in a certain 
basis. The classical information of how to measure subsequent qubits
must all be 
contained in the results of the already performed
measurements. Similarly, the final result of the computation must be
contained in all the measurement outcomes together.

In the following we will adopt the strategy that every cluster qubit
is measured at the earliest possible time. 
This means that each qubit is measured  as soon as the required
measurement results from other qubits 
which determine its measurement basis are known.  Let us denote by $Q_t$ the
set of qubits which can be measured at the same time in the
measurement round $t$. So, how can the sets $Q_t$ be determined? $Q_0$ is the set of qubits which are
measured in the first round. These are all the
qubits whose observables $\sigma_x$, $\sigma_y$ or $\sigma_z$ are
measured. The measurement bases for these qubits do not depend
on the results of any previous measurements. To determine the
subsequent set $Q_1$, one looks at which qubits can be measured with
the knowledge of the measurement results from the qubits in
$Q_0$. Next, one looks which qubits
can be measured with the measurement results from the qubits in $Q_0$
and $Q_1$ known. These qubits form the set $Q_2$. In this manner one
proceeds until the whole cluster ${\cal{C}}$ is divided into disjoint
 subsets $Q_t$.

As will become clear later, it is useful to introduce the sets
$Q^{(t)}$ of yet-to-be measured qubits. More precisely,      
$Q^{(t)}$ is the set of qubits which remain to be measured
after measurement round No. $t-1$,
\begin{equation}
    \label{Qupt}
     Q^{(t)}= \bigcup\limits_{i=t}^{t_{\mbox{\tiny{max}}}} Q_i.
\end{equation}
Mathematically, the sets $Q_t$ are derived from a
strict partial ordering in ${\cal{C}}$. The
strict partial ordering, in turn, is generated by forward cones which are
explained in the next section.

\subsection{The forward- and backward cones} 
\label{FBC}

Be $g$ a gate in the network ${\cal{N}}$ to be simulated and $k \in
{\cal{C}}(g)$ a cluster qubit that belongs to the implementation of
$g$. Further, be ${\cal{O}}$, $A$ and $\Omega$ three vertical
cuts through the network ${\cal{N}}$. A vertical cut is such that it
intersects each qubit line in a network only once and that it does not
intersect gates. ${\cal{O}}$ intersects ${\cal{N}}$ just after the
gate $g$, i.e. the byproduct operator $U_{\Sigma,g}$ caused by the
implementation of $g$, as given in (\ref{Byprod1}) and (\ref{CNOTbyprop}),
is located on ${\cal{O}}$. Note that the byproduct operators generated
on ${\cal{O}}$ depend on the measurement results obtained in course of
the gate implementation via ${\left(U_k\right)}^{s_k}$. $A$ intersects
${\cal{N}}$ just be fore the 
input, i.e. an operator propagated to $A$ acts on the input register
of ${\cal{N}}$, and $\Omega$ intersects ${\cal{N}}$ just before the
output such that an operator propagated to $\Omega$ acts on the output
register of ${\cal{N}}$. We can now define the forward- and backward
cones of the cluster qubits $k \in {\cal{C}}$.
\begin{definition}
    \label{fwc}
    The forward cone $\mbox{{\em{fc}}}(k)$ of a cluster qubit $k \in
    {\cal{C}}$ is the set of all those cluster qubits $j \in Q^{(1)}$
    whose measurement basis
    ${\cal{B}}(\varphi_{j,\mbox{{\em{\footnotesize{meas}}}}})$ 
    depends on the result $s_k$ of the measurement
    of qubit $k$ after the byproduct operator
    ${\left(U_k\right)}^{s_k}$ is propagated from ${\cal{O}}$ to $\Omega$. 
\end{definition} 

\begin{definition}
    \label{bwc}
    The backward cone $\mbox{{\em{bc}}}(k)$ of a cluster qubit $k \in
    {\cal{C}}$ is the set of all those cluster qubits $j \in Q^{(1)}$
    whose measurement basis
    ${\cal{B}}(\varphi_{j,\mbox{{\em{\footnotesize{meas}}}}})$ 
    depends on the result $s_k$ of the measurement
     of qubit $k$ after the byproduct operator
    ${\left(U_k\right)}^{s_k}$ is propagated from ${\cal{O}}$ to $A$. 
\end{definition} 
It will turn out that only the backward cones of the qubits $k \in
Q_0$ constitute part of the information specifying an algorithm on the
\QCns, but nevertheless all the backward- and forward
cones are important objects in the scheme. Either of the sets, the set
of the forward- and that of the backward cones, separately contains
the full information of the temporal structure of a computation
on the \QCns.

Let us examine the definitions \ref{fwc} and \ref{bwc} for a
particular example, the general one-qubit rotation (\ref{Euler}) as
implemented by the 
procedure (\ref{Rotproc}) modulo a byproduct operator
$U_{\Sigma,Rot}$ as given in (\ref{Byprod1}). The measurement result
$s_1$ of qubit 1 (cf. Fig.~\ref{Gates}) modifies the measurement angle
of qubit 2, which is 
responsible for implementing an 
$x$-rotation $U_x(\xi)$,  by a factor
${(-1)}^{s_1}$. Further, it causes a byproduct operator
${(\sigma_z)}^{s_1}$ at ${\cal{O}}$. If this byproduct operator is
propagated forward from ${\cal{O}}$ to $\Omega$ it has no effect on
qubit 2, because qubit 2 is behind ${\cal{O}}$. The dependence on
$s_1$ of the basis in which qubit 2 has to be measured persists and thus qubit
2 is in the forward cone of qubit 1, $\,\,2 \in \mbox{fc}(1)$. The
situation is different 
if the byproduct operator ${(\sigma_z)}^{s_1}$ is propagated backwards
from ${\cal{O}}$ to $A$: via the propagation relation (\ref{Rotprop})
the Euler angle $\xi$ is modified by a factor ${(-1)}^{s_1}$ which has
to be accounted for by multiplying the measurement angle
$\varphi_{2,{\mbox{\footnotesize{meas}}}}$ by a factor ${(-1)}^{s_1}$,
too. Thus, the factor ${(-1)}^{s_1}$ modifies the measurement angle
$\varphi_{2,{\mbox{\footnotesize{meas}}}}$ twice, once via the
procedure (\ref{Rotproc}) and once in backward propagation, and there
is no net effect. Qubit 2 is not in the backward cone of qubit 1, $\,\,2
\not\in \mbox{bc}(1)$.  

What does it mean that a cluster qubit $j$ is in the forward cone of
another cluster qubit $k$, $j\in \fc(k)$? According to the definition,
a byproduct operator created via the measurement at cluster qubit $k$
influences the measurement angle
$\varphi_{j,\mbox{\footnotesize{meas}}}$ at cluster qubit $j$. To
determine the measurement angle at $j$ 
one must thus wait for the measurement result at $k$ to see what the
byproduct operator created randomly by the measurement at $k$ is. If $j \in
\fc(k)$, the measurement at qubit $j$ is performed later than that at
qubit $k$. This we denote by $k \prec j$
\begin{equation}
    \label{coneImpl}
    j \in \fc(k) \Rightarrow k \prec j.
\end{equation}  
Please note that the converse of (\ref{coneImpl}) is not true. If $k
\prec j$ holds, still $j \in \fc(k)$ may not. This can be easily
verified for the example of a general rotation (\ref{Euler}). There,
according to the procedure for implementing such a rotation described
in Section~\ref{summary}, the result of the measurement of qubit 1
enters into in which basis qubit 2 has to be measured. Hence, $2 \in
\fc(1)$. By (\ref{coneImpl}), $1 \prec 2$ which means that the
measurement at qubit 2 has to wait for the result of the measurement
on qubit 1. Similarly, the measurement result on qubit 2 enters in the
choice of the measurement basis for the measurement on qubit 3. $3 \in
\fc(2)$ and thus $2 \prec 3$. Then  $1 \prec 3$ also holds as shown below in
(\ref{transitive}) 
, but $3 \not \in \fc(1)$, since the measurement result on
qubit 1 does not influence the choice of the measurement basis for the
measurement on qubit 3. 
  
The relation ``$\prec$'' is a strict partial ordering. Suppose,
that besides $k \prec j$, for another cluster qubit $l$ one had $l \in
\fc(j)$ and thus $j \prec l$. This would mean that the measurement at $l$
must wait for the measurement at $j$, which itself had to wait for the
measurement at $k$. Thus, the measurement at $l$ also had to wait for
the measurement at $k$. Therefore the relation ``$\prec$'' is transitive,
\begin{equation}
    \label{transitive}
    k \prec j \,\, \wedge \,\,j \prec l 
    \, \longrightarrow \, k \prec l.
\end{equation} 
Further, a measurement to implement a gate cannot and does not need to
wait for its own result.  Therefore the relation ``$\prec$'' is anti-reflexive,
\begin{equation}
    \label{irreflexive}
    \neg \exists j \in {\cal{C}}: j\prec j. 
\end{equation}
Let us now cast the procedure to construct the sets of simultaneously
measured qubits given above in more precise terms.
Be $Q_t \subset {\cal{C}}$ the set of cluster qubits measured
in measurement round $t$, and $Q^{(t)} \subset {\cal{C}}$ the
set of qubits which are to be measured in the measurement round
$t$ and all subsequent rounds, as defined in (\ref{Qupt}). Then, $Q_0$
is the set of qubits which are measured in the first round. These are
the qubits of which the 
observables $\sigma_x$, $\sigma_y$ or $\sigma_z$ are 
measured, so that the measurement bases are not influenced by other
measurement results. Further, $ Q^{(0)} = {\cal{C}}$. Now, the
sequence of sets $Q_t$ can be constructed using the following
recursion relation
\begin{equation}
    \label{Qrecur}
    \begin{array}{lcl}
        Q_t &=& \displaystyle{\left\{ q \in Q^{(t)} |
        \neg \exists p 
        \in {Q}^{(t)}: p \prec q \right\}} \\
        Q^{(t+1)} &=& \displaystyle{Q^{(t)} \backslash
        Q_t}.
    \end{array}
\end{equation}
All those qubits which have no precursors in some remaining set
$Q^{(t)}$ and thus do not have to wait for results of
measurements of qubits in $Q^{(t)}$ are taken out of this set
to form $Q_t$. The recursion proceeds until
${Q}^{(t_{\mbox{\footnotesize{max}}}+1)} = \emptyset$ for some maximal
value  $t_{\mbox{\footnotesize{max}}}$ of $t$.

Can it happen that the recursion does not terminate? That were the case
if for a number $m$ of qubits $j_1,...,j_m \in {\cal{C}}$ formed a
cycle $j_1 \prec j_2 \prec ... \prec j_m \prec j_1$. Then, none of the
qubits $j_1,..,j_n$ could ever taken out of the 
set. However, by transitivity (\ref{transitive}) we then had $j_1 \prec j_1$
which contradicts anti-reflexivity (\ref{irreflexive}). Hence, such a
situation cannot occur. 

Let us at the end of this section define the forward- and backward
cones $\mbox{fc}(g)$, $\mbox{bc}(g)$ of the gates $g$. In
eq.~(\ref{CNOTbyprop}) we have seen that the byproduct operator caused
by the implementation of a CNOT-gate contains a constant contribution
$U_0(CNOT)=\sigma_z^{(c)}$. This contribution to the byproduct
operator does not depend on any local variables such as the
measurement results and is thus attributed to the gate as a
whole. This byproduct operator is of the same form as those
depending on the individual measurement results and can influence measurement
angles when being propagated forward or backward. Thus we define the
forward- and backward cones of gates, in analogy to those of the
cluster qubits $k \in {\cal{C}}$, as follows: 
 
The forward cone $\mbox{fc}(g)$ of a gate $g \in
{\cal{N}}$ is the set of all those cluster qubits $j \in Q^{(1)}$
of which the measurement basis
${\cal{B}}(\varphi_{j,\mbox{{\em{\footnotesize{meas}}}}})$ 
is modified if the byproduct operator $U_0(g)$ is propagated
forward from ${\cal{O}}$ to $\Omega$. 

The backward cone $\mbox{bc}(g)$ of a gate $g \in
{\cal{N}}$ is the set of all those cluster qubits $j \in Q^{(1)}$
of which the measurement basis
${\cal{B}}(\varphi_{j,\mbox{{\em{\footnotesize{meas}}}}})$ 
is modified if the byproduct operator $U_0(g)$ is propagated
backward from ${\cal{O}}$ to $A$. 

The forward- and backward cones of gates contain some information
about the temporal structure of an algorithm on the \QCns, but --in
contrast to the backward- and forward cones of the cluster qubits--
not all. Also they do not form part of the information representing a
quantum algorithm on the \QCns, they will be absorbed into the
algorithm angles. Their role for the description of a computation on
the \QC is a technical one.

\subsection{The algorithm- and measurement angles}
\label{angles}

There are three different types of angles involved in the described
scheme of quantum computation of which the most prominent are the
algorithm angles and the measurement angles. 

The {\em{algorithm angles}}
$\big \{\varphi_{j,{\mbox{\footnotesize{algo}}}}, j\in Q^{(1)} \big
\}$ are part of the information that specifies an algorithm on the
\QCns. They are derived from the network angles $\big
\{ \varphi_{j,\mbox{\footnotesize{qln}}}, j \in Q^{(1)} \big \}$, i.e. the Euler angles of the
one-qubit rotations in the quantum logic network. Further, the
algorithm angles   
depend on the set $\big \{ \kappa_k, k \in {\cal{C}} \big \}$
characterizing the cluster state $|\phi\rangle_{\cal{C}}$ in
(\ref{EVeqn}), and on special properties of the measurement
pattern. We see that the network angles are absorbed into the
algorithm angles. They do not constitute part of the information
specifying a \QCns-algorithm. 

As described before, the process of computation with the \QC comprises
several measurement rounds. The first round, in wich the qubits in the
set $Q_0$ are measured, is somewhat different from the following
rounds. Therein, all gates of the circuit that belong to the Clifford 
group are implemented at the same time, no matter where they are
located in the corresponding quantum logic network and in which step
they would be carried out there. This results in byproduct operators
scattered all over the place. These byproduct operators are, according
to the scheme described in Section~\ref{model}, propagated
backwards. To account for the effect that the byproduct operators have
on the algorithm angles, these angles have to updated to the {\em{modified
algorithm angles}} $\big 
\{\varphi_{j,{\mbox{\footnotesize{algo}}}}^\prime, j\in Q^{(1)} \big
\}$. The modified algorithm angles
$\varphi_{j,{\mbox{\footnotesize{algo}}}}^\prime$  are calculated from
the respective algorithm angles
$\varphi_{j,{\mbox{\footnotesize{algo}}}}$ and the results obtained in
the first measurement round $\big \{ s_k, k \in Q_0 \big \}$. In the
subsequent measurement rounds no further update of the modified
algorithm angles occurs. Finally, each qubit $j \in Q_t \subset
Q^{(1)}$ is measured in some measurement round $t$ in the basis
${\cal{B}}(\varphi_{j,{\mbox{\footnotesize{meas}}}})$ where
$\varphi_{j,{\mbox{\footnotesize{meas}}}}$ denotes the measurement
angle of qubit $j$. The {\em{measurement angle}}
$\varphi_{j,{\mbox{\footnotesize{meas}}}}$ of a qubit $j \in Q_t$ is
calculated from the modified algorithm angle,
$\varphi_{j,{\mbox{\footnotesize{algo}}}}^\prime$ and the results
$\big \{ s_k, k \in \bigcup_{i=0}^{t-1} Q_i \big \}$ of the so far
obtained measurements.

Before a quantum algorithm is run on the \QCns, the algorithm angles
are determined from the cluster and the properties of the algorithm. During
runtime of the \QCns, in the first measurement round $(t=0)$, the algorithm
angles are replaced by the modified algorithm angles, i.e. only the
latter are kept while the former are erased. Then, in the measurement
round $t$ a qubit $j \in Q_t$ is measured in the basis
determined by the measurement angle
$\varphi_{j,{\mbox{\footnotesize{meas}}}}$. After the measurement of
qubit $j$ both $\varphi_{j,{\mbox{\footnotesize{algo}}}}^\prime$ and
$\varphi_{j,{\mbox{\footnotesize{meas}}}}$ can be erased.

Now there arises the question of how the measurement angles of the actual
measurements are calculated from the results of previous
measurements. This question will be answered in Section~\ref{model}.
The question which interests us most, of
course, is: ``How can 
the final result of the computation be determined from all the
measurement outcomes?'' It will turn out that the answers to both
questions are very much related. 

\subsection{Quantities for the processing of the measurement results}  

\subsubsection{The information vector}
\label{Ivect}

Let us again --as in
Section~\ref{sets}-- make the change of the viewpoint from an
explanation of the \QC within the network model to a more suitable
description, but now focusing on the quantities required to describe
the processing of the measurement results.   

Suppose in the simulation of a quantum logic network
${\cal{N}}$ on the \QC in a network manner --i.e. measuring the
``readout'' qubits at last-- the processing has reached the stage where
all but those  cluster qubits  have been
measured which form the output register. 

The accumulated byproduct operator $U_\Sigma$ to act upon the logical
output qubits $1,.., n$ is known. It has the form
\begin{equation}
    \label{byprod}
    U_\Sigma = \prod \limits_{i=1}^n
    \,\,{\left(\sigma_x^{(i)}\right)}^{x_i}
    {\left(\sigma_z^{(i)}\right)}^{z_i}, 
\end{equation}  
where $x_i,z_i \in \{0,1\}$ for $1\leq i \leq n$.
Let us now label the unmeasured qubits on the cluster in the same way
as the readout qubits on the quantum logic network are labelled. 

The ``readout'' qubits on the cluster are, at this point, in a state $U_\Sigma
|\mbox{out}\rangle$,  where $|\mbox{out}\rangle$ is the output state
of the corresponding quantum logic network. In the network
  picture, the computation is completed by measuring each qubit in
  the $\sigma_z$-eigenbasis, thereby obtaining the measurement results
  $\{s_i^\prime \}$, say. In the \QC scheme, one measures the state
  $U_\Sigma |\mbox{out}\rangle$ directly, whereby  outcomes
    $\{ s_i \}$ are obtained and the ``readout'' qubits are projected
    into the state $|{\cal{M}}\rangle =
  \prod_{i=1}^n \frac{1+{(-1)}^{s_i} \sigma_z^{(i)}}{2}\,
    \, U_\Sigma |\mbox{out}\rangle $. Depending on the known byproduct
    operator $U_\Sigma$, the set of measurement results $\{s_i\}$ in
    general has a 
    different interpretation from what the network readout
    $\{s_i^\prime\}$ would have. The measurement basis is the same. From
    (\ref{byprod}) one obtains
\begin{equation}
    \label{identification}
    \begin{array}{rcl}
    |{\cal{M}}\rangle =
  \prod_{i=1}^n \frac{1+{(-1)}^{s_i} \sigma_z^{(i)}}{2}\,
    U_\Sigma |\mbox{out}\rangle &=& U_\Sigma \left(U_\Sigma^\dagger \,
  \prod_{i=1}^n \frac{1+ 
  {(-1)}^{s_i} \sigma_z^{(i)}}{2}  U_\Sigma \right)
  |\mbox{out}\rangle\\
  &=& U_\Sigma \, \prod_{i=1}^n
  \frac{1+{(-1)}^{s_i+x_i} \sigma_z^{(i)}}{2}  |\mbox{out}\rangle
  \end{array}
\end{equation}
From (\ref{identification}) we see that a $\sigma_z$-measurement on the
state $U_\Sigma \, |\mbox{out}\rangle$ with results $\{s_i\}$
represents the same algorithmic output as a $\sigma_z$-measurement of
the state $|\mbox{out} \rangle$ with the results $\{s_i^\prime\}$, where the
sets $\{s_i\}$ and $\{s_i^\prime\}$ are related by
\begin{equation}
    \label{join}
    s_i^\prime \equiv s_i+x_i\; \mbox{mod} \,\, 2.
\end{equation}
The set $\{s_i^\prime\}$ represents the result of the computation. It can be
calculated from the results $\{ s_i \}$ of the $\sigma_z$-measurements
on the ``readout'' cluster qubits, and the values $\{x_i\}$ which can
be extracted from the byproduct operator $U_\Sigma$.
We see that both the contribution of the byproduct operator and the
result of the measurement on the ``readout'' qubits of the cluster  
enter expression (\ref{join})  in the same
way. Indeed, there is 
no need to distinguish between these two contributions. On the level of
the byproduct operators, the
readout measurement result is translated into an additional
contribution to the accumulated byproduct
operator. Both contributions to
the such extended byproduct operator 
\begin{equation}
    \label{ExtByprop}
    U_{\Sigma R}=U_\Sigma
    U_R,\; \mbox{with}\; U_R=\prod_{i=1}^n {(\sigma_x^{(i)})}^{s_i},
\end{equation}
stem from 
random measurement results. It is just that the contributions which
constitute 
$U_\Sigma$ must be propagated forward from where they originated and
the additional contributions from the readout measurements must be
propagated one step backwards.  Both the forward and
the backward propagated contributions to $U_{\Sigma R}$ are propagated
  to the same location in the network. Forward and backward propagation
are closely related. In fact, as the propagation relations
(\ref{Rotprop}), (\ref{CNTprop}) and (\ref{Hadaprop}) are their own
inverse, the rules are the same for both directions. The distinguished 
role of the readout qubits is only a remnant of the interpretation of
the \QC  as a quantum logic network. A more adequate description
will have the consequence that the cluster qubits on the ``output 
register'', for instance, will be measured during the initialization
of the \QC such that they are removed from the entangled quantum state even
before the main part of the computation starts. \medskip 

We now define the  the {\em{information vector}} ${\bf{I}}$, a
$2n$-component binary vector which is a function of the 
quantities $\{x_i,z_i\}$ and  the results $\{s_i\}$ of the measurements
on the cluster output register. Together, these quantities  determine
the extended byproduct operator $U_{\Sigma R}$ via (\ref{byprod}),
and(\ref{ExtByprop}). 
\begin{definition}
The information vector ${\bf{I}}$ is given by
\begin{equation}
    \label{Ivector}
    {\bf{I}}=\left(\begin{array}{c} \mbox{ } \\ {\bf{I}}_x \\ 
    \vspace*{0.3cm} \\ {\bf{I}}_z \\ \mbox{ }\end{array} \right),\;\;\;
    \mbox{with}\;\;{\bf{I}}_x=\left(\begin{array}{c} x_1+s_1 \\ x_2+s_2 \\
    . \\ . \\ . \\ x_n+s_n  \end{array} \right), \;\;\;
    {\bf{I}}_z=\left(\begin{array}{c} z_1 \\ z_2 \\
    . \\ . \\ . \\ z_n \end{array} \right).
\end{equation} 
\end{definition}
As can be seen from (\ref{join}) and (\ref{Ivector}), ${\bf{I}}_x$ is
a possible result of a readout measurement in  a corresponding quantum logic
network. ${\bf{I}}_z$ is redundant. However,
in Section~\ref{Ivonsk} the flow quantity ${\bf{I}}(t)$, the
{\em{information flow vector}}, 
will be defined for which
${\bf{I}}(t_{\mbox{\footnotesize{max}}})={\bf{I}}$, with 
$t_{\mbox{\footnotesize{max}}}$ the index of the final computational
step. For $t <
t_{\mbox{\footnotesize{max}}}$, in ${\bf{I}}(t)$ 
both the $z$-part ${\bf{I}}_z(t)$ and the $x$-part ${\bf{I}}_x(t)$ are 
required to determine the bases for the one-qubit measurements in
$Q_{t+1}$.  As ${\bf{I}}_z(t)$ is of equal importance as
${\bf{I}}_x(t)$ throughout the
process of computation we keep ${\bf{I}}_z$ in the definition of
${\bf{I}}$ as well.  

The set of possible information vectors ${\bf{I}}$ forms a $2n$ dimensional
vector space over $F_2$, ${\cal{V}}$. Let us consider the group
${\cal{U}}^{\mbox{\tiny{local}}}$ 
of all possible extended byproduct operators $U_{\Sigma R}$. If we divide
out the normal divisor $\{\pm 1\}$ of
${\cal{U}}^{\mbox{\tiny{local}}}$, the 
resulting factor  
group $\overline{\cal{U}}  = {\cal{U}}^{\mbox{\tiny{local}}} / \{ \pm
1 \}$ is isomorphic to 
${\cal{V}}$. From the viewpoint of physics, dividing out the
normal divisor $\{ \pm 1 \}$ means that we ignore a global phase.  The
isomorphism ${\cal{I}}$ 
which maps an ${\bf{I}} \in {\cal{V}}$ to the corresponding $U_{\Sigma R} \in
\overline{\cal{U}}$ is given by
\begin{equation}
    \label{Iso}   
        {\cal{I}}:\,\, {\cal{V}} \ni {\bf{I}} \longrightarrow
        U_{\Sigma R} =  \prod \limits_{i=1}^n
    \,\,{\left(\sigma_x^{(i)}\right)}^{{[I_x]}_i}
    {\left(\sigma_z^{(i)}\right)}^{{[I_z]}_i} \in \overline{\cal{U}}, 
\end{equation}
where ${[I_x]}_i$ and ${[I_z]}_i$ are the respective components of
${\bf{I}}_x$ and 
${\bf{I}}_z$. The component-wise addition of vectors in ${\cal{V}}$
corresponds, via the isomorphism ${\cal{I}}$, to the multiplication of
byproduct 
operators modulo a phase factor $\{ \pm 1\}$. The
procedure to implement this product is to first use the operator
product, then bring the factors into normal order according to
(\ref{Iso}) and finally drop the phase. Multiplication of vectors ${\bf{I}}
\in {\cal{V}}$ with the scalars 0,1 corresponds to raising the
byproduct operators $U_{\Sigma R} \in \overline{\cal{U}}$ to the
respective powers. One may switch
between the two pictures via the isomorphism (\ref{Iso}).
The algebraic structures involved will be more
apparent in the representation using the information vector
${\bf{I}}={\bf{I}}\left(\{x_i,z_i,s_i\}\right)$ than in
the formulation of the operator  $U_{\Sigma R}$. 

Now that we have defined the information vector ${\bf{I}}$ in
(\ref{Ivector}) and have seen that the result of the computation can
be directly read off from the $x$-part of ${\bf{I}}$, we would like to
find out how ${\bf{I}}$ depends on the measurement outcomes $\big\{
s_k \big \}$ and the set $\big\{
\kappa_k \big \}$ of binary numbers that determine the cluster state
$|\phi\rangle_{\cal{C}}$ in (\ref{EVeqn}). This task is
left until Section~\ref{Ivonsk}. Before we can accomplish it we need 
some further definitions. It will turn out that the information vector
${\bf{I}}$ can be written as a linear combination of the
{\em{byproduct images}} which are explained next.

\subsubsection{The byproduct images} 
\label{TheFs}

Be $\Omega$ the ``cut'' through a network ${\cal{N}}$
which intersects the qubit lines just before its output. This is the
cut at which the extended 
byproduct operator $U_{\Sigma R}$ is accumulated. Consider a qubit $k$
on the cluster
${\cal{C}}$ which is measured in the course of computation. Depending
on the result of the measurement on qubit $k$, a
byproduct operator ${\left(U_k\right)}^{s_k}$ is introduced  in
${\cal{N}}$ at the location of the logical output qubits of the gate
for whose implementation the cluster qubit $k$ was measured.  This byproduct
operator $U_k$ can  --by using the
propagation relations (\ref{Rotprop}), (\ref{CNTprop}) and
(\ref{Hadaprop})-- propagated from where it occurred to the cut
$\Omega$. There it appears as the forward propagated byproduct
operator $U_k|_\Omega$. Now we can define the {\em{byproduct image}}
${\bf{F}}_k$ of a cluster qubit $k \in {\cal{C}}$.
\begin{definition}
    Each cluster qubit $k \in {\cal{C}}$ has a byproduct image
    ${\bf{F}}_k$, which is the vector that
    corresponds via the isomorphism ${\cal{I}}^{-1}$ {\em{(\ref{Iso})}} to
    the forward propagated byproduct operator $U_k|_\Omega$,
    \begin{equation}
        \label{FI}
        {\bf{F}}_k = {\cal{I}}^{-1}\left(U_k|_\Omega \right).
    \end{equation}
\end{definition}
In the definition (\ref{FI}) of the byproduct image ${\bf{F}}_k$ it is
mentioned only implicitly that the image is evaluated on the cut
$\Omega$. Later in the discussion it will become apparent that we
could evaluate the byproduct image on every vertical cut
${\cal{O}}$.  Sometimes, if we compare to other vertical
cuts, we will explicitly write ${\bf{F}}_k |_\Omega$ for ${\bf{F}}_k$.
 
The set of byproduct images $\left\{ {\bf{F}}_k, k \in {\cal{C}}\right\}$ is an
important quantity for the scheme. It represents  part of the information
which is needed to run a quantum algorithm with the
$\mbox{QC}_{\cal{C}}$. 

In eq.~(\ref{CNOTbyprop}) the byproduct operator for the
CNOT-gate as realized according to Fig.~\ref{Gates} is given. This
byproduct operator contains a constant contribution
$U_0(CNOT)=\sigma_z^{(c)}$. As $U_0$ does not depend on any local
variables, neither on $\big \{ s_k \big\}$ nor on $\big \{ \kappa_k \big
  \}$, it makes no sense to attribute it to any of the cluster qubits
that were measured to realize the gate. Instead, it is attributed to
the part of the measurement pattern that implements the gate as a
whole, or --for 
simplicity-- to the gate itself. For any gate $g$, $U_0(g)$
can be propagated 
forward to the cut $\Omega$ to act upon the ``readout'' qubits. There
it appears as the forward propagated byproduct operator
$U_0(g)|_{\Omega}$. In analogy to the byproduct images of the cluster
qubits, we can now define the byproduct images of the gates $g$ of the
quantum logic network that is simulated on the \QCns.
For any such gate $g$ the byproduct image ${\bf{F}}_g$ is the vector
that corresponds to $U_0(g)|_{\Omega}$ via
\begin{equation}
    \label{Fg}
    {\bf{F}}_g = {\cal{I}}^{-1}(U_0(g)|_{\Omega}).
\end{equation}
Please note that in contrast to the byproduct images ${\bf{F}}_k$ of
cluster qubits $k \in {\cal{C}}$ the byproduct images ${\bf{F}}_g$ of gates
do not form a separate part of the information specifying a quantum
algorithm on the \QCns. They will be absorbed into the initialization value
${\bf{I}}_{\mbox{\footnotesize{init}}}$ of the information flow vector
defined in Section~\ref{Ivonsk} and
they are thus only a convenient tool in the derivation of the
computational model.

Via ${\cal{I}}^{-1}$ we map the multiplication of byproduct operators,
i.e. their accumulation, onto addition modulo 2 on the level of the
vectors in ${\cal{V}}$. Now there arises the question whether other
operations on the byproduct operators could be expressed in
terms of the corresponding vectors, too. Specifically, one may
ask how the byproduct operator propagation looks like on the level of
the ${\bf{I}} \in {\cal{V}}$.

\subsubsection{The propagation matrices}
\label{propmats}

The answer to this question is that on the level of the vector
quantities in ${\cal{V}}$  propagation is described by
multiplication with certain $2n \times 2n$-matrices $C$. Consider two
cuts  ${\cal{O}}_1$ and 
${\cal{O}}_2$  through a network which intersect each qubit
line only once. Further, be the two cuts such that they do not intersect
each other and that ${\cal{O}}_1$ is earlier than  ${\cal{O}}_2$. The part
of the quantum logic network between ${\cal{O}}_1$ and ${\cal{O}}_2$
is denoted by ${\cal{N}}_{{\cal{O}}_1\rightarrow {\cal{O}}_2}$. Be
${\bf{I}}_k|_{{\cal{O}}_1}$ and 
${\bf{I}}_k|_{{\cal{O}}_2}$ the vectors describing a byproduct operator
resulting from the measurement of qubit $k$, propagated to the cuts
${\cal{O}}_1$ and ${\cal{O}}_2$, respectively. Then we have
\begin{equation}
    \label{Iprop}
    {\bf{I}}_k|_{{\cal{O}}_2}= C({{\cal{N}}_{{\cal{O}}_1\rightarrow
    {\cal{O}}_2}})\, {\bf{I}}_k|_{{\cal{O}}_1}. 
\end{equation}
To any quantum logic network
${\cal{N}}$ a matrix $C_{\cal{N}}$ can be assigned. For a
network ${\cal{N}}_2 \circ {\cal{N}}_1$ composed of two subnetworks
${\cal{N}}_1$ and ${\cal{N}}_2$ (of which ${\cal{N}}_1$ is carried out
first) the propagation matrix is equal to the product of the
propagation matrices of the subnetworks
\begin {equation}
    \label{concatenat}
    C({{\cal{N}}_2 \circ {\cal{N}}_1}) = C({{\cal{N}}_2}) C({{\cal{N}}_1}).
\end{equation}  
Because of property (\ref{concatenat}) we only need to find the
propagation matrices for the general one-qubit rotations, the
CNOT-, the Hadamard- and the $\pi/2$-phase  gate. The one-qubit rotations and
the CNOT-gate alone form a universal set of gates. The reason why we
also include the Hadamard- and the $\pi/2$-phase gate is that here they are
treated  differently from the general rotations, as can be seen from
the propagation relations (\ref{Rotprop}) and (\ref{Hadaprop}). By
propagation through a Hadamard- or $\pi/2$-phase gate, the gate is
left unchanged while the byproduct operator changes; whereas for the
propagation through a general rotation, the  rotation changes and the 
byproduct operator stays the same.  Thus,
for finding the byproduct images the general rotations in ${\cal{N}}$ can
be replaced by the identity. Only the CNOT-, Hadamard and
$\pi/2$-phase gates have an effect. The special treatment of the Hadamard and
the $\pi/2$-phase gate is
advantageous with respect to the temporal complexity of a computation,
because if one uses the propagation relation (\ref{Hadaprop}) the
implementation of the Hadamard- and the $\pi/2$-phase  gate does not
need to wait for results 
of any previous measurements.  To sum up, to each possible ${\cal{N}}$
belongs a unitary operation $U({\cal{N}})$ in the Clifford
group and a corresponding matrix $C({\cal{N}})$, such that
\begin{equation}
    \label{correspondence}
    {\cal{I}}\left({C({\cal{N}}){\bf{I}}}\right) = U({\cal{N}})\,
    {\cal{I}}\left({\bf{I}}\right)\, U({\cal{N}})^\dagger\, , \;\;\;\;
    \forall {\bf{I}} \in {\cal{V}}.
\end{equation}
Let us now give the propagation matrices for propagation through
CNOT-, Hadamard and $\pi/2$-phase gates. The propagation matrices $C$ are
conveniently written in block form
\begin{equation}
    \label{blockform}
    C = \left(  
        \begin{array}{ccc|ccc}
            & \mbox{ } & & \mbox{ } &  \\
            & C_{xx} & & & C_{zx} & \\
            & \mbox{ } & & &\\ \hline
            & \mbox{ } & & &\\
            & C_{xz} & & & C_{zz} & \\
            & \mbox{ } & & & 
        \end{array}
    \right),
\end{equation} 
where $C_{xx}$, $C_{zx}$, $C_{xz}$ and $C_{zz}$ are $n \times n$
matrices with binary-valued entries. 

For the Hadamard gate $H^{(i)}$ on the logical qubit $i$ one
finds
\begin{equation}
    \label{Hadapropmat}
    \begin{array}{rcl}
        {\left[ C_{xx}(H^{(i)})\right]}_{kl} = {\left[
        C_{zz}(H^{(i)})\right]}_{kl}  &=& \delta_{kl} +
        \delta_{ki}\delta_{il}, \\
        {\left[ C_{zx}(H^{(i)})\right]}_{kl} = {\left[
        C_{xz}(H^{(i)})\right]}_{kl}  &=& \delta_{ki} \delta_{il},
    \end{array}
\end{equation}
where e.g. ${\left[ C_{xx}(H^{(i)})\right]}_{kl}$ denotes the entry of
row $k$ and column $l$ in $C_{xx}$. Note that the qubit index $i$ is not summed
over in  (\ref{Hadapropmat}) and that the addition is modulo 2. 

For the $\pi/2$-phase gate $U_z^{(i)}(\pi/2)$ on the logical qubit $i$ one
finds
\begin{equation}
    \label{Phaspropmat}
    \begin{array}{rcl}
        {\left[ C_{xx}(U_z^{(i)}(\pi/2))\right]}_{kl} &=&
        \delta_{kl},\\
        {\left[C_{zz}(U_z^{(i)}(\pi/2))\right]}_{kl}  &=& \delta_{kl},\\
        {\left[ C_{xz}(U_z^{(i)}(\pi/2))\right]}_{kl} &=& \delta_{ki}
        \delta_{il},\\
        {\left[ C_{zx}(U_z^{(i)}(\pi/2))\right]}_{kl} &=& 0.
    \end{array}
\end{equation}

For the CNOT-gate on control qubit $c$ and target qubit $t$ one finds
the propagation matrix $C(\CNOT(c,t))$ with
\begin{equation}
    \label{CNTpropmat}
    \begin{array}{rcl}       
        {\left[ C_{xx}(\CNOT(c,t))\right]}_{kl} & = & \delta_{kl}+
        \delta_{kt}\delta_{cl}, \\
        {\left[ C_{zz}(\CNOT(c,t))\right]}_{kl} & = & \delta_{kl}+
        \delta_{kc}\delta_{tl}, \\
        C_{zx}(\CNOT(c,t)) &=& 0,\\
        C_{xz}(\CNOT(c,t)) &=& 0.
    \end{array}
\end{equation}
We will make use of the propagation matrices in the discussion of
temporal complexity of algorithms on the \QC in Section~\ref{dg}.

For the action of the propagation matrices $C$ on the vectors
${\bf{I}} \in {\cal{V}}$ there exist
conserved quantities. One of them, ${\bf{I}}_{x,1}^T
{\bf{I}}_{z,2} + {\bf{I}}_{z,1}^T {\bf{I}}_{x,2}$, is discussed in
the next section.

\subsubsection{Conservation of the symplectic scalar product}
\label{ConservQt}

The symplectic scalar product 
\begin{equation}
    \label{sympscalp}
    ({\bf{I}}_1,{\bf{I}}_2)_S = {\bf{I}}^T_{x,1} {\bf{I}}_{z,2} +
    {\bf{I}}^T_{z,1} {\bf{I}}_{x,2}
    \;{\mbox{\small{mod}}}\; 2
\end{equation} 
is conserved. For any ${\bf{I}}_1,{\bf{I}}_2 \in {\cal{V}}$ and $C$
the identity 
\begin{equation}
    \label{conssympprod}
    \left({\bf{I}}_1,{\bf{I}}_2\right)_S = \left(C {\bf{I}}_1,C
    {\bf{I}}_2\right)_S 
\end{equation}
holds. Let us briefly explain why the symplectic scalar product
(\ref{sympscalp}) is conserved. First, note that the symplectic scalar
product tells whether two operators ${\cal{I}}({\bf{I}}_1)$,
  ${\cal{I}}({\bf{I}}_2)$ in the Pauli group commute or anti-commute,
\begin{equation}
    \label{acc}
    {\cal{I}}({\bf{I}}_1) {\cal{I}}({\bf{I}}_2)=
    {(-1)}^{{({\bf{I}}_1,{\bf{I}}_2)}_{\!S}} {\cal{I}}({\bf{I}}_2)
    {\cal{I}}({\bf{I}}_1).
\end{equation} 
Relation (\ref{acc}) is the only place in this paper where we
pay attention to the sign factor of a byproduct operator. There, the
product, e.g. ${\cal{I}}({\bf{I}}_1) {\cal{I}}({\bf{I}}_2)$, denotes the
usual operator product. However, everywhere else in this paper a
product ${\cal{I}}({\bf{I}}_1) {\cal{I}}({\bf{I}}_2)$ denotes operator
multiplication modulo a global phase factor $\pm 1$, i.e. the product
is normal ordered as in (\ref{Iso}) and the phase factor is dropped. 

Using relation (\ref{acc}), the invariance (\ref{conssympprod}) of the
scalar product (\ref{sympscalp}) is easily demonstrated. Consider the
quantity ${\cal{I}}(C{\bf{I}}_1) {\cal{I}}(C{\bf{I}}_2)$ with
${\cal{I}}(C {\bf{I}}_{1}) = U\, {\cal{I}}({\bf{I}}_{1})\, U^\dagger$
and ${\cal{I}}(C {\bf{I}}_{2}) = U\, {\cal{I}}({\bf{I}}_{2})\,
U^\dagger$ as in (\ref{correspondence}). Then, we find
\begin{equation}
    \label{inv1}
    \begin{array}{rcl}
        {\cal{I}}(C{\bf{I}}_1) {\cal{I}}(C{\bf{I}}_2) &=& U\,
        {\cal{I}}({\bf{I}}_{1})\, U^\dagger \, U\,
        {\cal{I}}({\bf{I}}_{2})\, U^\dagger \\
        &=& U\,
        {\cal{I}}({\bf{I}}_{1})\,
        {\cal{I}}({\bf{I}}_{2})\, U^\dagger \\
        &=& {(-1)}^{{({\bf{I}}_1, {\bf{I}}_2)}_{\!S}} U\,
        {\cal{I}}({\bf{I}}_{2})\,
        {\cal{I}}({\bf{I}}_{1})\, U^\dagger \\
        &=& {(-1)}^{{({\bf{I}}_1, {\bf{I}}_2)}_{\!S}} U\,
        {\cal{I}}({\bf{I}}_{2})\,U^\dagger \, U\,
        {\cal{I}}({\bf{I}}_{1})\, U^\dagger \\
         &=& {(-1)}^{{({\bf{I}}_1, {\bf{I}}_2)}_{\!S}}
        {\cal{I}}(C{\bf{I}}_2) {\cal{I}}(C{\bf{I}}_1), 
    \end{array}
\end{equation} 
where the third line holds by (\ref{acc}). On the other hand, as we can
see from (\ref{acc}) directly that
\begin{equation}
    \label{inv2}
    {\cal{I}}(C{\bf{I}}_1) {\cal{I}}(C{\bf{I}}_2) =
    {(-1)}^{{(C{\bf{I}}_1, C{\bf{I}}_2)}_{\!S}} 
    {\cal{I}}(C{\bf{I}}_2) {\cal{I}}(C{\bf{I}}_1). 
\end{equation}
From (\ref{inv1}) and (\ref{inv2}) together it follows that
$\left({\bf{I}}_1,{\bf{I}}_2\right)_S = \left(C {\bf{I}}_1,C
    {\bf{I}}_2\right)_S$, as stated in (\ref{conssympprod}).

The  symplectic scalar product (\ref{sympscalp}) will prove useful
in determining the measurement angles from previously obtained
measurement results.

\subsubsection{The cone test}
\label{conetest}

The cone test is used to find out whether two measurements, which
are part of  some gates of a circuit, influence each other,
i.e. whether 
one of the measurements has to wait for the result of the other. The
cone test does not reveal which of the two measurements has to be
performed first.  

Let $j,k$ be some cluster qubits $k\in{\cal{C}}$ and $j \in
 Q^{(1)}$. Qubit $j$ is not measured in the first measurement round
and thus the observable measured at qubit $j$ is a nontrivial linear
combination of $\sigma_x$ and $\sigma_y$, hence $j$ can be in the
forward and backward cones of some other cluster qubits. We would like
to find out whether $j$ is in the forward or backward cone of $k$. For
this question the cone test provides a necessary and sufficient
criterion. It reads
\begin{equation}
    \label{conecrit}
    \forall\;k \in {\cal{C}},j\in Q^{(1)}:\;\; j\in \fc(k) \vee
    j\in \bc(k) \Longleftrightarrow ({\bf{F}}_j,{\bf{F}}_k)_S=1. 
\end{equation}
To check whether a qubit lies in some other qubits backward or forward
cone we only need the two forward images and can use the symplectic
scalar product. 

We further observe that
\begin{equation}
    \label{fcbc}
    \forall\;j,k \in  Q^{(1)}:\;\; j \in \fc(k)
    \Longleftrightarrow k \in \bc(j). 
\end{equation}
If we confine $k$ to $k \in Q^{(1)} \subset {\cal{C}}$ 
we can insert (\ref{fcbc}) into (\ref{conecrit}) such that the expression on
the l.h.s. of (\ref{conecrit}) becomes symmetric with respect to $j$
and $k$. This fits in 
well since the r.h.s of (\ref{conecrit}) is also symmetric. 

Similar to (\ref{conecrit}) we can give a criterion for whether or not a
qubit $j \in Q^{(1)}$ is in the forward- or backward cone
$\mbox{fc}(g)$, $\mbox{bc}(g)$ of some gate $g$. It reads
\begin{equation}
    \label{gateconecrit}
    \forall g \in {\cal{N}},\,\, j \in Q^{(1)}: \; j \in
    \mbox{fc}(g) \vee j \in \mbox{bc}(g) \Longleftrightarrow
    ({\bf{F}}_j, {\bf{F}}_g)_S =1.
\end{equation}

The proofs
of the cone tests (\ref{conecrit}), (\ref{gateconecrit}) and the
forward-backward cone 
relation (\ref{fcbc}) are given in appendix~\ref{Coneproof}.

\subsection{To what a quantum logic network condenses}
\label{cond_mat}

Simulating a quantum logic network on a \QC is a two-stage
process. Before the genuine computation, we feed a classical computer
with the network to be
simulated. It returns the quantities needed to run the respective
algorithm on the \QCns. These quantities are the sets $Q_t$
of simultaneously measurable qubits,  the measurement bases of the
qubits $k \in Q_0$, the 
algorithm angles $\varphi_{l,\mbox{\footnotesize{algo}}}$ for $l \in
{\cal{C}} \backslash Q_0$,  the backward cones $\bc(k)$ of the
qubits $k \in Q_0$, the byproduct images ${\bf{F}}_j$ for $j
\in {\cal{C}}$ and the initialization value
${\bf{I}}_{\mbox{\footnotesize{init}}}$ of the information flow vector
${\bf{I}}(t)$. Together these quantities represent the program for the \QCns. 

In \cite{QCmeas} we wrote that the set of one-qubit measurements on a
cluster state represents the program. Now we can be more specific
about the measurement pattern representing the program for the
\QCns. The measurement pattern has both a temporal and a spatial
structure. The temporal structure is given by the sets $Q_t$ of
simultaneously measured qubits. The spatial structure consists of the
bases ($\sigma_x$-, $\sigma_y$- or
$\sigma_z$-) of the measurements in the first round and of the
measurement angles in the subsequent 
rounds. The measurement angles can be 
determined only run-time, since they involve the random outcomes of
previous measurements. The measurement angles are determined using the
algorithm angles and the byproduct images.

\section{Computational model for the \QC}
\label{about}

In the preceding two sections we have established the notions of the
sets of simultaneously measurable qubits, 
backward cones, byproduct  
images, measurement angles and the information vector. In this
section, the 
computational model underlying  the \QC is described in these
terms. First, we would like to give a summary of the characteristic
features of the model:

\begin{itemize}
    \item{The \QC has  no quantum input
          and no quantum output.}
    \item{For any given quantum algorithm, the 
          cluster ${\cal{C}}$ is divided into disjoint subsets $Q_t \subset
          {\cal{C}}$ of qubits, $t =
          0,1,\dots,t_{\mbox{\footnotesize{max}}}$, where $Q_p \cap Q_q =
          \emptyset$ for $p \neq q$ and
          $\bigcup_{t=0}^{t_{\mbox{\tiny{max}}}} Q_t = 
          {\cal{C}}$. These subsets are 
          measured one after the other in the order given by the index
          $t$. In measurement round $t$ the
          set $Q_t$ of qubits is measured.} 
    \item{The classical information gained by the measurements is
          processed within a flow scheme. The flow quantity is a
          classical $2n$-component binary vector ${\bf{I}}(t)$,
          where $n$ is the 
          number of logical qubits of a corresponding quantum  logic
          network and $t$ the number of the measurement round.}  
    \item{This vector ${\bf{I}}(t)$, the {\em{information flow vector}}, is
          updated after every 
          measurement round. That is, after the one-qubit measurements of
          all qubits of
          a set $Q_{t}$ have been performed
          simultaneously, ${\bf{I}}(t-1$) is updated to
          ${\bf{I}}(t)$ through the results of these measurements. In
          turn, ${\bf{I}}(t)$ determines  
          which one-qubit observables are
          to be measured of the qubits of the set $Q_{t+1}$.} 
    \item{The result of the computation is given by the information
          flow vector ${\bf{I}}(t_{\mbox{\footnotesize{max}}})$ 
          after the last measurement round. From this
          quantity the readout measurement result of the
          quantum register 
          in the corresponding quantum logic network can be read off directly
          without further processing.} 
\end{itemize}
\begin{figure}
    \begin{center}
    \epsfig{file=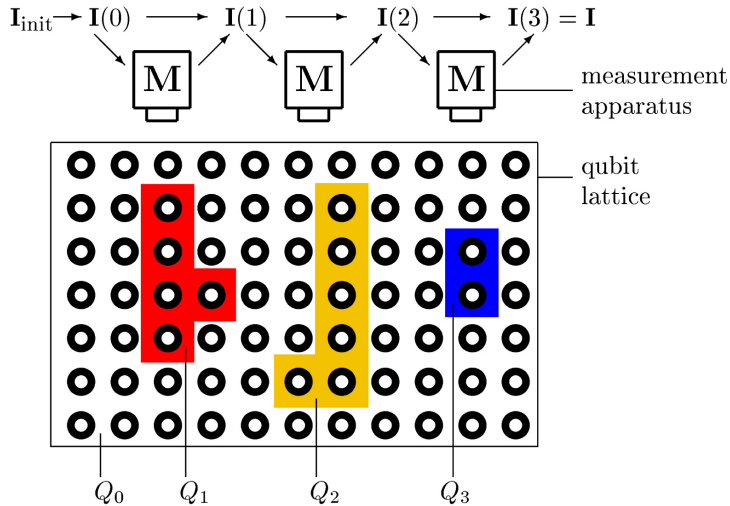,width=10cm}
    \parbox{0.83\textwidth}{\caption{\label{Schema}General scheme of
        the quantum computer via one-qubit measurements. The sets
    $Q_t$ of lattice qubits are measured one after the other. The
    results of earlier measurements determine the measurement bases of
    later ones. All classical information from the measurement results
    needed to steer the \QC is contained in the information flow vector
    ${\bf{I}}(t)$. After the last measurement round
    $t_{\mbox{\footnotesize{max}}}$,
    ${\bf{I}}(t_{\mbox{\footnotesize{max}}})$ contains the result of
    the computation.}}  
    \end{center}
\end{figure}
We should make a comment on the first point. The \QC has no quantum
output. Of course, the final result of any computation --including
quantum computations-- is a
classical number, but for the quantum logic network the state of the
output register before the readout measurements plays a distinguished
role. For the \QC this is not the case, there are just cluster qubits
measured in a certain order and basis. If, to perform a
particular algorithm on the \QCns, a quantum logic network is
implemented on a cluster state there is a subset of cluster qubits
which play the role of the output register. These
qubits are, however, not the final qubits to be measured, but among the
first (!).

The \QC has no quantum input. This means that the quantum input state
is {\em{known}} and can thus be created from some standard quantum state,
e.g. $|00...0\rangle$, by a circuit preceding the main part of the
computation. Shor's 
algorithm where one starts with an input state $\bigotimes_{i=1}^n
1/\sqrt{2}(|0 \rangle_i + |1 \rangle_i)$ is an example for such a
situation. Other scenarios are conceivable, e.g. where an unknown
quantum input is processed and the classical result of the computation
is retransmitted to the sender of the input state; or the unmeasured
network output register state is retransmitted. These scenarios would lead
only to slight modifications in the computational model. They are,
however, not in the focus of this paper. The reader
who is interested in  how to read in and  process an unknown quantum
state with the \QC is referred to \cite{QCmeas}. 

\subsection{Obtaining the computational result from the
  measurement outcomes}
\label{Ivonsk}

Now that we have defined the information vector ${\bf{I}}$ in
(\ref{Ivector}) and have seen that the result of the computation can
be directly read off from the $x$-part of ${\bf{I}}$, we will explain how ${\bf{I}}$ depends on the measurement outcomes $\big\{
s_k \big \}$ and the set $\big\{
\kappa_k \big \}$ of binary numbers that determine the cluster state
$|\phi\rangle_{\cal{C}}$ in (\ref{EVeqn}). For this purpose, we will
express $U_{\Sigma R}$ in terms of $\big\{
s_k \big \}$ and $\big\{
\kappa_k \big \}$, and use the isomorphism (\ref{Iso}) to obtain
${\bf{I}}$. As in (\ref{ExtByprop}), $U_{\Sigma R}$ can be decomposed
into two parts, the accumulated byproduct operator $U_\Sigma$ and the
contribution from the ``readout'' measurements, $U_R$. We will proceed
in three steps. First,
we will consider $U_\Sigma$ for the case where only the cluster qubits
for the ``readout'' and those of which observables $\mbox{cos}\varphi
\, \sigma_x + \mbox{sin}\varphi \, \sigma_y$ are measured are
present. Second, we will extend the 
obtained result to the case where 
both the relevant and the redundant qubits are present, i.e. where a
universal cluster ${\cal{C}}$ is used for computation. Third, we will
include the contribution  $U_R$ from the ``readout'' qubits.

To derive ${\bf{I}}$ as a function of $\big\{
s_k \big \}$ and $\big\{
\kappa_k \big \}$, we need to define the following sets. ${\cal{C}}$
is a universal cluster. Let $O \subset {\cal{C}}$ be the subset of the
cluster which, in the simulation of a quantum logic network on the
\QCns, consists of the readout qubits. Let ${\cal{C}}_N
\subset {\cal{C}}$
denote the cluster that contains only the relevant cluster qubits,
i.e. those which are measured in a direction in the equator of the
Bloch sphere, and the ``readout'' qubits. Be $Q_{0,z} \subset
{\cal{C}}$ the set of qubits of which the operator $\sigma_z$ is
measured. Among these sets, the following relations hold:
\begin{equation}
    \label{Sets1}
    \begin{array}{rcl}
        {\cal{C}}_N \cup Q_{0,z} &=& {\cal{C}}\\
         {\cal{C}}_N \cap Q_{0,z} &=& O.
    \end{array}
\end{equation}  
Further, let us denote a standard form of the cluster on which the
gate $g$ can be implemented as 
${\cal{C}}(g)$. The cluster ${\cal{C}}(g)$ shall consist only of essential
qubits, i.e. those which are measured in a direction in the 
equator of the Bloch sphere ($x$-$y$-plane). The
set ${\cal{C}}_N$ of qubits  is the union of all clusters
${\cal{C}}(g)$
\begin{equation}
    \label{Sets2}
    {\cal{C}}_N = \bigcup_{g \in {\cal{N}}} {\cal{C}}(g) .
\end{equation}
Each ${\cal{C}}(g)$ is
subdivided into an input zone ${\cal{C}}_I(g)$, an output zone
${\cal{C}}_O(g)$ and the body of the gate ${\cal{C}}_M(g)$;
${\cal{C}}_I(g) \cap {\cal{C}}_O(g) = \emptyset,
{\cal{C}}_M(g)={\cal{C}}(g) \backslash \left({\cal{C}}_I(g) \cup
    {\cal{C}}_O(g)\right)$. In the two cases described in
Section~\ref{summary}, the general rotation and the CNOT-gate, these
sets are 
${\cal{C}}(U_{Rot}(\xi,\eta,\zeta))=\{1,2,3,4,5\}$,
${\cal{C}}_I(U_{Rot}(\xi,\eta,\zeta))=\{1\}$,
${\cal{C}}_M(U_{Rot}(\xi,\eta,\zeta)) 
= \{2,3,4\}$, ${\cal{C}}_O(U_{Rot}(\xi,\eta,\zeta))=\{5\}$ and
${\cal{C}}(CNOT)=\{1,..,15\}$, ${\cal{C}}_I(CNOT)=\{1,9\}$,
${\cal{C}}_M(CNOT) 
= \{2,..,6,8, 10,..,14\}$, ${\cal{C}}_O(CNOT)=\{7,15\}$, with the labeling of qubit
sites according to 
Fig.~\ref{Gates}. 

Let us now explain in greater detail the notation
$\kappa_{k,I},\kappa_{l,O}$ which made use of in equations (\ref{Rotproc}),
(\ref{Byprod1}) and (\ref{CNOTbyprop}) in Section~\ref{summary}. 
Some of the sets ${\cal{C}}(g)$ have an overlap. If
a qubit $k$ in the output zone ${\cal{C}}_O(g)$, $k \in
{\cal{C}}_O(g)$ is not a readout qubit, it is also in the input zone
of some gate $\tilde{g}$ succeeding $g$, i.e. $k \in
{\cal{C}}_I(\tilde{g})$.    
 The procedures to implement the gates $g, \tilde{g}$ on
clusters ${\cal{C}}(g), {\cal{C}}(\tilde{g})$ depend on the
eigenvalues of the states
$|\phi\rangle_{{\cal{C}}(g)},|\phi\rangle_{{\cal{C}}(\tilde{g})}$ in the
respective eigenvalue equations of form (\ref{EVeqn}). Namely, the
procedures depend on 
the set $\{ \kappa_a^\prime, a \in {\cal{C}}(g) \}$ or  $\{
\kappa_a^\prime, a \in {\cal{C}}(\tilde{g}) \}$, respectively. In the
described case, we have
$k \in {\cal{C}}(g)$, $k \in {\cal{C}}(\tilde{g})$ and $k \in 
{\cal{C}}_N$. Thus, there are three eigenvalue equations associated
with $k$, one for the state $|\phi\rangle_{{\cal{C}}(g)}$, one
for $|\phi\rangle_{{\cal{C}}(\tilde{g})}$ and one for
$|\phi\rangle_{{\cal{C}}_N}$. In all the three cases the symbols
$K^{(k)}$ denote different operators, because the sets $ngbh(k)$ are
different. For ${\cal{C}}(g)$, $k\in {\cal{C}}_O(g)$ has a
left but no right 
neighbor. For ${\cal{C}}(\tilde{g})$, $k \in {\cal{C}}_I(\tilde{g})$
has a right but no left 
neighbor. For ${\cal{C}}_N$, $k$ has both a left and right
neighbor. We specify the eigenvalue in the relevant equation
(\ref{EVeqn}) for the 
state $|\phi\rangle_{{\cal{C}}(g)}$, where $k$ is an output qubit, by
$\kappa^\prime_{k,O}$, 
i.e. $K^{(k)}|\phi\rangle_{{\cal{C}}(g)} = 
{(-1)}^{\kappa^\prime_{k,O}} |\phi\rangle_{{\cal{C}}(g)}$. Similarly,
we specify the eigenvalue in the equation for
$|\phi\rangle_{{\cal{C}}(\tilde{g})}$, where $k$ is an input qubit, by
$\kappa^\prime_{k,I}$ and in the equation for the state
$|\phi\rangle_{{\cal{C}}_N}$ by $\kappa^\prime_k$. The
$\kappa^\prime_{k,I}$, $\kappa^\prime_{k,O}$ and
$\kappa^\prime_k$ are generally
different, but related in the following way
\begin{equation}
    \label{kapparel1}
    \begin{array}{lrcl}
        \displaystyle{\forall k\;|\;\exists \, g,\tilde{g} \in {\cal{N}}, \;
        \mbox{such that}\; k \in {\cal{C}}_O(g) \wedge k \in
        {\cal{C}}_I(\tilde{g}):} & \kappa^\prime_k &=&
        \kappa^\prime_{k,I}+\kappa^\prime_{k,O},\\
        \displaystyle{\forall k\;|\; \exists g \in {\cal{N}}, \;
        \mbox{s.th.}\; k \in {\cal{C}}_O(g) \,\, \wedge \,\,\neg
        \exists \tilde{g} \in {\cal{N}}, \; 
        \mbox{s.th.}\; k \in {\cal{C}}_I(\tilde{g}):} & \kappa^\prime_k &=&
        \kappa^\prime_{k,O}\\
        \displaystyle{\forall k\;|\; \neg \exists g \in {\cal{N}}, \;
        \mbox{s.th.}\; k \in {\cal{C}}_O(g) \,\, \wedge \,\, \exists
        \tilde{g} \in {\cal{N}}, \; 
        \mbox{s.th.}\; k \in {\cal{C}}_I(\tilde{g}):} & \kappa^\prime_k &=&
        \kappa^\prime_{k,I}.
    \end{array}
\end{equation}
The latter case of (\ref{kapparel1}) applies to cluster qubits which
belong to the cluster 
equivalent of the input- or output quantum register. The first line of
(\ref{kapparel1}) is proved in appendix~\ref{Kaprel}, the second
and third line are straightforward. 

For all qubits $k$ in the gate bodies, i.e. for which $\exists
  g\in{\cal{N}}\;\;\mbox{such 
  that}\; k\in {\cal{C}}_M(g) \subset {\cal{C}}_N$, the $\kappa^\prime_k$ that defines
$|\phi\rangle_{{\cal{C}}(g)}$ and the $\kappa^\prime_k$ that defines
$|\phi\rangle_{{\cal{C}}_N}$ are equivalent,
\begin{equation}
    \label{kapparel2}
    \forall k\,|\, \exists g \in {\cal{N}},\;
    \mbox{such that} \; k \in {\cal{C}}_M(g): \; \kappa^\prime_{k \in
    {\cal{C}}(g)} = \kappa^\prime_{k \in 
    {\cal{C}}_N} 
\end{equation}
and are therefore denoted by the same symbol. Provided with these
definitions and relations, we can now start to express ${\bf{I}}$ in
terms of $\big \{ s_k, k \in {\cal{C}} \big \}$ and  $\big \{ \kappa_k,
k \in {\cal{C}} \big \}$.

Let us --in the first step-- discuss the accumulated byproduct
operator $U_\Sigma$ for 
a computation on the special cluster ${\cal{C}}_N$. To
$U_\Sigma$ contribute all the byproduct operators $U_\Sigma(g)$ that
are created in the implementation of the gates $g$. For all necessary
cases, the general rotations (\ref{Byprod1}), the CNOT-gate
(\ref{CNOTbyprop}) and the special rotations
Hadamard-gate and $\pi/2$-phase-gate (\ref{Hadabyprop}), the byproduct
operators 
$U_\Sigma(g)$ can be written in the form 
\begin{equation}
    \label{Usg}
    U_\Sigma(g) = \left(\prod_{k \in {\cal{C}}_I(g)}
    {\left(U_k\right)}^{s_k+\kappa_{k,I}^\prime} \right)
    \left(\prod_{k \in {\cal{C}}_M(g)}
    {\left(U_k\right)}^{s_k+\kappa_k^\prime} \right)
    \left(\prod_{k \in {\cal{C}}_O(g)}
    {\left(U_k\right)}^{\kappa_{k,O}^\prime}\right) 
    U_0(g),
\end{equation}
where we have attributed the contributions to $U_\Sigma$ which depend
on $\kappa_{k,I}^\prime$, $\kappa_{k}^\prime$ or $\kappa_{k,O}^\prime$
to the qubit $k$. $U_0(g)$ is constant in the measurement outcomes
$\big \{ s_k, k \in 
{\cal{C}}_N \backslash O \big \}$ and $\big \{ \kappa_k^\prime, k \in
{\cal{C}}_N \backslash O \big \}$ and we therefore attribute it to the
gate $g$ as a whole rather than to a particular cluster qubit. For all
rotations we have 
$U_0(g)={\bf{1}}$, but for the CNOT-gate --if  realized
as depicted in Fig.~\ref{Gates}-- the contribution is nontrivial as
can be read off from (\ref{CNOTbyprop}),
$U_0(CNOT)=\sigma_z^{(c)}$. 

To determine the effect of $U_\Sigma(g)$ on $U_\Sigma$ we propagate, by use
of the propagation relations (\ref{Rotprop}), (\ref{CNTprop}) and
(\ref{Hadaprop}), the byproduct operators $U_\Sigma(g)$
forward to the cut $\Omega$ which intersects the corresponding network
${\cal{N}}$ just before the output. The forward propagated byproduct
operator that results from the byproduct operator $U_\Sigma(g)$ we
denote by $U_\Sigma(g)|_\Omega$. In the same way, the forward
propagated byproduct operator originating from $U_k$, the byproduct
operator generated via the measurement of qubit $k$, is denoted by
$U_k|_\Omega$ for all $k \in {\cal{C}}_N \backslash O$ . Finally, the forward
propagated byproduct operator 
originating from $U_0(g)$, the byproduct operator attributed to the
gate $g$ as a whole, is denoted by $U_0(g)|_\Omega$. To give an
explicit expression, be ${\cal{O}}$ the vertical cut through a network
${\cal{N}}$ at the output of a gate $g$ and $U({\cal{N}}_{{\cal{O}}
    \rightarrow \Omega})$ the unitary operation in the Clifford group
which corresponds to the part of the network ${\cal{N}}$ with all the
one-qubit rotations except for the Hadamard- and $\pi/2$-phase gates
replaced by the identity, as explained in
Section~\ref{propmats}. Then, the forward propagated byproduct
operators are given by
\begin{equation}
    \label{fpbop}
    \begin{array}{rcl}
        {U_\Sigma(g)|}_\Omega &=& U({\cal{N}}_{{\cal{O}}
    \rightarrow \Omega})\, U_\Sigma(g) \,U({\cal{N}}_{{\cal{O}}
    \rightarrow \Omega})^\dagger\\
    {U_k|}_\Omega &=& U({\cal{N}}_{{\cal{O}}
    \rightarrow \Omega})\, U_k  \,U({\cal{N}}_{{\cal{O}}
    \rightarrow \Omega})^\dagger\\
    {U_0(g)|}_\Omega &=& U({\cal{N}}_{{\cal{O}}
    \rightarrow \Omega})\, U_0(g)  \,U({\cal{N}}_{{\cal{O}}
    \rightarrow \Omega})^\dagger\\
    \end{array}
\end{equation}
The contribution $U_\Sigma(g)|_\Omega$  from the gate $g$ to $U_\Sigma$ is
\begin{equation}
    \label{Usomg}
    U_\Sigma(g)|_\Omega = \left(\prod_{k \in {\cal{C}}_I(g)}
    {U_k|_\Omega}^{s_k+\kappa_{k,I}^\prime} \right)
    \left(\prod_{k \in {\cal{C}}_M(g)}
    {U_k|_\Omega}^{s_k+\kappa_k^\prime} \right)
    \left(\prod_{k \in {\cal{C}}_O(g)}
    {U_k|_\Omega}^{\kappa_{k,O}^\prime}\right) 
    U_0(g)|_\Omega.
\end{equation}
The total byproduct operator $U_\Sigma$ is the product of all forward
propagated byproduct operators $U_\Sigma(g)|_\Omega$, $U_\Sigma =
\prod_{g \in {\cal{N}}} U_\Sigma(g)|_\Omega$, and
thus given by    
\begin{equation}
    \label{Us}
    U_\Sigma  = \prod_{g \in {\cal{N}}}
    \left(  U_0(g)|_\Omega \prod_{k \in {\cal{C}}_I(g)}
    {U_k|_\Omega}^{s_k+\kappa_{k,I}^\prime} 
    \prod_{k \in {\cal{C}}_M(g)}
    {U_k|_\Omega}^{s_k+\kappa_k^\prime} 
    \prod_{k \in {\cal{C}}_O(g)}
    {U_k|_\Omega}^{\kappa_{k,O}^\prime}
   \right).
\end{equation}
As can be seen from eq.~(\ref{Us}), for $k \in {\cal{C}}_M(g), \; g
\in {\cal{N}}$ the accumulated byproduct operator $U_\Sigma$
depends only on the combinations $s_k + \kappa_k^\prime$. This is, in
fact, true for all $k \in {\cal{C}}$ as will be shown below. That this
property of $U_\Sigma$ in its general form is not directly visible in
eq.~(\ref{Us}) is a disadvantage of this equation. A further
disadvantage of (\ref{Us}) is that it depends on the 
$\kappa_{k,I}^\prime, \; k \in {\cal{C}}_I(g)$, 
$\kappa_{l}^\prime, \; l \in {\cal{C}}_M(g)$ and 
$\kappa_{m,O}^\prime, \; m \in {\cal{C}}_O(g)$, but not on the
$\kappa_k,\; k \in {\cal{C}}$ which define the cluster state
$|\phi\rangle_{\cal{C}}$. 

To find a more convenient expression for $U_\Sigma$ we first
observe that by use of eq.~(\ref{Us}) $U_\Sigma$ can be
written in the form
\begin{equation}
    \label{Us2}
    U_\Sigma = \left( \prod_{k \in {\cal{C}}_N\backslash O}
    {U_k|_\Omega}^{s_k} \right) \cdot \mbox{const}\left(\big\{ \kappa_k^\prime,
    \,\, k \in {\cal{C}}_N \big \}\right).
\end{equation}
This is possible because in eq.~(\ref{Us}) for $k \in {\cal{C}}_O(g)$
the factors contributing to $U_\Sigma$ do not depend on
$\{s_k\}$. Thus, if one only picks  the $s_k$-dependent
contributions the 
product index variable $k$ runs over $\bigcup_{g \in
  {\cal{N}}} 
{\cal{C}}_I(g) \cup {\cal{C}}_M(g) = {\cal{C}}_N \backslash O$. Now,
 $U_\Sigma$ depends for all $k \in {\cal{C}}_N \backslash O$ only on
the combinations $s_k + \kappa_k^\prime$ which can be seen as follows:
Let us consider two cluster states $|\phi\rangle_{{\cal{C}}_N}$ and
$|\tilde{\phi}\rangle_{{\cal{C}}_N}$ on the cluster ${\cal{C}}_N$
which are related via $|\tilde{\phi}\rangle_{{\cal{C}}_N}=
\sigma_z^{(k)} |\phi\rangle_{{\cal{C}}_N}$, such that the respective
eigenvalues are related by 
$\kappa_k^\prime(|\tilde{\phi}\rangle_{{\cal{C}}_N}) =
\kappa_k^\prime(|\phi\rangle_{{\cal{C}}_N}) + 1 \; \mbox{mod}\,\,2$,
and  $\kappa_l^\prime(|\tilde{\phi}\rangle_{{\cal{C}}_N}) =
\kappa_l^\prime(|\phi\rangle_{{\cal{C}}_N})$ for all $l \neq
k$. Suppose one would use the cluster state
$|\phi\rangle_{{\cal{C}}_N}$ for a computation. For each
choice of $k \in {\cal{C}}_N \backslash O$ qubit $k$
is measured in some direction $\vec{r}$ on the equator of the Bloch
sphere, i.e. the 
operator $\vec{r} \cdot \vec{\sigma}^{(k)}$ with $\vec{r} \cdot
\vec{e}_z=0$ is measured, and the measurement result is $s_k$. For the
resulting state one finds
\begin{equation}
    \label{phiphiteq}
    \frac{1+{(-1)}^{s_k}\vec{r} \cdot \vec{\sigma}^{(k)}}{2}\,
    |\phi\rangle_{{\cal{C}}_N} = \sigma_z^{(k)} \frac{1+{(-1)}^{s_k+1}
      \vec{r} \cdot \vec{\sigma}^{(k)}}{2}\,
    |\tilde{\phi}\rangle_{{\cal{C}}_N}. 
\end{equation}
The state on the r.h.s. of eq.~(\ref{phiphiteq}) is modulo the posterior
$\sigma_z^{(k)}$ equal to the state into which one had projected if the
cluster state $|\tilde{\phi}\rangle_{{\cal{C}}_N}$ instead of
$|\phi\rangle_{{\cal{C}}_N}$ was used for computation and the result
$s_k+1\;\mbox{mod}\,\,2$ was obtained. Since we are only interested in
the measurement 
results but not the resulting quantum state, the posterior
$\sigma_z^{(k)}$ is irrelevant for the computation. Using a cluster state
$|\phi\rangle_{{\cal{C}}_N}$, characterized by $\kappa_k^\prime$, and
obtaining the measurement result $s_k$ is equivalent to using a
cluster state $|\tilde{\phi}\rangle_{{\cal{C}}_N}$, characterized by
$\kappa_k^\prime + 1 \; \mbox{mod}\,\,2$, and obtaining the
measurement result $s_k+1 \; \mbox{mod}\,\,2$. Thus, $U_\Sigma$ can
depend only on $s_k+\kappa_k^\prime \; \mbox{mod}\,\,2$ for all $k \in
{\cal{C}}_N \backslash O$, i.e.
\begin{equation}
    \label{Us3}
    U_\Sigma = \left( \prod_{k\in {\cal{C}}_N \backslash O}
    {U_k}^{s_k+\kappa_k^\prime} \right) \cdot \mbox{const.}
\end{equation}
Now, the constant operator can be identified via comparison of eqs.~(\ref{Us})
and (\ref{Us3}). For this purpose, we choose $\big \{\kappa_{k,I}^\prime=0,
\kappa_{l}^\prime=0, \kappa_{m,O}^\prime=0,\; \forall g \in
{\cal{N}}, k \in {\cal{C}}_I(g),  l \in {\cal{C}}_M(g),  m \in
{\cal{C}}_O(g) \big \}$ and insert it into eq.~(\ref{Us}). This choice
implies via eq.~(\ref{kapparel1}) that $\big \{ \kappa_k^\prime = 0,\;
\forall k \in {\cal{C}}_N \big \}$ which we insert into
eq.~(\ref{Us3}). In this way we find $\mbox{const} = \prod_{g \in
  {\cal{N}}} U_0(g)|_\Omega$ and thus
\begin{equation}
    \label{Us4}
     U_\Sigma = \prod_{k\in {\cal{C}}_N \backslash O}
    {U_k}^{s_k+\kappa_k^\prime}  \cdot \prod_{g \in
  {\cal{N}}} U_0(g)|_\Omega. 
\end{equation}

Let us now --in the second step-- include the effect of the qubits $l
\in Q_{0,z} \backslash O$ on $U_\Sigma$. These qubits are among the
redundant qubits which are measured in the first measurement
round. Redundant here always means redundant with respect to a given
quantum logic network to be simulated. If one starts with a
universal cluster state $|\phi\rangle_{\cal{C}}$ on a cluster
${\cal{C}}$ and projects out the qubits $l \in Q_{0,z} \backslash O$
the resulting state on the unmeasured qubits $k \in {\cal{C}}_N$ is
again a cluster state $|\phi\rangle_{{\cal{C}}_N}$. The eigenvalues
that specify the state $|\phi\rangle_{{\cal{C}}_N}$ in an equation
analogous to eq.~(\ref{EVeqn}) depend on the results of the
$\sigma_z$-measurements on the qubits  $l \in Q_{0,z} \backslash O$.
For the set $\big \{\kappa_k^\prime,\; k \in {\cal{C}}_N \big \}$ that
specifies  $|\phi\rangle_{{\cal{C}}_N}$ one finds
\begin{equation}
    \label{kappakappaprime}
    \forall k \in {\cal{C}}_N:\; \kappa_k^\prime = \kappa_k +
    \sum_{j \in \, nbgh(k) \cap Q_{0,z} \backslash O} s_j,
\end{equation} 
as can be easily derived from eq.~(\ref{EVeqn}). In
eq.~(\ref{kappakappaprime}) the $\kappa_k$ are those which specify the
universal cluster state  $|\phi\rangle_{\cal{C}}$ in (\ref{EVeqn}).
If one now inserts eq.~(\ref{kappakappaprime}) into (\ref{Us4}) one
obtains
\begin{equation}
    \label{Us5}
    U_\Sigma = \left( \prod_{k \in {\cal{C}}_N \backslash O}
    {U_k|_\Omega}^{s_k + \kappa_k} \right) \left( \prod_{k \in
    {\cal{C}}_N \backslash O} \prod_{ \footnotesize{\begin{array}{rl}
    j|& \!\!\!\!j \in nbgh(k)\, \wedge \\ & \!\!\!\! j \in Q_{0,z} \backslash O
    \end{array}}} {U_k|_\Omega}^{s_j} \right) \left( \prod_{g \in
  {\cal{N}}} U_0(g)|_\Omega \right).
\end{equation}
Therein and in the following it should be understood that a product of
operators is set equal to the unity operator if the index variable runs over an empty set. The second factor in eq.~(\ref{Us5}) can now be rewritten in the
following way
\begin{equation}
    \label{F2}
    \begin{array}{rcl}
        \displaystyle{\prod \limits_{k \in
            {\cal{C}}_N \backslash O} \prod \limits_{
            \footnotesize{\begin{array}{rl} 
                  j|& \!\!\!\!j \in nbgh(k)\, \wedge \\ & \!\!\!\! j
                  \in Q_{0,z} \backslash O \end{array}}}
            {U_k|_\Omega}^{s_j}} &=&  \displaystyle{\prod
            \limits_{j \in {\cal{C}}_N \backslash O} \prod \limits_{
              \footnotesize{\begin{array}{rl}  k|& \!\!\!\!j \in nbgh(k)\,
                    \wedge \\ & \!\!\!\! k \in Q_{0,z} \backslash O
                \end{array}}} {U_j|_\Omega}^{s_k}} \vspace{0.2cm}\\  
            &=&  \displaystyle{\prod \limits_{ \footnotesize{\begin{array}{rl}
                      (j,k) |& \!\!\!\! j \in {\cal{C}}_N \backslash
                      O\, \wedge \\
                      & \!\!\!\!j \in nbgh(k)\, \wedge \\ & \!\!\!\! k \in
                      Q_{0,z} \backslash O \end{array}}}
                      {U_j|_\Omega}^{s_k} } \vspace{0.2cm}\\
            &=& \displaystyle{\prod \limits_{k \in
                      Q_{0,z}\backslash O} \prod \limits_{
                      \footnotesize{\begin{array}{rl} 
                            j|& \!\!\!\!j \in nbgh(k)\, \wedge \\ & \!\!\!\! j
                            \in {\cal{C}}_N \backslash O \end{array}}}
                    {U_j|_\Omega}^{s_k}}. 
    \end{array}
\end{equation}
In the first line of (\ref{F2}) the labels $j$ and $k$ were
interchanged and the relation $j \in nbgh(k) \; \Longleftrightarrow \;
k \in nbgh(j)$ was used. In the second and third line the order of the
products over $k$ and $j$ was interchanged.

We now define the forward propagated byproduct operators $U_k|_\Omega$
for qubits $k \in Q_{0,z} \backslash O = {\cal{C}} \backslash {{\cal{C}}_N}$ as
\begin{equation}
    \label{Bypop0z}
    U_k|_\Omega = \prod \limits_{
      \footnotesize{\begin{array}{rl} j|& \!\!\!\!j \in nbgh(k)\,
      \wedge \\ 
      & \!\!\!\! j \in {\cal{C}}_N \backslash O \end{array}}}
      {U_j|_\Omega}\,,\;\;\;\;\; \forall  k \in Q_{0,z} \backslash O.
\end{equation}
In this way, we have traced back the forward propagated byproduct
operators for qubits $ k \in Q_{0,z} \backslash O$ to those for qubits
$j \in {\cal{C}}_N \backslash O$ which are already known. On the level
of the corresponding byproduct images we find via the isomorphism
(\ref{Iso}) 
\begin{equation}
    \label{F0z}
    {\bf{F}}_k = \sum \limits_{
      \footnotesize{\begin{array}{rl} j|& \!\!\!\!j \in nbgh(k)\,
      \wedge \\ 
      & \!\!\!\! j \in {\cal{C}}_N \backslash O \end{array}}}
      {\bf{F}}_j\,,\;\;\;\;\; \forall  k \in Q_{0,z} \backslash O.
\end{equation}
If we insert the definition (\ref{Bypop0z}) into (\ref{F2}) we obtain
\begin{equation}
    \label{F2-2}
    \displaystyle{\prod \limits_{k \in
            {\cal{C}}_N \backslash O} \prod \limits_{
            \footnotesize{\begin{array}{rl} 
                  j|& \!\!\!\!j \in nbgh(k)\, \wedge \\ & \!\!\!\! j
                  \in Q_{0,z} \backslash O \end{array}}}
            {U_k|_\Omega}^{s_j}} = \displaystyle{\prod_{k \in Q_{0,z}
            \backslash O} {U_k|_\Omega}^{s_k}.}
\end{equation}
Thus, with $Q_{0,z} \backslash O = {\cal{C}} \backslash {\cal{C}}_N$
from eq.~(\ref{Sets1}) and substituting (\ref{F2-2}) into (\ref{Us5}),
one finds
\begin{equation}
    \label{Us6}
    U_\Sigma = \left( \prod \limits_{k\in {\cal{C}}\backslash O}
    {U_k|_\Omega}^{s_k} \right)  \left( \prod \limits_{k\in
    {\cal{C}}\backslash Q_{0,z}} {U_k|_\Omega}^{\kappa_k} \right)
    \left( \prod_{g \in {\cal{N}}} U_0(g)|_\Omega
    \right). 
\end{equation}

Finally --in the third step-- we investigate the contribution from $U_R$ to
$U_{\Sigma R}$ which comes from the set $O$ of ``output''
qubits. With eq.~(\ref{ExtByprop}) we have $U_R = \prod_{k=1}^n 
    \left( \sigma_x^{(k)} \right)^{s_k}$ where the $n$ cluster qubits in the
set $O$ shall be labeled in the same way as the $n$ logical
qubits. Then, we can define the propagated byproduct operators
$U_k|_\Omega$ for $k \in O$ as
\begin{equation}
    \label{BypropO}
    U_k|_\Omega = \sigma_x^{(k)}\, ,\;\;\;\; \forall k \in O,
\end{equation}
and the corresponding byproduct images via (\ref{FI})
\begin{equation}
    \label{ByimO}
    {\bf{F}}_k =
    {(0_1,0_2,..,0_{k-1},1_k,0_{k+1},..,0_n;0_{n+1},..,0_{2n})}^T\,
    ,\;\;\;\; \forall k \in O, 
\end{equation}
where e.g. $1_k$ denotes a 1 at the $k$th position of ${\bf{F}}_k,\; k \in 0$. 
Combining eqs.~(\ref{Us6}) and (\ref{BypropO}) in (\ref{ExtByprop}),
the extended byproduct operator $U_{\Sigma R} = U_{\Sigma} U_R$
becomes 
\begin{equation}
    \label{Usr}
    U_{\Sigma R} = \left( \prod \limits_{k\in {\cal{C}}}
    {U_k|_\Omega}^{s_k} \right)  \left( \prod \limits_{k\in
    {\cal{C}}\backslash Q_{0,z}} {U_k|_\Omega}^{\kappa_k} \right)
    \left( \prod_{g \in {\cal{N}}} U_0(g)|_\Omega
    \right). 
\end{equation}
Via the isomorphism (\ref{Iso}) and using the definitions of the
byproduct images (\ref{FI}) and (\ref{Fg}) one can now express the
information vector ${\bf{I}} = {\cal{I}}^{-1} (U_{\Sigma R})$ as a
function of the measurement results $\{ s_k,\; k \in {\cal{C}} \}$ and
the $\{ \kappa_k, \; k \in {\cal{C}} \}$ defining the cluster state
$|\phi\rangle_{\cal{C}}$ on the cluster ${\cal{C}}$,
\begin{equation}
    \label{Ivsk}
    {\bf{I}} = \sum_{k \in {\cal{C}}} s_k {\bf{F}}_k + \sum_{k \in
    {\cal{C}}\backslash Q_{0,z}} \kappa_k {\bf{F}}_k + \sum_{g \in
    {\cal{N}}} {\bf{F}}_g
\end{equation} 
To derive the expression (\ref{Ivsk}) for the information vector
has been the primary purpose of this section.

With the expression (\ref{Ivsk}) at hand we are finally able to define
the quantity which carries the algorithmic information
during the computational process and which has already been mentioned
on earlier occasions in this paper, the {\em{information flow 
    vector}} ${\bf{I}}(t)$.  
\begin{definition}
The information flow vector ${\bf{I}}(t)$ is given by
\begin{equation}
    \label{Iflow}
    {\bf{I}}(t)= \sum \limits_{k \in \bigcup\limits_{i=0}^t Q_i}
    s_k\,{\bf{F}}_k + \sum_{k \in
    {\cal{C}}\backslash Q_{0,z}} \kappa_k {\bf{F}}_k + \sum_{g \in
    {\cal{N}}} {\bf{F}}_g
\end{equation}
\end{definition}
The quantity
${\bf{I}}(t)$ is similar to ${\bf{I}}$ as given in (\ref{Ivsk}), but to
${\bf{I}}(t)$ only contribute the byproduct images of qubits from a subset 
$\bigcup\limits_{i=1}^t Q_i$ of ${\cal{C}}$.
The information flow vector ${\bf{I}}(t_{\mbox{\footnotesize{max}}})$ after
the final measurement round $t_{\mbox{\footnotesize{max}}}$ equals the
information vector ${\bf{I}}$, 
\begin{equation}
    \label{result}
    {\bf{I}} = {\bf{I}}(t_{\mbox{\footnotesize{max}}}).
\end{equation}
As will be shown later, during all steps of the computation, except for
after the final one, the  
information flow vector determines the measurement bases for the
cluster qubits that are to be measured in the next round. After the final round
it contains the result of the computation. Thus, it has a meaning in
every step of the computation. No further information obtained from the
measurements is needed. In this sense, the information
flow vector can be regarded as the carrier of the {\em{algorithmic
information}} on the 
\QCns.\footnote{The way we use the term ``algorithmic information''
  has nothing to do with the --in general non-computable-- algorithmic
  information content of an object as it is defined in Kolmogorov
  complexity theory \cite{Kolm}.}  

The information flow vector has a constant part which does not depend
on the measurement results $\big \{ s_k \big \}$. This part alone
forms its initialization value ${\bf{I}}_{\mbox{\footnotesize{init}}}$,
\begin{equation}
    \label{I_init}
    {\bf{I}}_{\mbox{\footnotesize{init}}} = \sum_{k \in
    {\cal{C}}\backslash Q_{0,z}} \kappa_k {\bf{F}}_k + \sum_{g \in
    {\cal{N}}} {\bf{F}}_g,
\end{equation}
such that ${\bf{I}}(t)$ becomes
\begin{equation}
    \label{Iflow2}
    {\bf{I}}(t) = {\bf{I}}_{\mbox{\footnotesize{init}}} + \sum
    \limits_{k \in \bigcup\limits_{i=0}^t Q_i} s_k\,{\bf{F}}_k.  
\end{equation}
From eq.~(\ref{I_init}) we see that the byproduct images of the gates
${\bf{F}}_g$ do not form an independent part of the information
specifying a quantum algorithm on the \QCns. Instead, they are absorbed
into the initialization value $ {\bf{I}}_{\mbox{\footnotesize{init}}}$
of ${\bf{I}}(t)$.

The measurement bases in which the results $s_k$ are obtained --referred
to implicitly in (\ref{Ivsk}) and (\ref{Iflow})--
are not fixed a priori, but must be determined during the computation. They
will be calculated using the byproduct images $\{{\bf{F}}_k, k \in
{\cal{C}}\}$ and ${\bf{I}}(t)$, as explained in Sections~\ref{model} and
  \ref{proof}. Besides the byproduct images, the algorithm angles
$\varphi_{j,\mbox{\footnotesize{algo}}},\; j \in Q^{(1)}$  are also
needed to determine the appropriate measurement 
bases. They are related to the network angles $\varphi_{j,\mbox{\footnotesize{qln}}},\; j \in
Q^{(1)}$ that specify the one-qubit rotations in the corresponding
quantum logic network via
\begin{equation}
    \label{Algoang}
    \varphi_{j,\mbox{\footnotesize{algo}}} = {(-1)}^{\eta_j}\,
    \varphi_{j,\mbox{\footnotesize{qln}}}\, , \;\;\;\; j \in Q^{(1)},
\end{equation}
where $\eta_j$ is given by
\begin{equation}
    \label{eta}
    \eta_j = \sum \limits_{
            \footnotesize{\begin{array}{rl} 
                  k|& \!\!\!\!k \in {\cal{C}} \backslash Q_{0,z}\, ,
             \\ & \!\!\!\! j \in \mbox{bc}(k) \end{array}}}
    \kappa_k +  \sum \limits_{
      \footnotesize{\begin{array}{rl} 
            g|& \!\!\!\!g \in {\cal{N}} \, ,
            \\ & \!\!\!\! j \in \mbox{bc}(g) \end{array}}} 1 .
\end{equation} 
The pair of equations (\ref{Algoang}), (\ref{eta}) is, for the moment,
just a definition of the algorithm angles. It will become apparent in
Sections~\ref{model} and \ref{proof} that this definition is indeed useful.

\subsection{Description of the model}
\label{model}

As already listed in Section~\ref{cond_mat}, a quantum algorithm on
the \QC is specified by the sets $Q_t$
of simultaneously measured qubits, the backward cones $\bc(k)$ of the
qubits $k \in Q_0$, the measurement bases of the
qubits $k \in Q_0$, the byproduct images ${\bf{F}}_j$ for $j
\in {\cal{C}}$, the 
algorithm angles $\varphi_{l,\mbox{\footnotesize{algo}}}$ for $l \in
Q^{(1)}$ and the initialization value
${\bf{I}}_{\mbox{\footnotesize{init}}}$ of the information flow vector
${\bf{I}}(t)$. If an algorithm is not given in this form but rather as
a quantum logic network composed of CNOT-gates and one-qubit
rotations, the above quantities can be derived from the network as
explained in the previous sections. 

Let us summarize this step of
classical pre-processing. First, the measurement pattern is obtained
--if one has no better idea-- by patching together the measurement
patterns for the individual gates displayed in Fig.~\ref{Gates}. This
gives the measurement directions for the qubits $k \in Q_0$. The
network angles $\varphi_{j,\mbox{\footnotesize{qln}}}$ for the qubits $j \in Q^{(1)}$ are taken
from the quantum logic network to be simulated. To determine the sets $\{Q_t,
t=0..t_{\mbox{\footnotesize{max}}} \}$, we need the forward cones. The
forward cones $\mbox{fc}(k)$ for all qubits $k \in {\cal{C}}$ can be
obtained using the
expressions (\ref{Byprod1}), (\ref{CNOTbyprop}) for the
byproduct operators and the propagation relations (\ref{Rotprop}),
(\ref{CNTprop}) and (\ref{Hadaprop}).
From the forward cones we derive a strict partial ordering ``$\prec$''
(\ref{coneImpl}) among the cluster qubits, and from the strict partial
ordering we derive the sets $Q_t \subset {\cal{C}}$ via
(\ref{Qrecur}). The byproduct images ${\bf{F}}_k$ for the qubits $k
\in {\cal{C}}\backslash Q_{0,z}$ 
are obtained from their definition (\ref{FI}) once the 
corresponding forward propagated byproduct operators are obtained 
from (\ref{fpbop}). The byproduct
images of the qubits $k \in Q_{0,z} \backslash O$ are traced back to
those in the set ${\cal{C}}\backslash Q_{0,z}$ via
eq.~(\ref{F0z}). The byproduct images of the remaining qubits in
${\cal{C}}$, $k \in O$, are given by (\ref{ByimO}). To determine the
algorithm angles we need the backward 
cones $\mbox{bc}(k)$ for the qubits $k \in Q_0$ and the backward cones
of gates $\mbox{bc}(g)$. Then, the algorithm angles are given by
(\ref{Algoang}), 
(\ref{eta}). Finally, for the initialization value
${\bf{I}}_{\mbox{\footnotesize{init}}}$ of the information flow vector
we need the byproduct images ${\bf{F}}_g$ of the gates $g$ which we
obtain from eq.~(\ref{Fg}). 
${\bf{I}}_{\mbox{\footnotesize{init}}}$ is set via
(\ref{I_init}). All the pre-processing required to extract the listed
quantities from a quantum logic network can be performed efficiently
on a classical computer, see Appendix~\ref{preproc}.

The scheme of quantum computation on the \QC comprises several
measurement rounds in which the following steps have to be performed:   
\begin{enumerate}
\item{First measurement round.
      \begin{enumerate}
          \item{Measure all qubits $k \in Q_0$. Obtain
                measurement results $\{s_k | k \in Q_0 \}$.}
          \item{Modify the angles $\varphi_{j,\mbox{\footnotesize{algo}}}$ for
                the continuous gates
                \begin{equation}
                    \label{modfd_angl}
                    \varphi_{j,{\mbox{\footnotesize{algo}}}} \longrightarrow
                    \varphi^\prime_{j,{\mbox{\footnotesize{algo}}}} =
                    \varphi_{j,{\mbox{\footnotesize{algo}}}} \, {(-1)}
                    ^{\eta_j^\prime},
                \end{equation}
                with
                \begin{equation}
                \label{eta_pr}
                    \eta_j^\prime = \sum_{k \in Q_0 | j \in 
                    \mbox{\small{bc}}(k)} s_k 
                \end{equation}}
                for all $j \in  Q^{(1)}$.
          \item{Update the information flow vector from
                ${\bf{I}}_{\mbox{\footnotesize{init}}}$ to 
                ${\bf{I}}(0)$ 
                \begin{equation}
                    \label{I0}
                    {\bf{I}}(0)={\bf{I}}_{\mbox{\footnotesize{init}}}
                    + \sum_{k \in Q_0} s_k {\bf{F}}_k .
                \end{equation}}
                      
      \end{enumerate}}
\item{Subsequent measurement rounds.\vspace{0.2cm}\\
      Perform the following three steps (\ref{a}) - (\ref{c}) for all
      qubit sets 
      $Q_t \subset  {\cal{C}} \backslash Q_0$ in ascending order,
      beginning with $Q_1$. In the measurement round $t$,
      \begin{enumerate}
          \item{\label{a}Determine the measurement bases for ${j \in
                  Q_t}$ according to
                \begin{equation}
                    \label{setmeasang}
                    \varphi_{j,\mbox{\footnotesize{meas}}} =
                    \varphi^\prime_{j,\mbox{\footnotesize{algo}}}\,
                    {(-1)} ^{\left({\bf{I}}(t-1),{\bf{F}}_j\right)_S}
                \end{equation}
                }
          \item{Perform the measurements on the qubits $j \in
                Q_t$. Thereby obtain the measurement results
                $\left\{s_j \in \{0,1\} \; | \; j \in Q_t
                \right\}$.}
          \item{\label{c}Update the information flow vector ${\bf{I}}$
                \begin{equation}
                    \label{Iup}
                    {\bf{I}}(t) = {\bf{I}}(t-1) + \sum\limits_{j \in Q_t}
                    s_j\,{\bf{F}}_j. 
                \end{equation}}
      \end{enumerate}
}    
\end{enumerate}
The information flow vector ${\bf{I}}(t_{\mbox{\footnotesize{max}}})$ after
the final measurement round $t_{\mbox{\footnotesize{max}}}$ 
equals the  information vector ${\bf{I}}$, as can be
seen from (\ref{result}). At the end of the computation, from ${\bf{I}}$ we can
directly read off the result ${\bf{I}}_x$ of the computation. ${\bf{I}}_x$ is
identical to the readout of the corresponding quantum logic network.

{\em{Remark 1.}} Note that in the first measurement round the byproduct
operators created 
by the measurements on qubits in $Q_0$ have been propagated
{\em{backwards}} to set the angles
  $\{\varphi^\prime_{j,\mbox{\footnotesize{algo}}}\}$. There is also a
scheme in which the byproduct operators caused by the measurements in
the initialization round are propagated forward to set the modified
algorithm angles $\{ \varphi^\prime_{j\mbox{\footnotesize{algo}}}\}$. In that
scheme, the update of the information flow vector ${\bf{I}}(t)$ and
the rule to determine the measurement angles
$\varphi_{j,{\mbox{\footnotesize{meas}}}}$ are the same as in the described
scheme, given by (\ref{setmeasang}) and (\ref{Iup}). What is different
is the initialization and the appearance of a step of
post-processing. In the modified scheme, in eqs.~(\ref{eta}) and
(\ref{eta_pr}) the backward cones $\mbox{bc}(k)$ are replaced by the
respective forward cones $\mbox{fc}(k)$ and
${\bf{I}}_{\mbox{\footnotesize{init}}}$ is set to zero. The quantity
which was ${\bf{I}}_{\mbox{\footnotesize{init}}}$ in 
(\ref{I_init}) is computed as well but now stored as an auxiliary
quantity $\Delta{\bf{I}}$
until the end of the computation. After the last measurement round
$t_{\mbox{\footnotesize{max}}}$, the information vector ${\bf{I}}$
then is obtained by the relation 
${\bf{I}}={\bf{I}}(t_{\mbox{\footnotesize{max}}}) + \Delta {\bf{I}}$,
which requires the extra post-processing step and extra memory during
the computation. We have
chosen to present the scheme with backward propagation of byproduct
operators  in order to avoid this superfluous
post-processing. This way, the quantity ${\bf{I}}(t)$ which steers the
computational process directly displays the result of the computation
after the final 
update to ${\bf{I}}(t_{\mbox{\footnotesize{max}}})$. 

{\em{Remark 2.}} This  comment concerns the choice ${\cal{O}}=\Omega$
of the cut on 
which the byproduct images ${\bf{F}}_k$ and  ${\bf{F}}_g$ are evaluated. In the
visualization of the \QC as an implementation of a quantum logic
network the cut $\Omega$ plays a distinguished role. The byproduct
operators accumulated at $\Omega$ determine how the ``readout''
measurements have to be interpreted. In the computational model
underlying the \QCns, however, the former readout qubits are just qubits
to be measured like any other cluster
qubits. Here, the cut $\Omega$ is not
distinguished. Due to the invariance
(\ref{conssympprod}) of the symplectic 
scalar product (\ref{sympscalp}) the byproduct images ${\bf{F}}_k$, which
enter the expression (\ref{setmeasang}) for the
$\varphi_{k,{\mbox{\footnotesize{meas}}}}$ directly and via
(\ref{I_init}) and (\ref{Iup}), can be evaluated with respect to
{\em{any}} vertical cut ${\cal{O}}$ through the corresponding quantum logic
network. The information vector ${\bf{I}}$ which displays the result
of the computation in its $x$-part ${\bf{I}}_x$ would then be related
to the information flow vector after the final measurement round
${\bf{I}}(t_{\mbox{\footnotesize{max}}})$ via ${\bf{I}} =
C({\cal{N}}_{{\cal{O}}\rightarrow \Omega})\,
{\bf{I}}(t_{\mbox{\footnotesize{max}}})$. Thus, the particular vertical cut
${\cal{O}} = \Omega$ was chosen just to avoid an additional step
of post-processing.  The dependence on
the cut ${\cal{O}}$ would vanish altogether if one would write the $n$
output bits of the quantum computation in the form ${\left[I_x\right]}_i =
{({\bf{I}}|_{\cal{O}},f_i|_{\cal{O}})}_S$ for suitably chosen $\{f_i
\in {\cal{V}}, 
i=1,..,n\}$, e.g. for the case ${\cal{O}}=\Omega$, $f_1 =
(0,..,0;1,0,..,0)^T$, $f_2 =
(0,..,0;0,1,0,..,0)^T$,  and the other $f_i,\; i\leq n$ accordingly.   

\subsection{Proof of the model}
\label{proof}

In this section it is shown that if we run the \QC according to the
scheme described in Section~\ref{model}, we obtain the same result as
in the corresponding quantum logic network. This requires to prove
that (a) one does indeed choose all
the measurement angles correctly and (b) obtains at the end of the
computation the result ${\bf{I}}_x$, the $x$-part of the information
vector ${\bf{I}}$ as given in (\ref{Ivsk}).

To show point (b), we use (\ref{I_init}), (\ref{I0}) and (\ref{Iup})
and obtain for the information vector  
\begin{eqnarray}
    {\bf{I}} = \sum_{k \in {\cal{C}} \backslash Q_{0,z}} \kappa_k
    {\bf{F}}_k + \sum_{g \in 
    {\cal{N}}} {\bf{F}}_g +\sum_{k \in Q_0} s_k
    {\bf{F}}_k + \sum_{k \in \bigcup_{i=1}^{t_{\mbox{\tiny{max}}}}
    Q_i} s_k {\bf{F}}_k \nonumber 
\end{eqnarray}
which coincides with (\ref{Ivsk}). This ensures that we obtain the right
vector ${\bf{I}}$ at the end of the computation, provided the
measurement bases were chosen appropriately, as required for (a). This
is checked below. 
 
First we observe that the measurement angle
$\varphi_{j,{\mbox{\footnotesize{meas}}}}$ and the network angle
$\varphi_{j,\mbox{\footnotesize{qln}}}$ are for all $j \in Q^{(1)}$ related in the following way
\begin{equation}
    \label{algomeas}
    \varphi_{j,{\mbox{\footnotesize{meas}}}} =
    {(-1)}^{\vartheta_j}\,\varphi_{j,\mbox{\footnotesize{qln}}} \,,
\end{equation}
with
\begin{equation}
    \label{theta}
    \vartheta_j = \sum\limits_{k \in {\cal{C}} | j
    \in \mbox{\small{fc}}(k)} s_k +  \sum\limits_{k \in
    {\cal{C}}\backslash Q_{0,z} | j
    \in \mbox{\small{fc}}(k)} \kappa_k +  \sum\limits_{g \in
    {\cal{N}} | j
    \in \mbox{\small{fc}}(g)} 1  \;\;\; \mbox{mod} \; 2. 
\end{equation}
Why does the pair of equations~(\ref{algomeas}), (\ref{theta}) hold?
As can be seen from the propagation relation for rotations
(\ref{Rotprop}), the network and the measurement angle of a qubit $j \in
Q^{(1)}$ can differ only by a sign factor $\pm1$ and can therefore
always be related as in (\ref{algomeas}). The first and the third sum
in (\ref{theta}) follow from the 
definition of the forward cones of the cluster qubits and of gates in 
Section~\ref{FBC}. The measurement angle at $j$ acquires a factor
${(-1)}^{s_k}$ if 
$j \in \fc(k)$ and a factor of $(-1)$ for each gate $g$ with $j \in
\mbox{fc}(g)$. The $\kappa$-dependent part can be derived in the same
way as it was derived for $U_\Sigma$ through equations~(\ref{Us2}) -
(\ref{Us6}). Note that $\vartheta_j$ depends, similar to the
information vector ${\bf{I}}$,  only on the combination
$s_k+\kappa_k$ for $k \in {\cal{C}}\backslash Q_{0,z}$ and on $s_k$
for $k \in Q_{0,z}$.  

Now, we rewrite  the quantity $\vartheta_j$ in the following way
\begin{eqnarray}
    \label{7terms}
    \vartheta_j &=&  \sum\limits_{k \in {\cal{C}} | j
    \in \mbox{\small{fc}}(k)} s_k +  \sum\limits_{k \in
    {\cal{C}}\backslash Q_{0,z} | j
    \in \mbox{\small{fc}}(k)} \kappa_k +  \sum\limits_{g \in
    {\cal{N}} | j
    \in \mbox{\small{fc}}(g)} 1  \;\;\; \mbox{mod} \; 2 \nonumber\\
  &=& \sum\limits_{k \in {\cal{C}} | j
    \in \mbox{\small{fc}}(k)} s_k +  \sum\limits_{k \in
    {\cal{C}}\backslash Q_{0,z} | j
    \in \mbox{\small{fc}}(k)} \kappa_k +  \sum\limits_{g \in
    {\cal{N}} | j
    \in \mbox{\small{fc}}(g)} 1 + \nonumber \\
  & & + 2 \left( \sum\limits_{k \in Q_0 | j
    \in \mbox{\small{bc}}(k)} s_k  + \sum\limits_{k \in
    {\cal{C}}\backslash Q_{0,z} | j
    \in \mbox{\small{bc}}(k)} \kappa_k +  \sum\limits_{g \in
    {\cal{N}} | j
    \in \mbox{\small{bc}}(g)} 1  \right)\;\;\;\mbox{mod}\,\,2
    \nonumber \\
  & =& \underbrace{\sum\limits_{k \in Q_0 |  j \in \mbox{\footnotesize{fc}}(k)
    \vee j \in \mbox{\footnotesize{bc}}(k)}  s_k}_{S_1} \;\;+ 
    \underbrace{\sum\limits_{k \in Q^{(1)} |  j \in
    \mbox{\footnotesize{fc}}(k)}
    s_k}_{S_2}\;\; + 
   \underbrace{\sum\limits_{k \in Q_0 |  j \in
    \mbox{\footnotesize{bc}}(k)}  s_k}_{S_3} \; + \\
  & & + \underbrace{\sum\limits_{k \in {\cal{C}}\backslash Q_{0,z} |
    j \in \mbox{\footnotesize{fc}}(k) 
    \vee j \in \mbox{\footnotesize{bc}}(k)}  \kappa_k}_{S_4} \;\;+
    \underbrace{\sum\limits_{k \in  {\cal{C}}\backslash Q_{0,z} |  j \in
    \mbox{\footnotesize{bc}}(k)}
    \kappa_k}_{S_5}\; + \nonumber \\
  & & + \underbrace{\sum\limits_{g \in
    {\cal{N}} | j
    \in \mbox{\small{fc}}(g) \vee j \in \mbox{\small{bc}}(g)} 1}_{S_6}
    \;\; +  \underbrace{\sum\limits_{g \in 
    {\cal{N}} | j \in \mbox{\small{bc}}(g)} 1}_{S_7} \;\;\;\mbox{mod}\,\,2.
    \nonumber  
\end{eqnarray} 
 We now discuss the seven terms
$S_1,\, .. \, ,S_7$. All sums are evaluated modulo 2. 

Term $S_1$ of (\ref{7terms}):
\begin{equation}
    \label{T1}
    S_1= \sum\limits_{k \in Q_0 |  j \in \mbox{\footnotesize{fc}}(k)
    \vee j \in \mbox{\footnotesize{bc}}(k)} s_k = \sum\limits_{k \in Q_0} s_k
    \,({\bf{F}}_k, {\bf{F}}_j)_S,
\end{equation}
where the last identity  holds by the cone test (\ref{conecrit}).

Term $S_2$ of (\ref{7terms}):\\
Let be $j \in Q_t$ and $k \in Q_i$. Qubit $j$ can only then be in the
forward cone of $k$, $j \in \fc(k)$, if $i < t$. Hence
\begin{eqnarray}
    \label{T2}
    S_2 &=& \sum\limits_{\left.{k \in Q^{(1)}}\right|  j \in
    \mbox{\footnotesize{fc}}(k)}
    s_k \nonumber \\
    &=&\sum\limits_{k \in \bigcup\limits_{i=1}^{t-1}
    \left.{Q_i}\right|  j \in 
    \mbox{\footnotesize{fc}}(k)} 
    s_k \nonumber \\
    &=& \sum\limits_{k \in \bigcup\limits_{i=1}^{t-1} Q_i} 
    s_k \, ({\bf{F}}_k, {\bf{F}}_j)_S .
\end{eqnarray}  
In (\ref{T2}) the last line again follows by using the cone test
(\ref{conecrit}).

Term $S_3$ of (\ref{7terms}):\\
\begin{equation}
    \label{T3}
    S_3 = \sum\limits_{k \in Q_0 |  j \in
    \mbox{\footnotesize{bc}(k)}}  s_k = \eta_j^\prime.
\end{equation}
This equity follows by the definition of $\eta_j^\prime$ in
(\ref{eta_pr}). Thus, the term $S_3$ is 
the contribution to $\vartheta_j$ coming from the first measurement
round  where the algorithm angles $\{
\varphi_{j,\mbox{\footnotesize{algo}}} \}$ are changed to the modified
algorithm angles $\{\varphi^\prime_{j,\mbox{\footnotesize{algo}}} \}$. 

Term $S_4$ of (\ref{7terms}):\\
\begin{equation}
    \label{T4}
    S_4 = \sum\limits_{k \in {\cal{C}}\backslash Q_{0,z} |
    j \in \mbox{\footnotesize{fc}}(k) 
    \vee j \in \mbox{\footnotesize{bc}}(k)}  \kappa_k =
    \sum\limits_{k \in {\cal{C}}\backslash Q_{0,z}}  \kappa_k \left(
    {\bf{F}}_k, {\bf{F}}_j \right)_S , 
\end{equation}
which follows by the cone test (\ref{conecrit}). 

Terms $S_5+S_7$ of (\ref{7terms}):\\
\begin{equation}
    \label{T5T7}
    S_5+S_7 = \sum\limits_{k \in  {\cal{C}}\backslash Q_{0,z} |  j \in
    \mbox{\footnotesize{bc}}(k)}
    \kappa_k \;+ \sum\limits_{g \in
    {\cal{N}} | j \in \mbox{\small{bc}}(g)} 1 = \eta_j,
\end{equation}
via the definition (\ref{eta}) of the $\eta_j$.

Finally, term $S_6$ of (\ref{7terms}):\\
\begin{equation}
    \label{T6}
    S_6 = \sum\limits_{g \in
    {\cal{N}} | j
    \in \mbox{\small{fc}}(g) \vee j \in \mbox{\small{bc}}(g)} 1 =
    \sum\limits_{g \in {\cal{N}}} \left({\bf{F}}_g,
    {\bf{F}}_j \right)_S, 
\end{equation}
which follows by the cone criterion (\ref{gateconecrit}) for gates. 

Now we combine these seven terms $S_1,\, ..\,, S_7$. By
(\ref{7terms}) - (\ref{T6}) we obtain
\begin{eqnarray}
    \label{expo}
    \vartheta_j &=& \eta_j + \eta_j^\prime + \sum\limits_{g \in
    {\cal{N}}} \left({\bf{F}}_g, 
    {\bf{F}}_j \right)_S \; +\sum\limits_{k \in
    {\cal{C}} \backslash Q_0} \kappa_k
    \,({\bf{F}}_k, {\bf{F}}_j)_S \; + \sum\limits_{k \in
    \bigcup\limits_{i=0}^{t-1} Q_i}  
    s_k \, ({\bf{F}}_k, {\bf{F}}_j)_S  \nonumber \\
    &=& \eta_j + \eta_j^\prime + ({\bf{I}}(t-1),{\bf{F}}_j)_S. 
\end{eqnarray} 
The last line follows from the definition (\ref{Iflow}) of the
information flow vector. If we consider the relations 
(\ref{Algoang}), (\ref{modfd_angl}) and (\ref{algomeas})  between the angles
$\varphi_{j,\mbox{\footnotesize{algo}}}$,
$\varphi_{j,\mbox{\footnotesize{algo}}}^\prime$ and
$\varphi_{j,\mbox{\footnotesize{meas}}}$, we find 
\begin{equation}
    \label{almea}
    \varphi_{j,\mbox{\footnotesize{meas}}} = {(-1)}^{\vartheta_j -
    \eta_j - \eta^\prime_j}  \varphi_{j,\mbox{\footnotesize{algo}}}^\prime.
\end{equation}
Now we insert (\ref{expo})
into (\ref{almea}) and obtain 
\begin{eqnarray}                   
     \varphi_{j,\mbox{\footnotesize{meas}}} &=&
     \varphi^\prime_{j,\mbox{\footnotesize{algo}}}\,
     {(-1)} ^{\left({\bf{I}}(t-1),{\bf{F}}_j\right)_S} \nonumber,
\end{eqnarray}
which proofs that the assignment of the measurement angles
(\ref{setmeasang}) is correct, and thereby concludes the proof of
the computational model described in Section \ref{model}.
 
\section{Logical depth and temporal complexity}
\label{depth}

The logical depth has, to our knowledge, only been defined in the context of
quantum logic networks, but it can straightforwardly be generalized to
the \QCns. In networks one groups gates which can be performed in
parallel to layers. The logical depth of a quantum logic network then
is the minimum number of its layers. Similarly in case of the \QCns,
one can group the cluster qubits which can be measured simultaneously
to sets $Q_t$. There, the logical depth of 
the \QCns-realization of an algorithm is the minimal number of such
sets.

Since the one-qubit measurements on the cluster state mutually
commute, one may be led to think that they can always be performed all
in parallel. They could, but then the measurements
would in general not drive a deterministic computation.  

In the following, we will denote the logical depth in the context of
the \QC by $D$ and the logical depth of a quantum logic network by
$D_{\cal{N}}$. 

\subsection{$D=1$ for circuits in the Clifford group}
\label{Clifford}

The Clifford group of gates is generated by the CNOT-gates, the
Hadamard-gates and the $\pi/2$-phase shifts. 
In this section it is proved that the logical depth of
such circuits is $D=1$ on
the \QCns, independent of the number of
logical qubits $n$. For a subgroup of the Clifford
group, the group generated by the CNOT- and Hadamard gates alone we can
compare the result to the best known upper bound for quantum logic 
networks, where the logical depth $D_{\cal{N}}$ scales like $O(\log n)$
\cite{M&N}.   

The elementary gates we use are the Hadamard gate $H=1/\sqrt{2}
(\sigma_x + \sigma_z)$, the $\pi/2$-phase gate
$U_z(\pi/2)=\mbox{exp}(-i\pi/4\,\sigma_z)$, and 
the CNOT-gate between neighbouring qubits, whose realization on the
\QC is
depicted in Fig.~\ref{Gates}. Out of the latter we 
construct the CNOT-gate between arbitrary qubits via the swap-gate
composed of three CNOT-gates. Hence, out of the elements displayed in
Fig.~\ref{Gates} any circuit in the Clifford group can be
composed. At this point we must
emphasize that in a practical realization of a \QC we would 
not perform a general CNOT in the described manner using
    the swap gate. There is a more efficient realization for the
general CNOT, whose spatial resources scale more favourably. This
gate will be displayed elsewhere. The purpose here  is to keep the
argument compact rather than the gates. 

It is possible to measure all qubits at once. This works since, as
shown in Fig.~\ref{Gates}, all cluster qubits necessary for the
realization of the CNOT-, Hadamard- and $\pi/2$-phase gates
are measured either 
in the eigenbasis of $\sigma_x$ or 
of $\sigma_y$. The redundant qubits are measured in $\sigma_z$, as
explained in \cite{QCmeas}. Thus, none of the qubits is measured in a
basis whose proper adjustment requires classical information from
measurement results at other qubit sites. This concludes the proof of
$D=1$ for circuits in the Clifford group. In a
computation, of course all the measurement results obtained must be
interpreted. Therefore, there exists a contribution to the computation
time from classical post-processing.  The connection between logical
depth and computation time is discussed in section \ref{Tcomp}.  

\subsection{$D=2$ for circuits of CNOT-gates and a ${U(1)}^{\otimes
    n}$-subgroup of rotations} 
\label{dg}

In this section we prove that the logical depth $D$ of
a circuit composed of either CNOT-gates and rotations about the
$x$-axis or of CNOT-gates and rotations about the
$z$-axis is $D=2$. This set of circuits
contains all circuits of diagonal 2-qubit gates as a special case.
For circuits of diagonal 2-qubit gates we can compare our result $D=2$
to the best known result \cite{M&N} for quantum logic networks where
the logical depth scales logarithmically in the number of gates. 

Here we give the proof for circuits of CNOT-gates and rotations about
the $z$-axis $U_z(\alpha)= e^{-i\alpha  \frac{\sigma_z}{2}}$.
The elementary gates used are (a) the rotations about the $z$-axis
$U_z(\alpha)= e^{-i\alpha \frac{\sigma_z}{2}}$, and (b) the CNOT-gate
between neighbouring logical qubits. The realization of the rotation
$U_z$ is depicted in 
Fig.~\ref{Gates}. Of the CNOT-gate 
between neighbouring qubits we construct the swap-gate between
neighbouring qubits and by that the general CNOT-gate, as in
section \ref{Clifford}. The strategy to implement the circuit is then:
(1) to measure all those qubits on ${\cal{C}}$ which are to be
measured in the eigenbases of $\sigma_x$, $\sigma_z$ or
$\sigma_y$; and (2) to measure the remaining qubits, i.e. the ones which are
measured in a direction in the $x-y$-plane.

The result that the measurements in step (1) can be performed in one step has
already been shown in section \ref{Clifford}. It remains to be shown
that the measurements in the tilted measurement directions of step 
(2) can also be performed in parallel. 
Let $j$ and $l$ be two cluster qubits which are measured in a tilted 
basis in step 2 in order to implement the rotations. Using
(\ref{Qrecur}) one finds
\begin{equation}
    \label{Impl1}
    D > 2 \Longrightarrow \exists \;j,l \in Q^{(1)} : \; l \prec j
    \;\; \mbox{(that is, $Q^{(2)} \neq \emptyset$)}.
\end{equation}
Further holds
\begin{equation}
    \label{Impl2}
    l \prec j \Longrightarrow \exists \; k \in Q^{(1)} :\; j \in \fc(k), 
\end{equation}
because the strict partial ordering ``$\prec$'' is generated by the
forward cones, 
i.e. $l \prec j \Longleftrightarrow \mbox{either}\; j \in \fc(l),\;
\mbox{or}\; \exists (k_1,\dots\,k_r):\; k_1 \in \fc(l) \;\wedge\;
\big\{ k_s
\in \fc(k_{s-1})|\, 2 \leq s \leq r \big\}\;\wedge\; j \in
\fc(k_r)$. 

Moreover, from the criterion (\ref{conecrit}) one derives
\begin{equation}
    \label{Impl3}
    j \in \fc(k) \Longrightarrow ({\bf{F}}_j,{\bf{F}}_k)_S = 1.
\end{equation}
Now, by putting the implications (\ref{Impl1}),  (\ref{Impl2}) and
(\ref{Impl3}) together we obtain
\begin{equation}
    \label{aa}
    D >2 \Longrightarrow \exists \;j,k \in Q^{(1)}:\; ({\bf{F}}_j,{\bf{F}}_k)_S = 1,
\end{equation} 
which we negate to obtain 
\begin{equation}
    \label{conecritrepit}
    \forall j,k\in Q^{(1)}: ({\bf{F}}_j,{\bf{F}}_k)_S = 0
    \Longrightarrow D \leq 2.
\end{equation}
Next it is proved that $({\bf{F}}_j,{\bf{F}}_k)_S = 0$ does indeed
hold for all $j,k \in Q^{(1)}$.  

A measurement of a qubit at site $k$ , which is part of the 
implementation of a rotation about the $z$-axis (central qubit 3 in
Fig.~\ref{Gates}c), generates a
byproduct operator ${(U_k)}^{s_k}={(\sigma_z)}^{s_k}$. This can be
seen from equations 
(\ref{Euler}), (\ref{Rotprime}) and (\ref{Byprod1}). Note that in
Fig.~\ref{Gates}c, qubits 1,2,4 are measured in the
$\sigma_x$-eigenbasis, they belong to the set $Q_0$. Now let be $i$
the number of the {\em{logical}} qubit on which the 
rotation $U_z(\varphi_k)$  is performed by the measurement of cluster
qubit $k$. 
Further, let ${\cal{O}}$ be a vertical cut through the network
simulated by the \QCns. ${\cal{O}}$ intersects each qubit line only once. In
particular, it shall intersect the qubit line $i$ just at the output side
of the rotation $U_z(\varphi_k)$. Thus, the image
${\bf{F}}_k|_{\cal{O}}$ of $U_k$ on the cut ${\cal{O}}$ is
\begin{equation}
    \label{FatO}
    {\bf{F}}_k|_{\cal{O}} = \left( \begin{array}{c} \mbox{ } \\ 0
    \\ \mbox{ } \\ {{\bf{F}}_k}_z|_{\cal{O}} \\ \mbox{ } \end{array}
    \right), \;\;\mbox{with}\;{F_k}_{z,l}= \delta_{il}. 
\end{equation}
What we see from (\ref{FatO}) is that
${{\bf{F}}_k}_x|_{\cal{O}}=0$. Be ${\cal{N}}_{{\cal{O}} \rightarrow
  \Omega}$ the part of the network ${\cal{N}}$ which is located
between the two cuts ${\cal{O}}$ and $\Omega$. The byproduct image
${\bf{F}}_k$ corresponding to $U_k$ is then given by
\begin{equation}
    \label{Ez}
    {\bf{F}}_k \equiv {\bf{F}}_k |_\Omega =  C({\cal{N}}_{{\cal{O}}
    \rightarrow \Omega})\, 
    {\bf{F}}_k|_{\cal{O}}. 
\end{equation} 
The only gates that contribute to $C({\cal{N}}_{{\cal{O}} \rightarrow
  \Omega})$ are the CNOT gates, as described in section
\ref{propmats}. The propagation matrices for CNOT gates
(\ref{CNTpropmat}) have block-diagonal form. Hence, using
  (\ref{concatenat}) the propagation
matrix for the network ${\cal{N}}_{{\cal{O}} \rightarrow
  \Omega}$ has block-diagonal form
\begin{equation}
    \label{C-CNT}
    C({\cal{N}}_{{\cal{O}} \rightarrow
  \Omega}) = \left(  
        \begin{array}{ccc|ccc}
            & \mbox{ } & & \mbox{ } &  \\
            & C_{xx}({\cal{N}}_{{\cal{O}} \rightarrow
              \Omega}) & & & 0 & \\
            & \mbox{ } & & &\\ \hline
            & \mbox{ } & & &\\
            & 0 & & & C_{zz}({\cal{N}}_{{\cal{O}} \rightarrow
              \Omega})  & \\
            & \mbox{ } & & & 
        \end{array}
    \right).
\end{equation}
From (\ref{FatO}), (\ref{Ez}) and (\ref{C-CNT}) it follows that the
$x$-part of the 
byproduct image vector ${\bf{F}}_k$ vanishes for all $k$
\begin{equation}
    \label{ab}
    {\left[{{\bf{F}}_x}\right]}_k=0\;\; \forall k \in Q^{(1)}.
\end{equation} 
Hence by the definition of the symplectic scalar product
(\ref{sympscalp}), we obtain  $({\bf{F}}_j,{\bf{F}}_k)_S=0$ for all
$j,k \in Q^{(1)}$. This proves via (\ref{conecritrepit}) $D \leq
2$.  The measurements to implement the one-qubit rotations can thus
all be performed 
at the same time. In (\ref{conecritrepit}) the case $D=1$ can be easily be 
excluded for all interesting cases such that only $D=2$ remains. This
concludes the proof of $D=2$ for circuits of 
CNOT-gates and rotations of the form $e^{-i \varphi
  \frac{\sigma_z}{2}}$. The
proof for circuits of CNOT-gates and rotations $e^{-i \varphi
  \frac{\sigma_x}{2}}$ runs analogously.
\begin{figure}
    \begin{center}
        \setlength{\unitlength}{0.6cm}
        \begin{picture}(10,2.5)

            \definecolor{lightorange}{cmyk}{0,0.35,0.5,0}
            \definecolor{lightblue}{rgb}{0.5,0.6,1}
            \definecolor{orange}{cmyk}{0,0.7,1,0}

            \color{black}
            \linethickness{0.25mm}
            \put(4,-0.47){\line(0,1){2.47}}
            \put(8.5,-0.47){\line(0,1){2.47}}
            \put(0.0,2){\line(1,0){9.5}}
            \put(0.0,0){\line(1,0){9.5}}

            \put(4,2){\circle*{0.25}}
            \put(8.5,2){\circle*{0.25}}
            \put(4,0){\circle{1}}
            \put(8.5,0){\circle{1}}

            \put(0.9,1.85){\fcolorbox{black}{white}{$U_z(\alpha)$}}
            \put(0.9,-0.15){\fcolorbox{black}{white}{$U_z(\beta)$}}
            \put(5.4,-0.15){\fcolorbox{black}{white}{$U_z(\gamma)$}}

            \put(-1.9,0.9){\large{${|\psi\rangle}_{\mbox{\small{in}}}$}}
            \put(9.9,0.9){\large{${|\psi\rangle}_{\mbox{\small{out}}}$}}

        \end{picture}\\ \vspace*{0.4cm} 
     
        \parbox{0.8\textwidth}{\caption{\label{networks} Network for a
              diagonal gate composed of rotations $U_z$ and
              CNOT-gates.}}
    \end{center}
\end{figure}
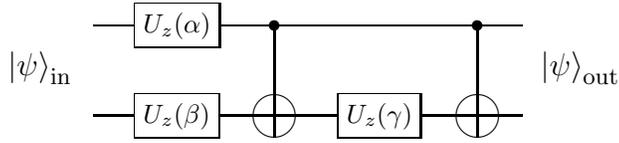
Now let us discuss the special case of circuits composed of diagonal
two-qubit gates. A diagonal gate $G_d$ in the computational basis is
of the form 
\begin{equation}
    \label{Gd}
    G_d = \left( \begin{array}{cccc} e^{i\varphi_1} \\ &
    e^{i\varphi_2} \\ & & e^{i\varphi_3} \\ & & & 1 \end{array} \right) ,
\end{equation} 
modulo a possible global phase which is not relevant.

The network of rotations about the $z$-axis and of a CNOT-gate
shown in Fig.~\ref{networks} realizes a general diagonal two-qubit gate.  
In order to obtain the angles $\varphi_1$, $\varphi_2$ and $\varphi_3$
specifying the diagonal gate $G_d$ in (\ref{Gd}), one chooses the
following angles for the three $z$-rotations in this network
\begin{equation}
    \label{ac}
    \begin{array}{rcl}
        \alpha &=& \displaystyle{\frac{1}{2}\left( -\varphi_1
        -\varphi_2 + \varphi_3\right)}, \vspace*{2mm}\\
         \beta &=& \displaystyle{\frac{1}{2}\left( -\varphi_1
        +\varphi_2 - \varphi_3\right)}, \vspace*{2mm}\\
         \gamma &=& \displaystyle{\frac{1}{2}\left( -\varphi_1
        +\varphi_2 + \varphi_3\right)}. 
    \end{array}
\end{equation}
Thus, a circuit of diagonal two-qubit gates can also be regarded as a
circuit of $z$-rotations and CNOT-gates. Therefore we find $D=2$ for
circuits of diagonal two-qubit gates on the \QCns. This result can be
compared to the best known upper bound \cite{M&N} for quantum logic
networks where the logical depth is of the order ${\cal{O}}(\log n_G)$
with $n_G$ the number of two-qubit gates. 

\subsection{The logical depth $D$ is a good measure for temporal
  complexity}
\label{Tcomp}

In this section, we will express the computation time as a function of
the logical depth. 
The computational model described in section \ref{model} consists of
an alternating series  of measurement rounds and classical processing
of the thereby obtained measurement results. The classical processing
contributes to the duration of the computation and will therefore
enter into
the relation between the computation time and the logical depth. For the
computation time, this results in a correction logarithmic in the
number $n$ of logical qubits involved, and thus the computation time
is no longer the logical depth times a constant. For all practical
purposes, however, this logarithmic correction is small compared to
the time required for the genuine quantum part of the computation,
consisting of the measurements. 

Let $\Delta_Q$ be the
time required to perform the simultaneous measurements in one
measurement round and $\Delta_{cl}$ the time required for the
elementary steps of classical processing: say, addition modulo 2 or
multiplication of two bits. The time $T_{cl}(t)$ required for
classical processing after 
each measurement round has two contributions. First, the time
$T_{cl,{\bf{I}}}(t)$ to update the information flow vector ${\bf{I}}(t)$
and second, the time $T_{cl,\pm}(t)$ to determine the signs of the
measurement angles of all measurements in the next round. The total
computation time $T_{\mbox{\footnotesize{comp}}}$ is given by
\begin{equation}
    \label{Tcomp_cl/qu}
    T_{\mbox{\footnotesize{comp}}}=D \Delta_Q +\sum\limits_{t=0}^{D-1}
    T_{cl,{\bf{I}}}(t) + T_{cl,\pm}(t)
\end{equation}
The update
of the information vector ${\bf{I}}(t)$ according to (\ref{Iup}) can
be done for all $2n$ components in parallel. The update 
${\bf{I}}(t-1) \longrightarrow {\bf{I}}(t)$ following measurement round
$t$ requires the time that it takes to add up $\| Q_t \|$ bits
modulo 2. As the drawing  below illustrates, $T_{cl,{\bf{I}}}(t)$ is
logarithmic in $\| Q_t \|$.
\begin{center}
    \setlength{\unitlength}{0.8cm}
    \begin{picture}(9,4.4)

        \put(4,3.5){\line(2,-1){2}}
        \put(4,3.5){\line(-2,-1){2}}
        \put(4,3.5){\vector(0,1){0.7}} 

        \put(2,2.5){\line(1,-1){1}}
        \put(2,2.5){\line(-1,-1){1}}

        \put(6,2.5){\line(1,-1){1}}
        \put(6,2.5){\line(-1,-1){1}}

        \put(1,1.5){\line(1,-2){0.5}}
        \put(1,1.5){\line(-1,-2){0.5}}

        \put(3,1.5){\line(1,-2){0.5}}
        \put(3,1.5){\line(-1,-2){0.5}}

        \put(5,1.5){\line(1,-2){0.5}}
        \put(5,1.5){\line(-1,-2){0.5}}

        \put(7,1.5){\line(1,-2){0.5}}
        \put(7,1.5){\line(-1,-2){0.5}}

        \put(0.5,0.5){\circle*{0.7}}
        \put(1.5,0.5){\circle*{0.7}}
        \put(2.5,0.5){\circle*{0.7}}
        \put(3.5,0.5){\circle*{0.7}}
        \put(4.5,0.5){\circle*{0.7}}
        \put(5.5,0.5){\circle*{0.7}}
        \put(6.5,0.5){\circle*{0.7}}
        \put(7.5,0.5){\circle*{0.7}}

        \put(1.0,1.5){\circle*{0.7}}
        \put(3.0,1.5){\circle*{0.7}}
        \put(5.0,1.5){\circle*{0.7}}
        \put(7.0,1.5){\circle*{0.7}}

        \put(2,2.5){\circle*{0.7}}
        \put(6,2.5){\circle*{0.7}}

        \put(4,3.5){\circle*{0.7}}

        \color{white}
        \put(0.5,0.5){\circle*{0.55}}
        \put(1.5,0.5){\circle*{0.55}}
        \put(2.5,0.5){\circle*{0.55}}
        \put(3.5,0.5){\circle*{0.55}}
        \put(4.5,0.5){\circle*{0.55}}
        \put(5.5,0.5){\circle*{0.55}}
        \put(6.5,0.5){\circle*{0.55}}
        \put(7.5,0.5){\circle*{0.55}}

        \put(1.0,1.5){\circle*{0.55}}
        \put(3.0,1.5){\circle*{0.55}}
        \put(5.0,1.5){\circle*{0.55}}
        \put(7.0,1.5){\circle*{0.55}}

        \put(2,2.5){\circle*{0.55}}
        \put(6,2.5){\circle*{0.55}}

        \put(4,3.5){\circle*{0.55}}

        \color{black}

        \put(3.833,3.40){$+$}

        \put(1.833,2.40){$+$}
        \put(5.833,2.40){$+$}

        \put(0.833,1.40){$+$}
        \put(2.833,1.40){$+$}
        \put(4.833,1.40){$+$}
        \put(6.833,1.40){$+$}

        \put(0.34,0.43){$s_0$}
        \put(1.34,0.43){$s_1$}
        \put(2.34,0.43){$s_2$}
        \put(3.34,0.43){$s_3$}
        \put(4.34,0.43){$s_4$}
        \put(5.34,0.43){$s_5$}
        \put(6.34,0.43){$s_6$}
        \put(7.34,0.43){$s_7$}

    \end{picture}
\end{center}
The number of qubits in the set $Q_t$ is bounded from above by $\|{\cal{C}}\|$
since $Q_t
\subset {\cal{C}}$. Here, ${\cal{C}}$ is any cluster sufficiently
large to carry the network to be simulated. Thus
\begin{equation}
    \label{T_Iup}
    T_{cl,{\bf{I}}}(t) \leq \Delta_{cl}\, \log \|{\cal{C}}\|.
\end{equation}
To determine the proper measurement angle
$\varphi_{k,\mbox{\footnotesize{meas}}}$ for  the measurement on qubit $k \in Q_{t+1}$ in the next measurement
round requires, according to (\ref{setmeasang}), the evaluation of the
symplectic scalar product $({\bf{I}}(t),F_k)_S$. This requires 1 step
for multiplication and $\log 2n$ steps for addition modulo 2. Thus,
\begin{equation}
    \label{T_mang}
    T_{cl,\pm} = \Delta_{cl} \left(\log n +2 \right).
\end{equation}
Combining (\ref{Tcomp_cl/qu}), (\ref{T_Iup}) and (\ref{T_mang}), the total computation time
$T_{\mbox{\footnotesize{comp}}}$ is bounded from above by
\begin{equation}
    \label{T_comp}
    T_{\mbox{\footnotesize{comp}}}\leq
    D\,\Delta_Q\left(1+\frac{\Delta_{cl}}{\Delta_Q} \left[
    \log\|{\cal{C}}\| + \log n +2\right]\right).
\end{equation} 
We see that, although the computation time
$T_{\mbox{\footnotesize{comp}}}$ is linear 
in the logical depth $D$, it contains contributions logarithmic in the
number $n$ of logical 
qubits and in the cluster size $\|{\cal{C}}\|$. These logarithmic
contributions are, however, suppressed by the ratio
between the characteristic  time for classical processing and the
characteristic time for the von-Neumann measurements,
${\Delta_{cl}}/{\Delta_Q}$. This ratio can, in 
practice, be very small. Therefore, the logarithmic corrections become
important only in the limit of large clusters and large $n$. As will
be argued below, even in the regimes where a quantum computer is
believed to become useful, say $n \approx 10^5$, the logarithmic
corrections have only a minor influence on the total computation time. 

We now eliminate the dependence of the total computation time on the
cluster size $\| {\cal{C}} \|$. For this we assume that on the \QC we
simulate a quantum logic network with the network logical
depth $D_{\cal{N}}$. Now, we
give an upper bound on $\| {\cal{C}} \|$ as a function of $n$ and
$D_{\cal{N}}$. As displayed in Fig.~\ref{Gates}, a single CNOT gate 
has height 3 and width 6 on the cluster
${\cal{C}}$. Here we do not count the output qubits of the gates since
they also form the input qubits of the gates in the next slice. As in
Fig.~\ref{Gates}, the rotation 
has height 1  and width 4, if the output qubit is again
not counted for the width. The wires for the 
logical qubits on the cluster can be arranged with distance 2. Each set of
parallelized gates will at most require a slice of width 6 on the
cluster. The circuit as a whole
requires an additional slice of width 
1 for the output.  
A swap-gate that is composed of three CNOT-gates, requires an array of
$3\times18$ qubits on the cluster. If a general CNOT-gate on the
cluster were composed of a next-neighbour CNOT-gate and swap-gates (in
practice it is not), then it would require at most an array of [height] $\times$ [width] =$[2n-1]\,\,\times\,\,[18 (n-2) +6]$ qubits. Hence, a CNOT
gate would, to leading order, consume at most $36n^2$ cluster qubits. Each
rotation would require at most --in the worst case where on the network it
could not be 
performed in parallel with other gates-- a slice of width 4 on
the cluster, so it consumes, to leading order, at most $8n$ cluster
qubits. The total 
number of gates in the network is bounded from above by 
$n\,D_{\cal{N}}$. The simulation of each gate costs at most
$\mbox{max}(36n^2,8n) = 36n^2$ cluster qubits. Hence, the size of the 
required cluster is bounded by
\begin{equation}
    \label{clustsiz}
    \|{\cal{C}}\| \leq 36n^3D_{\cal{N}}.
\end{equation} 
If we now use the assumption about a good quantum
algorithm that the logical depth scales polynomially in the number of
qubits $n$,
\begin{equation}
    \label{goodalg}
    D_{\cal{N}} = c\, n^p,
\end{equation}
and insert (\ref{clustsiz}) and (\ref{goodalg}) into (\ref{T_comp}), we
obtain
\begin{equation}
    \label{specT_comp}
    T_{\mbox{\footnotesize{comp}}} \leq D \, \Delta_Q \left( \left[ 1 +
    \frac{\Delta_{cl}}{\Delta_Q} (4 + \log 9c) \right] +
    \frac{\Delta_{cl}}{\Delta_Q} (p+4) \log n\right).
\end{equation}
From a 
practical point of view, we find that the logarithmic corrections --even
for numbers $n$ of logical qubits in the range of $10^5$-- play a minor
role since they are suppressed by the ratio
${\Delta_{cl}}/{\Delta_Q}$. We could plug in some typical numbers, say 
$\Delta_Q = 1 \;\mu\mbox{s}$, $\Delta_{cl} = 1 \;\mbox{ns}$, $p=3$
and $n=10^5$, to obtain ${\Delta_{cl}}/{\Delta_Q} \,(p+4) \log
n  \approx 0.12$ (or $\approx 0.24$ for $n=10^{10}$). 

The spatial overhead $\|{\cal{C}}\|$ is polynomial in the number $n$
of logical qubits. But,
if one adopts this more practical viewpoint one may 
not be satisfied by the mere result that the spatial overhead scales
polynomially, but might want to
know what the scaling power actually is. Above we found that
$\|{\cal{C}}\|$ scales with the $(p+3)$th power of $n$. However, in the
above argument, we focused on the computation time where the
precise value of exponent for the spatial scaling did not play an
important role, thus
have been extremely wasteful with spatial resources. A more careful discussion
yields a more favourable scaling of the spatial 
overhead. 

From a strict scaling point of view, we find in (\ref{specT_comp}) that
the computation time 
is no longer equal to the logical depth $D$ times a constant, but there are
$\log n$ -corrections due to the classical processing. This is,
as the above numbers illustrate, of little relevance for practical
purposes. The classical processing can be parallelized to such a
degree that it increases the total computation time only marginally.

\section{Discussion}
\label{discussion}

The discussion about the logical depth of certain algorithms with
the \QC in Section~\ref{depth} showed that there are ways of 
information processing possible with the \QC which cannot be explained
adequately in network model terms. This made a computational model
appropriately 
describing the \QC desirable. The computational model underlying the
\QC that we found does not seem to have much in common with the
network model. It is based on objects of a different sort which require
an interpretation.  In this section, we attempt to clarify the
role of the binary valued  information flow vector ${\bf{I}}(t)$ and
that of the stepwise measured quantum state.

What is the role of the information flow vector ${\bf{I}}(t)$? 
In every computational step except the final one the information flow
vector ${\bf{I}}(t)$ is completely random. So one might ask whether it contains
information at all. It does, since in every step except the last one
it tells what has to be done next. After the final computational step at time
$t_{\mbox{\footnotesize{max}}}$ the quantity
${\bf{I}}(t_{\mbox{\footnotesize{max}}})$ contains  
the result of the computation. Thus, the quantity ${\bf{I}}(t)$ has a
meaning in every computational step. In this sense, it represents the
algorithmic information  in the described scheme of quantum computation.

What is the role of the stepwise measured quantum state? To see that
explicitly, let us consider the scenario where a 
quantum computation is halted in the middle and continued 
at a later time by another person who only knows which
steps of the 
computation are left to perform but does not know what has been done
so far. In analogy to a teleportation protocol where both the
result from the Bell measurement and the quantum state at the
receivers side are required to reconstruct the initial state, the
halted computation can be successfully completed
only if both pieces --the intermediate information flow vector
${\bf{I}}(t_i)$ and the half-measured quantum state-- are stored until
the computation proceeds. Thus, the quantum state cannot be neglected
just because it does not appear in the formal description of the
computational model.
The quantum correlations in the stepwise measured state are what
basically enables the described way of quantum information
processing. However, the role of this state
is a passive one. It serves as a resource that is used up during the
course of computation. 

For the description of the \QC there is no necessity to refer
to the ``qubit'' as a basic notion of quantum information theory. As
described above, the stepwise modified quantum state 
attains the role of a consumable resource but not that of the carrier of
algorithmic information. It is thus sufficiently described in
terms of standard quantum mechanics, namely as a quantum state of
entangled two-level systems. The processed algorithmic information is
classical. 

Let us at the end of this discussion come back to the role which the
randomness of the individual measurement results plays for the
\QCns. It may surprise that a set of classical binary numbers
represents the algorithmic information in a scheme of quantum
computation. In the network model the quantum state (of the quantum
register) is usually considered to represent the processed
information. For the \QCns, the situation is different. There, the
randomness of the individual measurement results makes it necessary to
store classical steering information. The need to process this
information has called for a novel information carrying
quantity. What, in a network-like description of the \QCns, has been
regarded as a mere byproduct turns out to be the central quantity of
information processing with the \QCns.
   
\section{Conclusion}
\label{conclusion}

We have described the computational model underlying the
one-way quantum computer, which is very different from the quantum logic
network model.

The formal description of the \QC is based on primitive
quantities of which the most important are the sets $Q_t \subset
{\cal{C}}$ of cluster 
qubits defining the temporal ordering of measurements on the cluster
state, and the binary valued information flow vector ${\bf{I}}(t)$
which is the carrier of the algorithmic information.  

The information processing with the \QC goes beyond mere 
emulation of quantum gates by  sequences of measurements. The
complete description of the computational process on the 
\QC contains the temporal order in which the measurements
are performed, and the most efficient temporal order does not follow
from the rules that apply for the temporal ordering of gates 
in the network model. Thus, the network picture
is insufficient to describe computation with the \QC and is therefore
abandoned. In the proposed scheme the unitary gates
from some universal set can not be taken for the 
elementary building blocks of quantum computation.

As a practical implication of the new model, the logical depth of
certain algorithms 
on the \QC is lower than for their corresponding network
realizations. As shown for circuits of diagonal gates or of CNOT-,
Hadamard- and $\pi/2$-phase gates, the logical depths for the \QC are
constant in the number of gates and logical qubits, namely one and
two. The best  bounds for   
networks that have been known so far scale logarithmically. It
therefore seems that, at least for  
the \QCns, the question of temporal
complexity must be revisited. The tools for this discussion together
with first results have been provided in this paper. 

\section*{Acknowledgements}

This work has been supported by the Deutsche Forschungsgemeinschaft
(DFG) via the \mbox{Schwerpunkt}\-programm QIV. We thank
D. E. Browne, O. Forster, D. P. DiVincenzo, D. W. Leung, A. Winter and
R. Cleve for helpful discussions. 

\newpage

\appendix

\section{Proof: cone test and forward-backward cone relation}
\label{Coneproof}

Here we prove the cone test
(\ref{conecrit}) and the forward-backward cone relation
(\ref{fcbc}). Considering the cone test, first note that whether a one-qubit
rotation  at some position in the network 
is about the $z$-axis or a about the $x$-axis can be identified by the
potential byproduct operator produced when the rotation is
implemented. This can be seen by inspecting (\ref{Euler}),
(\ref{Rotprime}), (\ref{Byprod1}) and the procedure to implement a
general rotation as described in section \ref{summary}. The
$x$-rotations $U_x(\xi)$ and $U_x(\zeta)$ of $U_R(\xi,\eta,\zeta)$ in
(\ref{Euler}) are implemented by measurements on the qubits 2 and 4 of
a 5-qubit chain. As can be seen from (\ref{Byprod1}), they contribute to
the byproduct operator $U_\Sigma$ of the rotation $U_R$ with
${\sigma_x}^{s_2+s_4}$ where $s_2$ and $s_4$ are the results of the
measurements on qubits 2 and 4. Further, the rotation about the
$z$-axis, $U_z(\zeta)$, is implemented by measurement of qubit 3. The
contribution to the byproduct operator which is thereby generated is, from (\ref{Byprod1}), ${\sigma_z}^{s_3}$. We see that $x$-rotations
only generate byproduct operators $\sigma_x$ and $z$-rotations only
generate byproduct operators $\sigma_z$.

A byproduct operator generated via the measurement on the cluster
qubit $k$ must  be propagated either forward or backward to possibly
reach the rotation  
on the logical qubit $i$ implemented via the measurement on the
cluster qubit $j$. Let be ${\cal{O}}_K$ and ${\cal{O}}_J$ 
 two  cuts  
through the network which intersect each logical qubit line only
once. More specifically, ${\cal{O}}_K$ intersects the qubit line $i$
just before the rotation implemented by the measurement at cluster
qubit $k$. ${\cal{O}}_J$ intersects the qubit line $i$ just before the
rotation implemented by the measurement at cluster qubit $j$.

There are two cases which can occur. Either the cut ${\cal{O}}_K$ is
before the cut ${\cal{O}}_J$ in the network ${\cal{N}}$ which we
denote by ${\cal{O}}_K \leq {\cal{O}}_J$, or ${\cal{O}}_J$ is
before the cut ${\cal{O}}_K$ which we
denote by ${\cal{O}}_J \leq {\cal{O}}_K$. It can also be that both is
true at the same time but it cannot be that neither of the two
relations hold.

Case I:  ${\cal{O}}_K \leq {\cal{O}}_J$.\\
The byproduct operator generated via the measurement at qubit $k$ must
be propagated forward to possibly affect the measurement at qubit
$j$. It is not possible that the result of the measurement on qubit
$j$ has an effect on the measurement basis chosen at $k$.

Let us introduce a further cut 
${\cal{O}}_{J^\prime}$ which is the same as ${\cal{O}}_J$, except for
that it  intersects the line of the logical qubit $i$ in the network
${\cal{N}}$ just after the rotation implemented via the measurement on
the cluster qubit $j$. 
The
potential byproduct operator which is generated via the measurement on
cluster qubit $k$ and then propagated forward to the cuts
${\cal{O}}_J$ and ${\cal{O}}_{J^\prime}$, is denoted by
$U_k|_{{\cal{O}}_J}$ and 
$U_k|_{{\cal{O}}_{J^\prime}}$, respectively (the byproduct
operators which are actually generated are
${\left(U_k|_{{\cal{O}}_J}\right)}^{s_k}$ and
${\left(U_k|_{{\cal{O}}_{J^\prime}}\right)}^{s_k}$). Further, we denote the
restriction of the byproduct operators  $U_k|_{{\cal{O}}_J}$ and
$U_k|_{{\cal{O}}_{J^\prime}}$ to the logical qubit $i$ by
$\left[U_k|_{{\cal{O}}_J}\right]_i$ and
$\left[U_k|_{{\cal{O}}_{J^\prime}}\right]_i$. The
two cuts differ only on the logical qubit $i$, and there only by the
side of the  
respective cut on which the rotation is located. Therefore, using
(\ref{Rotprop}), 
it follows  that $U_k|_{{\cal{O}}_J} =
U_k|_{{\cal{O}}_{J^\prime}}$. Hence also 
\begin{equation}
    \label{through}
     \left[U_k|_{{\cal{O}}_J}\right]_i =
     \left[U_k|_{{\cal{O}}_{J^\prime}}\right]_i. 
\end{equation}
If the rotation implemented via the measurement on cluster qubit $j$
is about the $x$-axis, then the measurement on qubit $j$ has to wait
for the measurement on cluster qubit $k$ iff
$\left[U_k|_{{\cal{O}}_J}\right]_i$ contains a contribution
$\sigma_z$. The measurement on $j$ itself produces a potential byproduct
operator $\left[U_j|_{{\cal{O}}_{J^\prime}}\right]_i=\sigma_x$. Similarly, if
the rotation implemented via the measurement on $j$
is about the $z$-axis then the measurement on $j$ has to wait
for the measurement on $k$ iff
$\left[U_k|_{{\cal{O}}_J}\right]_i$ contains a contribution
$\sigma_x$. The measurement on $j$ itself produces a potential byproduct
operator $\left[U_j|_{{\cal{O}}_{J^\prime}}\right]_i=\sigma_z$. 

Because of (\ref{through}) $U_k$ can as well be evaluated at the cut
${\cal{O}}_{J^\prime}$ instead of ${\cal{O}}_J$. The byproduct operator on the
intersection of qubit line $i$ and cut ${\cal{O}}_{J^\prime}$
resulting from the measurement on qubit $j$ can be written in the
form 
\begin{equation}
    \label{Jby}
    \left[U_j|_{{\cal{O}}_{J^\prime}}\right]_i={\left(\sigma_x^{(i)}\right)}^{x_{j,i}}
    {\left(\sigma_z^{(i)}\right)}^{z_{j,i}} \; \mbox{with}\;\;
    \left(\begin{array}{c} x_{j,i} \\ z_{j,i} \end{array} \right) =
    \left\{ \begin{array}{rr} \left(\begin{array}{c} 0 \\
    1 \end{array} \right) & \mbox{for  $z$-rotations} \\
    \left(\begin{array}{c} 1 \\ 
    0 \end{array} \right) & \mbox{for $x$-rotations}
    \end{array}\right. .
\end{equation}
The byproduct operator on the intersection of qubit line $i$ and cut
${\cal{O}}_{J^\prime}$ resulting from the measurement on qubit $k$ reads
\begin{equation}
    \label{Kby}
    \left[U_k|_{{\cal{O}}_{J^\prime}}\right]_i={\left(\sigma_x^{(i)}\right)}^
    {x_{k,i}} {\left(\sigma_z^{(i)}\right)}^{z_{k,i}}.
\end{equation}
One can now easily check for both the cases of an $x$- and a
$z$-rotation implemented by the measurement on qubit $j$ that the
measurement of qubit $j$ must wait for the result of the measurement
of qubit $k$ iff
\begin{equation}
    \label{localtest}
    x_{j,i}z_{k,i} + z_{j,i}x_{k,i} = 1 \; \mbox{mod} \; 2.
\end{equation}
Now we note that the correspondence between $\left(\begin{array}{c}
        x_{j,i} \\ z_{j,i} \end{array} \right)$ and
        $\left[U_j|_{{\cal{O}}_{J^\prime}}\right]_i$; and between
        $\left(\begin{array}{c} 
        x_{k,i} \\ z_{k,i} \end{array} \right)$ and
        $\left[U_k|_{{\cal{O}}_{J^\prime}}\right]_i$ is via the restriction of
        the isomorphism (\ref{Iso}) on qubit $i$. Thus, $x_{j,i},
        z_{j,i}$ are just the $i$-components of
        ${\bf{I}}_x|_{{\cal{O}}_{J^\prime}}$ 
        and ${\bf{I}}_z|{{\cal{O}}_{J^\prime}}$,
        respectively. Equivalent relations hold 
        for $x_{k,i},z_{k,i}$. One finds 
\begin{equation}
    \label{ae}
    \begin{array}{lcl}
        x_{j,i} = \left[I_{x,j}|_{{\cal{O}}_{J^\prime}}\right]_i &,&  z_{j,i} =
        \left[I_{z,j}|_{{\cal{O}}_{J^\prime}}\right]_i \\ 
        x_{k,i} = \left[I_{x,k}|_{{\cal{O}}_{J^\prime}}\right]_i &,&  z_{k,i} =
        \left[I_{z,k}|_{{\cal{O}}_{J^\prime}}\right]_i
    \end{array}.
\end{equation}
Further we observe that
\begin{equation}
    \label{JByislocal}
    \left[I_{x,j}|_{{\cal{O}}_{J^\prime}}\right]_l = 0, \;\;
    \left[I_{z,j}|_{{\cal{O}}_{J^\prime}}\right]_l = 0 \; 
    \mbox{for all}\; l \neq i, 
\end{equation}
since the byproduct operator introduced by the implementation of the
rotation acts, at the cut ${\cal{O}}_{J^\prime}$, non-trivially only on the
logical qubit $i$. Thus we can write
\begin{equation}
    \label{globaltest}
    \begin{array}{rcl}
          x_{j,i}z_{k,i} + z_{j,i}x_{k,i} &=& \displaystyle{\sum\limits_{l=1}^n
          x_{j,l}z_{k,l} + z_{j,l}x_{k,l}}\\ 
          &=& \displaystyle{\left({\bf{I}}_j|_{{\cal{O}}_{J^\prime}},
              {\bf{I}}_k|_{{\cal{O}}_{J^\prime}}  \right)_S} \\
          &=& \displaystyle{\left({\bf{I}}_j|_\Omega,
          {\bf{I}}_k|_\Omega  \right)_S} 
          \\ 
          &=& \displaystyle{\left({\bf{F}}_j,{\bf{F}}_k  \right)_S}, 
   \end{array}
\end{equation}
where the second line holds by the definition (\ref{sympscalp}) and
the third by (\ref{Iprop}) and the 
conservation (\ref{conssympprod}) of the symplectic scalar product. 
Inserting (\ref{globaltest}) into (\ref{localtest}) yields
\begin{equation}
    \label{case1}
    {\cal{O}}_K \leq {\cal{O}}_J:\; \;j \in \fc(k) \Longleftrightarrow
    \left({\bf{F}}_j,{\bf{F}}_k\right)_S=1 .
\end{equation}
For ${\cal{O}}_K \leq {\cal{O}}_J$, $j \in \mbox{bc}(k)$ cannot occur,
hence with (\ref{case1}),
\begin{equation}
    \label{case1e}
    {\cal{O}}_K \leq {\cal{O}}_J:\; \;j \in \fc(k) \vee j \in \bc(k)
    \Longleftrightarrow 
    \left({\bf{F}}_j,{\bf{F}}_k\right)_S=1 .
\end{equation}

Case II: ${\cal{O}}_J \leq {\cal{O}}_K$.\\
First we observe that $j$ can only be in the backward cone of $k$, but
not in the forward cone. Thus, the
byproduct operator generated via the measurement on $k$ must be
propagated backwards in the network to reach the gate for whose
implementation qubit $j$ is to be measured.  The reasoning is
completely analogous to case I, up to the fact that the potential
byproduct operator generated via the measurement of cluster qubit $k$
is in this case propagated backwards onto the cut
${\cal{O}}_{J^\prime}$. Qubit $j$ is in 
the backward cone of qubit $k$ iff the quantity
$\left({\bf{I}}_j|_{{\cal{O}}_{J^\prime}}, 
    {\bf{I}}_k|_{{\cal{O}}_{J^\prime}}  \right)_S$ is equal to 1. Again, by
conservation (\ref{conssympprod}) of the symplectic scalar product
follows
\begin{equation}
    \label{case2}
    {\cal{O}}_J \leq {\cal{O}}_K :\; \; j \in \bc(k) \Longleftrightarrow 
    ({\bf{F}}_j,{\bf{F}}_k)_S=1.
\end{equation}
For  ${\cal{O}}_J \leq {\cal{O}}_K$, $j \in \mbox{fc}(k)$ cannot
occur, and therefore with (\ref{case2}),
\begin{equation}
    \label{case2e}
    {\cal{O}}_J \leq {\cal{O}}_K:\; \;j \in \fc(k) \vee j \in \bc(k)
    \Longleftrightarrow 
    \left({\bf{F}}_j,{\bf{F}}_k\right)_S=1 .
\end{equation} 
Now we combine the two cases and with (\ref{case1e}) and
(\ref{case2e}) we obtain 
\begin{eqnarray}
    k \in {\cal{C}}, j \in  Q^{(1)}: \;\; j \in \fc(k) \; \vee
    \; j \in \bc(k) \Longleftrightarrow 
    \left({\bf{F}}_j,{\bf{F}}_k\right)_S=1,  \nonumber
\end{eqnarray}
which proves the cone test (\ref{conecrit}). 

The proof of the cone test for gates (\ref{gateconecrit}) goes along
the same lines, only the byproduct operator $(U_k)^{s_k}$ generated
via the measurement at cluster qubit $k \in {\cal{C}}$ has to be
replaced with the byproduct operator $U_0(g)$ of the gate $g$. 

Finally, the proof the forward-backward cone relation shall be
outlined.
Suppose that $j \in \fc(k)$. With the same methods as used in the
proof of (\ref{conecrit}) one can derive that
\begin{equation}
    \label{2cuts}
    \begin{array}{lcl}
        j \in \fc(k) &\Longleftrightarrow&
        ({\bf{I}}_k|_{{\cal{O}}_J},{\bf{I}}_j|_{{\cal{O}}_J})_S =1, \\
        k \in \bc(j) &\Longleftrightarrow& 
        ({\bf{I}}_k|_{{\cal{O}}_K},{\bf{I}}_j|_{{\cal{O}}_K})_S =1.
    \end{array}
\end{equation}
Then, with (\ref{2cuts}) and the invariance (\ref{conssympprod}) of
the symplectic scalar product
\begin{eqnarray}
    j \in \fc(k) &\Longleftrightarrow&  k \in \bc(j), \nonumber
\end{eqnarray}
which proves (\ref{fcbc}).

\section{Relation among $\kappa_{k,I}^\prime$, $\kappa_{k,O}^\prime$
  and $\kappa_{k}^\prime$}
\label{Kaprel}   

Let us briefly explain
why the first line of eq.~(\ref{kapparel1}) holds. Be the left
neighbor of the qubit $k \in 
{\cal{C}}_O(g) \cap {\cal{C}}_I(\tilde{g})$ in question
denoted by $k-1$ and the right neighbor by $k+1$. The interaction
$S_N$ takes a product state
$|P\rangle_{{\cal{C}}_N}=\bigotimes_{i \in {\cal{C}}_N} 
|+\rangle_i$ to a cluster state $|\phi\rangle_{{\cal{C}}_N}$. Since all
pair-interactions in the Hamiltonian generating $S_N$ commute, $S_N$ can
be written 
in the form $S_N=S_{k-1,k}\,S_{k,k+1}\, S^\prime$, where $S_{k-1,k}$
describes the pair-interaction between $k-1$ and $k$,  $S_{k,k+1}$
the pair-interaction between $k$ and $k+1$, and $S^\prime$
all the remaining pair-interactions. One can easily check
that 
\begin{equation}
    \label{konj}
    \begin{array}{rcl}
        S_N \,\sigma_x^{(k)}\, S_N^\dagger &=& {(-1)}^\alpha \sigma_z^{(k-1)}
        \sigma_x^{(k)} \sigma_z^{(k+1)}\\
        S_{k-1,k}\,\sigma_x^{(k)}\, S_{k-1,k}^\dagger &=& {(-1)}^\beta
        \sigma_z^{(k-1)} \sigma_x^{(k)} \\ 
        S_{k,k+1}\,\sigma_x^{(k)}\, S_{k,k+1}^\dagger &=&
        {(-1)}^\gamma  \sigma_x^{(k)} \sigma_z^{(k+1)}\\
         S_{k-1,k}\,\sigma_z^{(k+1)}\, S_{k-1,k}^\dagger &=&
        \sigma_z^{(k+1)},
    \end{array}
\end{equation}    
with $\alpha,\beta,\gamma \in \{0,1\}$. Since the state
$|P\rangle_{{\cal{C}}_N}$ obeys 
the eigenvalue equation $|P\rangle_{{\cal{C}}_N} = \sigma_x^{(k)}
|P\rangle_{{\cal{C}}_N} $ the
state $|\phi\rangle_{{\cal{C}}_N}$ obeys $|\phi\rangle_ {{\cal{C}}_N}= S_N
    \,\sigma_x^{(k)}\, S_N^\dagger \, |\phi\rangle_{{\cal{C}}_N}$. Thus,
via eqs.~(\ref{EVeqn}) and (\ref{konj}) we can identify
$\alpha=\kappa_k^\prime$. Further holds
\begin{equation}
    \label{decompS}
    S_N\,\sigma_x^{(k)}\,
    S_N^\dagger=S_{k-1,k} S_{k,k+1}  \,\sigma_x^{(k)}\, S_{k+1,k}^\dagger
    S_{k,k-1}^\dagger = {(-1)}^{\beta+\gamma} \sigma_z^{(k-1)}
    \sigma_x^{(k)} \sigma_z^{(k+1)},
\end{equation} 
such that
$\alpha=\kappa_k^\prime=\beta + \gamma$. For the cluster state
$|\phi\rangle_{{\cal{C}}(g)}$ on the cluster ${\cal{C}}(g)$ where only
the qubits $k-1$ and $k$ are present, but not qubit $k+1$, there holds
$|\phi\rangle_{{\cal{C}}(g)}= S_{k-1,k} \sigma_x^{(k)}
S_{k-1,k}^\dagger \, |\phi\rangle_{{\cal{C}}(g)}$. Via
eqs.~(\ref{EVeqn}) and (\ref{konj}) we can thus identify
$\beta=\kappa_{k,O}^\prime$. In the 
same way, if we look at the cluster state
$|\phi\rangle_{{\cal{C}}(g^\prime)}$ on the cluster ${\cal{C}}(g^\prime)$
we can identify $\gamma=\kappa_{k,I}^\prime$. Combining the relations
for $\alpha,\beta,\gamma$, we finally obtain
$\kappa_k^\prime=\kappa_{k,I}^\prime + \kappa_{k,O}^\prime$ as in
eq.~(\ref{kapparel1}). 

\section{Temporal complexity of the classical pre-processing}
\label{preproc}

The time that it takes for a classical computer (i.e. a compiler) to
translate an algorithm into a machine-specific set of operations
(i.e. the machine code) is usually not
regarded as to count for the temporal complexity of that
algorithm. For quantum logic networks this viewpoint is certainly
justified because   
there the complexity to compute the circuit layout
is well understood and known not to exceed the complexity of the
quantum logic network itself. 

For the \QC however, the situation is less clear. For the time being,
we do not know of any other way to obtain the quantities
characterizing a 
\QCns-computation than to derive them from the
network formulation of the respective algorithm. Therefore, we
must exclude the possibility that for the \QC the algorithmic
complexity of a quantum computation is shuffled from the
genuine quantum part of the computation to the classical
pre-processing, and that this classical pre-processing may be
exponentially hard. As will be
shown below, such a case does not occur. All 
the classical pre-processing can be done in polynomial time.

To see this, we assume that the quantum algorithm on $n$ logical
qubits is given as a
sequence of $\|{\cal{N}}\|$ elementary gates. For good quantum
algorithms, $\|{\cal{N}}\|$ is polynomial in $n$, as is
$\|{\cal{C}}\|$, the number of physical qubits in the cluster
${\cal{C}}$ required to run the algorithm (see Section~\ref{Tcomp}).

The layout of the measurement pattern requires to assign $\|Q_0\|$
measurement bases and $\|{\cal{C}} \backslash Q_0\|$ angles. Creating
the pattern is for itself not a problem since it can be
obtained by patching together the measurement patterns of the
elementary gates which are available in block form. The temporal complexity for
this step is thus $O(\|{\cal{C}}\|)$. 

To obtain the byproduct images we introduce $\|{\cal{N}}\|$ vertical cuts
${\cal{O}}_i,\; i=1,..,\|{\cal{N}}\|$ to the network, one after each
gate (such that ${\cal{O}}_{\|{\cal{N}}\|} = \Omega$) and compute the
  $2n\times2n$-matrices $C({\cal{N}}_{{\cal{O}}_i \rightarrow
    \Omega})$ for $i=1,..,\|{\cal{N}}\|-1$, starting with
  $i=\|{\cal{N}}\|-1$. The
  operational effort for 
  this is of the order $O(n^3\|{\cal{N}}\|)$. By use of these matrices
  the byproduct images for cluster qubits $k \in {\cal{C}}\backslash
  Q_{0,z}$ can now be obtained via (\ref{Iprop}), which requires
  $O(n^2)$ elementary operations per byproduct image. The way to obtain the
  byproduct images ${\bf{F}}_g$ of the gates is the same. For $k \in
  Q_{0,z}\backslash O$ at most four byproduct images have to be added
  in (\ref{F0z}), which requires $O(n)$ operations. The computation of
  ${\bf{F}}_k$ for $k\in O$ is trivial. Thus, to compute a byproduct
  image requires at most $O(n^2)$ operations per cluster qubit or gate
  such that the complexity
  to compute all of them is at most $O\left(n^2 ( \|{\cal{C}}\| +
      \|{\cal{N}}\|)\right)$.

The backward- and forward cones of the cluster qubits $k \in {\cal{C}}$ are
computed using the temporal ordering of gates in a sequence representing
the quantum logic network and the cone test (\ref{conecrit}). The
number of cone tests that have to be performed in each case is $\|{\cal{N}}\|
(\|{\cal{N}}\|-1)/2$ where the computational effort for each test
scales like $O(n)$. Thus, the complexity of this step is $O(n
\|{\cal{N}}\|^2)$.  

The forward cones generate the anti-reflexive semi ordering
``$\prec$''. The semi ordering can be computed from them in
$O(\|{\cal{C}}\|^5)$ steps.

For each set $Q_t$ there have to be $\|Q^{(t)}\| \leq \|{\cal{C}}\|$
test of the relation $j \prec k,\,\, j \in Q^{(t)}$ performed to check
whether some qubit $k \in Q^{(t)}$ is in 
$Q_t$. Also, $\|Q^{(t)}\|$ qubits have to be checked for each
$Q_t$. At most $\|{\cal{C}}\|$ sets $Q_t$ exist such that the
operational effort to obtain the these sets is $O(\|{\cal{C}}\|^3)$.

As far as the stated upper bounds are conclusive, it looks as if the
computation of the anti-reflexive semi ordering is the toughest
part. However, as elementary a relation between the cluster qubits
``$\prec$'' is, for the conversion of a quantum logic network into a
\QCns-algorithm it needs not be computed. Please note that the semi
ordering is finally only needed to compute the sets $Q_t$ via
(\ref{Qrecur}). But instead of computing ``$\prec$'' from the forward
cones and the sets $Q_t$ from ``$\prec$'', the sets $Q_t$ can also be
computed from the forward cones directly. For this, please note that
$\exists j \in Q_t |\,\, j \prec k \in Q^{(t)} \Longleftrightarrow
\exists j^\prime \in Q_t |\,\, k \in \mbox{fc}(j^\prime)$. The
direction ``$\Longleftarrow$'' is obvious with $j=j^\prime$. The
opposite direction, ``$\Longrightarrow$'', holds by an argument
analogous to the one justifying (\ref{Impl2}). In fact, statement
``$\Longrightarrow$'' is the same as (\ref{Impl2}) with $Q^{(1)}$
replaced by $Q^{(t)}$. Thus, eq. (\ref{Qrecur}) can be replaced by 
\begin{equation}
    \label{replQrecur}
    Q_t = \left\{k \in Q^{(t)}|\,\,\neg \exists j^\prime \in
    Q^{(t)}: \,\,k \in \mbox{fc}(j^\prime) \right\}.
\end{equation}  
To set the algorithm angles via (\ref{Algoang}), (\ref{eta}) requires
at most $\|{\cal{C}}\| + \|{\cal{N}}\|$ additions per angle and there
are at most $\|{\cal{C}}\|$ such angles. Hence, in total it takes
$O\left( \|{\cal{C}}\| (\|{\cal{C}}\| + \|{\cal{N}}\|) \right)$ operations
to set them. Finally, to initialize the information flow vector via
(\ref{I_init}) requires $O\left( n (\|{\cal{C}}\| + \|{\cal{N}}\| )
\right)$ operations. Thus we see that all classical processing
requires only a polynomial overhead of elementary operations and can
therefore be
done in polynomial time.

\end{document}